\documentstyle[epsfig]{espart}

\newcounter{bild}
\font\menge msbm10 at12pt

\def\3{\ss}

\newcommand{\av}{\mathop{\rm av}}
\newcommand{\trafo}{\mathop{\longrightarrow}}

\newcommand{\A}{{\cal A}}
\newcommand{\D}{{\cal D}}
\newcommand{\R}{{\cal R}}
\newcommand{\T}{{\rm T}}

\newcommand{\be}{\begin{eqnarray}}
\newcommand{\ee}{\end{eqnarray}}
\newcommand{\nn}{\nonumber \\}

\newcommand{\dmu}{\partial_\mu}
\newcommand{\dnu}{\partial_\nu}
\newcommand{\drho}{\partial_\rho}

\begin{document}

%
{\pagestyle{empty}
\begin{center}
  \vspace*{1cm}
  {\bf 
  {{\huge Blockspin transformations \\
     for finite temperature field theories\\ 
     with gauge fields\\}
  \vspace*{5cm}
  {\large Dissertation\\
  zur Erlangung des Doktorgrades\\
  des Fachbereichs Physik\\
  der Universit\"at Hamburg\\[3cm]
  vorgelegt von\\
  Ute Kerres\\
  aus Dortmund\\[3cm]
  Hamburg\\
  1996\\ \quad}}}
\end{center}

\newpage  

\vspace*{\fill}
\begin{tabbing}
\hspace*{7cm}\= \\
Gutachter der Dissertation: 
\> Prof.\,Dr.\,G.\,Mack\\
\> Prof.\,Dr.\,W.\,Buchm\"uller\\
\> \\
Gutachter der Disputation: 
\> Prof.\,Dr.\,G.\,Mack\\
\> PD.\,Dr.\,K.\,Jansen\\
\\
Datum der Disputation: \> 11.\,Juli 1996\\
\> \\
Sprecher des Fachbereichs Physik\\
und Vorsitzender des \\
Promotionsausschusses:\> Prof.\,Dr.\,B.\,Kramer
\end{tabbing}
\newpage

\centerline{\bf Abstract}

A procedure is proposed to study quantum field theories at zero or at 
finite temperature by a sequence of real space renormalization group (RG) or 
blockspin transformations. 
They
transform to effective theories on coarser and coarser lattices.
The ultimate aim is to compute constraint effective 
potentials, i.e. the free energy as a function of suitable order parameters.
From the free energy one can read off the thermodynamic behaviour 
of the theory, in particular the existence and nature of phase transitions. 

In a finite temperature field theory one begins with either one or a sequence 
of transformations which transform the original theory into an effective 
theory on a three-dimensional lattice. 
Its effectiv action has temperature dependent coefficients. 
Thereafter one may proceed with further blockspin transformations of
the three-dimensional theory. 
Assuming a finite volume,
this can in principle be continued until one ends with a lattice with a single
site. 
Its effective action is the constraint effective potential. 

In each RG-step, an integral over the high frequency part of the field, 
also called the fluctuation field, has to be performed. This is done 
by perturbation theory. It requires the knowledge of bare fluctuation field 
propagators and of interpolation operators which enter into the vertices. 
A detailed examination of these quantities is presented for scalar 
fields, abelian gauge fields and for Higgs fields, finite
temperature is admitted.

The lattice perturbation theory is complicated because the bare lattice
propagators are complicated. 
This is due to
a partial loss of translation invariance in each step.  
Therefore the use of translation invariant cutoffs 
in place of a lattice is also discussed. 
In case of gauge fields this is only possible as a continuum version
of the blockspin method.

\newpage

\centerline{\bf Zusammenfassung}

Es wird ein Verfahren vorgestellt, Quantenfeldtheorien auch bei endlicher
Temperatur durch eine Folge von Blockspin Renormierungsgruppen 
Transformationen zu studieren.
Dabei erh\"alt man effektive Theorien, die auf immer gr\"oberen Gittern leben.
Das letztendliche Ziel ist die Berechnung des {\it constraint
effective potential},
d.h. der freien Energie als Funktion von geeigneten Ordungsparametern.
An der freien Energie l\"asst sich das thermodynamische Verhalten 
der Theorie ablesen, insbesondere die Existenz und Ordnung von 
Phasen\"uberg\"angen.

Bei endlicher Temperatur startet man mit einer Transformation
oder einer Folge von Transformationen,
die die urspr\"ungliche Theorie in eine effektive drei-dimensionale 
Gittertheorie \"uberf\"uhrt. 
Deren effektive Wirkung hat temperaturabh\"angige 
Koeffizienten.
Anschlie\3end kann man mit weiteren Blockspin Transformationen fortfahren.
Im Fall endlichen Volumens kann dies solange fortgesetzt werden bis
man bei einem Gitter mit nur noch einem Gitterpunkt landet.
Die zugeh\"orige  effektive Wirkung ist das {\it constraint
effective potential}.

In jedem RG-Schritt m\"ussen die hochfrequenten Feldanteile, das
sog. Fluktuationsfeld, im Funktionalintegral ausintegriert werden.
Dies wird hier st\"orungs\-theoretisch gemacht, und erfordert die Kenntnis
der nackten Feldpropagatoren und des Interpolationsoperators, der in die
Vertices eingeht.
Eine detailierte Untersuchung dieser Gr\"o\3en wird f\"ur das skalare
Feld, das abelsche Eichfeld und f\"ur das Higgs-Feld vorgestellt,
einschlie\3lich der Situation bei endlicher Temperatur.

Die Gitterst\"orungstheorie ist kompliziert, weil die nackten
Gitterpropagatoren kompliziert sind.
Dies resultiert aus dem Verlust der vollen Translationsinvarianz bei
den einzelnen RG-Schritten.
Es werden deshalb auch translationsinvariante {\it cutoff}-Verfahren
anstelle der Gitterregularisierung untersucht.
Im Falle von Eichfeldern ist das nur als eine Kontinuumsversion des
Blockspinverfahrens m\"oglich.

\newpage}

\tableofcontents

\newpage

\section{Introduction}
\label{secintro}

The possibility of restoration of spontaneously broken symmetry in the
electroweak theory at high temperature has recently led to a renewed
interest in the electroweak phase transition \cite{kapusta}.

Since the seminal and pioneering paper by Kirzhnits and Linde
\cite{kirzhnits}, considerable effort has been devoted to describe the
rather involved processes occurring very close to the critical
temperature.  The effective potential has shown great usefulness.  It
gives the free energy as a function of the magnetization.  There are
perturbative computations combined with $1/N$ expansions for the
effective potential \cite{dolan}, and also numerical results based on
Monte Carlo simulations and on three dimensional reduced actions
\cite{baig}.

Nevertheless the perturbative calculations have shown to be plagued
with problems which are to a large extent due to infrared divergences
and are manifested in the appearance of spurious complex terms in the
expansion for the effective potential. A great amount of work has been
done to obtain meaningful expressions very close to the critical
surface. Most of these attempts are related to the resummation of an
infinite number of daisy and superdaisy diagrams which have to be
taken into account in a consistent perturbative expansion, because
they are of the same order of magnitude in the coupling constants
\cite{takahashi}.

Another very successful approach was proposed by Buchm\"uller et al.
\cite{buchmuller}, who use an improved perturbation expansion for the
effective potential, where the dynamically generated plasma masses are
already included in the corresponding expressions from the beginning.
These plasma masses are computed up to one loop order by
self-consistent gap-equations, and have the effect of damping the
infrared divergences. In this context, in \cite{hebecker} the absence
of linear terms in the scalar field of the effective potential was
explicitly verified for the abelian Higgs model, up to the order $e^4$
and $\lambda^2$, and proved to remain valid at higher orders. Such a
linear term contains spurious infrared divergent contributions.

Despite the important and promising contributions to the understanding
of the nature of the phase transition that have already been made,
there are still ambiguities to be explained and higher order
corrections to be included consistently to ensure the survival of the
predictions to all orders.

We propose an alternative variational two-step method to study phase
transitions at finite temperature in the electroweak theory, and in
related models. It consists first in the calculation of a coarse
grained free energy by perturbative methods, obtaining a three-dimensional
perfect lattice action as a function of a block spin field.  Their
coefficients depend explicitly on the temperature. In a second step,
the Feynman-Bogoliubov method can be used with this three-dimensional
lattice action to obtain the best quadratic approximation to the
perfect action, and therefore the best split into a free part and an
interaction.  Using this, the constraint effective potential can be
computed.  To lowest order, the masses are obtained in this way as
solutions of a gap equation whose formal solution is the sum of
superdaisy diagrams for the three-dimensional lattice theory.  They depend
on temperature.  Higher order corrections can be computed in
principle.  These masses are to be inserted into the expression for
the constraint effective potential.

This procedure separates the UV- and IR-problems in a clean way.  It
allows to study the perfect three-dimensional lattice action directly by
numerical simulations.  One could for instance compute the constraint
effective potential numerically.

The paper is organized as follows. 
In section \ref{secprinciples} a general introduction to the idea of
effective actions is given.
In section \ref{secscalar} we introduce and compare three methods to
separate a scalar field in a high and a low frequency part.
The hard-soft invariance fixing method and the block spin
transformation for the scalar fields at zero temperature are discussed
in detail.
The evaluation of the perfect scalar action at T=0 is explained.
In section \ref{secmaxwell} we extend the
procedures to the Maxwell theory at zero temperature.  
The block spin of Ba{\l}aban and Jaffe for the abelian gauge field is
used \cite{balaban}.
The corresponding averaging operator and gauge fixing  are discussed.
A definition of the perfect lattice
action appropriate for perturbative evaluation is given.
In section \ref{secsqed} the procedure is extended to scalar
electrodynamics.
In section \ref{sectemper} the extension to
finite temperature is outlined. 
The connection to dimensional reduction is explained. 
In section \ref{secfourier} most of the 
quantities which are needed for perturbative calculations are
exhibited in momentum space.
In section \ref{secperturbation} an explicit
computation of the leading terms of the perfect action for the 
$\phi^4$-theory is shown.
It is explained how to extend the results to finite temperature. 
The Feynman-Bogoliubov method is introduced and applied to the just
obtained effective action. 
In section \ref{secconclusion} and \ref{secoutlook} we close with conclusion
and outlook.

\section{Effective actions}
\label{secprinciples}

A low energy effective action is an approximative action designed to
describe low energy phenomena.
For some effective actions it is not even possible to calculate
high energy observables, since the theory developes divergencies.
In any case from a certain energy onwards the predictions of an
effective action cannot be expected to agree with the experimental results.
This sets the range of validity of the effective action.
In the following we will always speak of a cutoff energy above which
the effective action ceases to be valid.

One example for an effective action is the Fermi theory of weak interaction.
It is valid only for energies so low, that one does not see the
intermediate vector bosons, i.e. the cutoff is of the order of the W
mass.

As one can see at this example an effective theory does not need to
be renormalizable since it is not expected that the theory is
valid for all energies.
In general the cutoff can not be sent to infinity, because that
would mean that the effective action describes the physics correctly
for all energies.
The degree of non-renormalizability of an effective action sets
inherently the scale of its validity.
A renormalizable effective action could in principle be valid up to
infinite energies.
It could be a fundamental action.
A non-renormalizable effective action will develop divergencies in
loop-integrals.
If the cutoff is very low the structure of the theory should already
be revealed in tree approximation because the phase space for the loop
integrals is small.

An effective action needs to be strictly local.
This shows that the appropriate degrees of freedom are chosen to
describe the physical phenomena.
The notion of locality depends on the energy scale via Heisenberg's
uncertainty principle.
If higher energies are used to probe the objects, the resolution in
space is increased.
This is an experimental possibility to set the scale of validity. 
Fermi's theory of weak interaction makes sense up to those energies,
where one starts to observe non-local interactions of the fermions.
Then one has to change the description, i.e. to introduce new degrees
of freedom, the intermediate vector bosons, which propagate  
and mediate an interaction over a distance which is given by the 
inverse of their mass.

The new action can again be an effective action valid up to a higher
cutoff.
This process can be iterated and by no experiment one can be forced to
choose an action which is valid for all energies since experiments are
always done at finite energy.
So the concept of an effective action is very natural.

There are two pricipally distinct ways to obtain a low energy
effective action.
The above description shows how effective actions are set up to explain 
the experimental data up to a cutoff.
New experiments at higher energies require new effective actions valid
up to a higher cutoff.
An example for this approach is the developement of a theory for the 
weak interaction.

Our aim is to proceed from larger to smaller energies.
We want to derive an effective action from a more fundamental one.
This more fundamental action can be an effective action with a higher
cutoff or a truly fundamental action valid for all energies.
Then it must be renormalizable and live on the continuum.
In the following we will call the lower energy action effective and
the higher energy action fundamental, not distinguishing whether it is
truly fundamental or itself an effective action.
 
In theories which are not asymptotically free, a genuine continuum theory may
not exist. In this case one has to start with a theory with a UV-cutoff.
But if the cutoff is high enough, it is still possible to consider
effective actions
which describe the physics at much larger length scales than the inverse
UV-cutoff, and which are approximately independent of the original
UV-cutoff. 

The process of lowering the energy consists in two parts.
First we have to decide which degrees of freedom we want to use in
the low energy action. 
Afterwards we have to get rid of all the others.

This approach to effective field theories was introduced into physics
by Wilson \cite{wilson}.
It was first applied to statistical physics problems and afterward to
Euclidian field theories.
The blockspin approach introduced by Kadanoff is a very intuitive way to
introduce the low energy degrees of freedom.
Blockspins are averages of the fundamental fields over a certain
space-time volume.
They are defined on a (coarser) lattice than the fundamental fields.
By this averaging process small scale fluctuations which correspond to
high frequencies are eliminated.

If the fundamental theory is a
continuum theory blockspins can be defined on the continuum, following
Wetterich.
The resulting effective action is then called an average action
\cite{wetterich}.

First we concentrate on the problem of choosing the low energy
degrees of freedom.
One has to determine the energy separating the two regimes and to
choose the type of low energy field.
This determines of course the type of the high energy field.

As mentioned above the effective action must be local.
Grabowski has shown that the adequate choice of the low energy
degrees of freedom is crucial to obtain locality. 
To choose the low energy field of the same type as the fundamental
field is not necessarily a good choice.
If the low energy behaviour of the theory is characterized by bound
states of the fundamental degrees of freedom, the low energy field
must be such a bound state.
If one ignores this the low energy effective action will 
develop nonlocalities.
This nonacceptable behaviour is a hint to the wrong choice of the low
energy field.
Therefore bound states are offered as low energy degrees of freedom to
the theory at each separation step.
If the effective action stays local if the bound states are integrated
out again, one removes them.
Otherwise they are kept \cite{grabowski}.

On the other hand we want to be able to have low energy degrees of
freedom of the same type as the fundamental fields.
This means especially that an internal symmetry of the fundamental
action must not be broken by the choice of the low energy fields.
This is of special importance if one starts with a truly fundamental
theory.
The very first step of cutoff `lowering' is here the intoduction of a
cutoff at all, i.e. a regularization of the theory.
To be forced to change the type of field in this process means
that the fundamental theory cannot be regularized.
 
Connected with these considerations is the question whether the
separation of the field in low energy part and fluctuation part is
performed in momentum or in coordinate space.
In momentum space a sharp or mollified momentum cutoff is introduced.
An infinitesimal cutoff lowering is possible thereby
introducing a small parameter in which one can expand.
In real space the low energy degrees of freedom are blockspins
living on a lattice.
Of course it is possible to go from momentum to coordinate space and
vice versa with Fourier transformations.
But in one of them the low energy field definition looks more natural.
For chiral fermions it is not possible to use the lattice as a
regulator without breaking the symmetry or introducing doublers.
Here a momentum space formulation is adequate.
 
After we have decided which part of the fundamental field we want to
keep, we need to integrate the other degrees of freedom.

Let us first suppose that this can be done exactly.
Assuming that we introduce also effective observables 
the effective theory is equivalent to the fundamental one.
It does not matter whether a low energy observable is calculated
with the effective or fundamental action.
There is no need to remove the momentum cutoff or to take the
continuum limit.
By integration of the fluctuation field the effective theory will gain
new interactions
which take into account the effect of the high frequency modes no more
explicitly present in the effective theory.
All UV-divergencies of the fundamental theory will show up in this
part of the calculation.
Therefore some of the effective couplings could be divergent.
This must be avoided by adjustment of the bare parameters of the
fundamental theory in such a way that the observables of the effective
theory are finite.
The fact that the adjustment of a finite number of bare parameters
ensures the convergence of all observables is due to the
renormalizability of the fundamental theory.

All effects of the high energy physics are now contained in the infinitely
many coupling constants of the effective action.
Since we can not demand the effective action to be renormalizable none
of these couplings are ruled out for dimensional reasons.
This does not result in divergencies, since the effective action 
is regulated by the cutoff.
But any symmetry of the fundamental action which is preseved in the
choice of the low energy field restricts the possible couplings in the
effective action.

The procedure to calculate an effective action from a (more)
fundamental one is called a renormalization group step.
If one repeats these steps one obtains a renormalization group flow of
effective actions. 
The physics is invariant under these renormalization group
transformations.
The renormalization group comes in different disguises.
If one separates high and low energy regime in real space with the
help of coarser and coarser lattices, it is called Wilson's real space
renormalization group \cite{wilson}.

In momentum space the cutoff can be lowered by an infinitesimal
amount.
The fluctuation integral can then be approximated by the leading term
of a loop expansion.
The RG equation can be differentiated with respect to
the cutoff lowering and one
obtains Polchinski's renormalization group equation \cite{polchinski}.

Starting with a continuum theory exact integration results in a
lattice theory without lattice artefacts.
This is called a perfect lattice action.
Observables calculated with the perfect lattice action are the
continuum values.
One does not have to take the continuum limit anymore.

We use the terminology `perfect action' as a synonym for any
accurate and manageable approximation to effective actions in 
the sense of Wilson.
Hasenfratz and Niedermayer consider special approximations 
which are designed to be accurate near a fixed point. 

In interacting theories it is not possible to do the
fluctuation integral exactly.
There are three steps to approximate it.

The first is very crude.
One ignores the fluctuation integral completely.
This is an approximation to zeroeth order in the
fluctuation propagator.
It is sufficient if one wants to use the effective theory as a regulated
version of the fundamental action to perform calculations, and if one
intends to remove the regulator afterwards.

An example for this is lattice regularization.
The low energy degrees of freedom are the blockspins living on a lattice.
The Laplacian is discretized using lattice derivatives, thereby next neighbour
interactions are introduced.
Lattice observables can be calculated numerically.
They differ from their continuum counterparts by lattice artefacts.
This happens, because the fluctuation field is ignored. 
If one uses the lattice as a method of regularization, one has to take
the continuum limit afterwards.
The physical correlation length is finite and so is the lattice
spacing, however small it is.
Hence one has to choose such lattice couplings for the simulation
which render an infinite lattice correlation length.
But with growing lattice correlation length it takes longer and longer
to obtain
statistically independent configurations for Monte Carlo measurements.
One is plagued by critical slowing down.

Another example of the same kind is
regularization by a momentum cutoff.

The next level of sophistication is to treat the fluctuation integral 
perturbatively.
Thereby high energy effects are introduced to the effective action
as modifications of the values of the coupling constants or as
additional couplings.
For lattice actions this is in the spirit of Symanzik's improvement
program which is designed to remove the lowest order lattice artefacts
from the observables by adding corrections to the lattice Laplacian.
In the continuum these alterations amount to the addition of higher
derivatives to the kinetic operator.

Expansion can be done in the fundamental couplings, in the
fluctuation propagator or in loop order.

In place of perturbation theory as will be used here, 
one can also use saddle point
approximations which amount to a partial resummation of perturbation theory.
This method is described in \cite{yoma} for zero temperature, and can be
adapted to the finite temperature situation in the same way as in the present
paper. 
Some numerical results are also found in \cite{yoma}.
 
The fluctuation integral is IR-regulated by the same cutoff which
UV-regulates the effective action.
This facilitates the application of non-perturbetive methods.
In particular one may sum certain classes of subdiagrams to infinite order.
For example the Feynman-Bogoliubov method sums up all superdaisy
diagrams.

\section{Scalar theory}
\label{secscalar}

\subsection{The scalar action}
\label{ssscalaraction}

Before we start this program we review various procedures to introduce
the low energy field.
After explaining the relations between the different methods we 
study which procedures can be generalized to gauge fields.
A scalar theory
without internal symmetries, i.e. $\phi^4$-theory, serves as a playground.
The Euclidian action we consider in our short review is 
\be
  S[\phi] = \frac12 \langle \phi,v^{-1} \phi\rangle  +V[\phi] \quad ,
\ee
where $v^{-1}$ is the inverse free propagator and $V$ a self-interaction.
To be specific we choose $v^{-1} = -\Delta^{-1} + m^2$.
The mass term is needed to render $v^{-1}$ invertible.
The interaction may contain another mass term.
 
In pertubation theory, calculations are based on the use of 
the generating functional
\be
  Z_0[j] &=& \sqrt{\det v^{-1}} \int \D \phi e^
          {-\frac12 \langle \phi,v^{-1} \phi\rangle  
                  + \langle \phi,\,j\rangle} 
\nn
  &=& \int d\mu_v[\phi] e^{\langle \phi,\, j\rangle} 
\nn
  &=& e^{\frac12 \langle j,vj\rangle}  \quad .
\label{sZ0}
\ee 
$ d\mu_v[\phi]= \sqrt{\det v^{-1}} 
\D \phi e^{- \frac12 \langle \phi,v^{-1} \phi\rangle}$ 
is the normalized Gaussian measure with covariance $v$ and $j$ the current.

\subsection{The convolution theorem for Gaussian measures}
\label{ssconvolution}

Since the propagator $v$ exists, we can apply the
convolution theorem for Gaussian measures.
The field split can be 
induced by a propagator split into a soft and a hard part
\be
  v = v^s + v^h \quad .
\ee 
Both propagators $v^s$ and $v^h$ must be invertible.
In addition both must be positive in momentum space to ensure the
convergence of the functional integrals.

The actual split can be of Pauli-Villars type 
\be
  \frac1{k^2 + m^2} 
  = \frac{M^2 -m^2}{(k^2 + m^2)(k^2 + M^2)} + \frac1{k^2 + M^2}
\ee
or a  mollified momentum cutoff following
Rivasseau \cite{rivasseau}
\be
  \frac1{k^2 + m^2} 
  = \frac1{k^2 + m^2} e^{-\frac{k^2}{M^2}}
  + \frac1{k^2 + m^2} (1-e^{-\frac{k^2}{M^2}}) 
\ee
where the mass $M$ sets the scale of the split.

A sharp momentum cutoff will not do because then the hard and soft
propagator are not invertible anymore.

With any of these splits the generating functional of 
eq.(\ref{sZ0}) factorizes
\be
  Z_0[j] &=& e^{\frac12 \langle j,vj\rangle} 
  = e^{\frac12 \langle j,v^s j \rangle} e^{\frac12 \langle j,v^hj\rangle} 
  \quad .
\ee
From the different forms of $Z_0[j]$ we see
\be
  \frac{\delta^n}{\delta j^n} Z_0[j] \Bigr |_{j=0} 
  = \langle \phi^n \rangle 
  = \langle (\phi^s+\phi^h)^n \rangle \quad .
\ee
That means for the fields itself $\phi=\phi^s+\phi^h$.
To deduce the field split from the propagator split both $v^s$ and $v^h$
must be invertible.
This is the only condition on the propagator split.
\be
  e^{\frac12 \langle j,v^sj\rangle}  e^{\frac12 \langle j,v^hj\rangle} 
  &=& \sqrt{\det {v^s}^{-1}} \int \D \phi^s 
  e^{-\frac12 \langle \phi^s,v^{s^{-1}} \phi^s\rangle  
             + \langle \phi^s,\,j\rangle} 
\nn
  & & \sqrt{\det {v^h}^{-1}} \int \D \phi^h 
  e^{-\frac12 \langle \phi^h,v^{h^{-1}} \phi^h\rangle  
             + \langle \phi^h,\,j\rangle} 
\nn
  &=& \int d\mu_{v^s}[\phi^s] e^{\langle \phi^s,\,j\rangle} 
      \int d\mu_{v^h}[\phi^h] e^{\langle \phi^h,\,j\rangle} 
\ee
The free part is now separated into contributions of the hard
respective soft field only.
The interaction connects both fields, since the same source $j$ is
coupled to the hard and the soft field. 

No knowledge of the functional relation of $\phi^s$ to $\phi$ is
obtained in this method.

\subsection{The blockspin method}
\label{ssscalarblock}

The second method to split the field is the blockspin method
\cite{mackgroup}
used in real space renormalization group calculations
\cite{wilson}.
The fundamental field $\phi$ lives either in the continuum or on a
fine lattice.
The block spin $\Phi$ is a block average of $\phi$.
It lives on a coarse lattice.
\be
  \Phi = C \phi
\ee
where $C$ is called the averaging operator.
During the averaging procedure all fine scale information of the
fundamental field is lost.
This implies that the averaging operator $C$ has no inverse.

To perform the field split of the fundamental field $\phi$ we have to
go back to the space it lives on, be it continuum or fine lattice.
Following Gawedzki and Kupiainen \cite{gawedzki} one splits the fundamental 
field $\phi$ into a low frequency part $\phi^s$ determined by the block
spin $\Phi$, and a high frequency or fluctuation field $\phi^h$ which
has vanishing block average.
This is done with the interpolation operator $\A$.
The soft field is
\be
  \phi^s = \A \Phi = \A C \phi
\ee
and the hard field contains the remainder, especially the fluctuations
\be
  \phi^h = \phi - \phi^s = (1-\A C) \phi \quad .
\ee
$\phi^h$ contains only fine scale information if its block average
vanishes
\be 
  C \phi^h = 0 \quad .
\ee
Then 
\be
  C \phi^s = \Phi 
  \quad \mbox{i.e.} \quad
  C \A = 1
  \quad \mbox{i.e.} \quad
  \A C \A C = \A C \quad .
\ee
The operator $\A C$ is the projector to the space of low frequency
fields, whereas
$(1-\A C)$ is the projector to the space of high frequency fields.

The use of projectors makes the blockspin method very clear as
far as the fields are concerned.
But projectors cannot be inverted so that the kinetic
terms of the hard and soft field cannot be inverted either.
They still posess a pseudoinverse, which is  sufficient for our present
purpose, but complicates the following calculations.

The hard and soft fields are defined as functions of the
fundamental field.
They are introduced into the generating functional
via delta-functionals
\be
  Z_0[j] = \sqrt{\det v^{-1}} \int \D \phi \D \Phi \D \phi^s \D \phi^h
  & & \delta(\Phi-C\phi) \delta(\phi^s-\A \Phi) \delta(\phi-\phi^s-\phi^h)
\nn
  & & e^{- \frac12 \langle \phi,v^{-1}\phi\rangle  
              + \langle \phi,j\rangle} 
  \quad .
\ee
Now we have to decide whether to use the blockspin $\Phi$ or
the background field $\phi^s$ as the low frequency variable.
Both of them contain the same information.
To keep $\Phi$ is more convenient for numerical simulations since 
$\Phi$ lives on a coarser lattice and one needs less computing time
to sweep through it.
For analytical calculations $\phi^s$ has the advantage to live on
the same space as the fundamental field $\phi$ and the fluctuation
field $\phi^h$.
We first keep the blockspin.
\be
  Z_0[j] = \sqrt{\det v^{-1}} \int \D \Phi \D \phi^h
  & & \delta(\Phi - C \A \Phi - C \phi^h) 
\nn
  & & e^{- \frac12 \langle (\A \Phi+\phi^h),v^{-1}(\A \Phi+\phi^h)\rangle
              + \langle (\A \Phi+\phi^h),j\rangle}
\ee
The kinetic term contains the mixed term
\be
  \langle \phi^h,v^{-1}\A \Phi\rangle  \quad .
\ee
This contradicts the idea of a field split because 
it is possible that a fluctuation field turns into a
blockspin or vice versa without involving an 
interaction.
With the proper choice of $\A$ one can ensure that this term
vanishes.
We recall that we have demanded $C \A=1$ which made $\A C$ and $(1-\A C)$
projectors.
To avoid the mixed kinetic term they should be orthogonal with
respect to the scalar product $\langle \cdot,v^{-1}\cdot\rangle $.

The kinetic term for the blockspin is
\be
  \frac12 \langle \Phi,\underbrace{\A^\dagger v^{-1}\A }_{=:u^{-1}}\Phi\rangle 
\ee
with the blockspin propagator
\be
  u = CvC ^\dagger \quad.
\label{suDef}
\ee
The interpolation kernel $\A$ which gives the projectors the correct
othogonality is defined by
\be
  v^{-1} \A = C^\dagger u^{-1}
\label{sinterpol}
\ee
This is the Gawedzki-Kupiainen interpolation kernel, which minimizes 
the kinetic part of the action for a given blockspin \cite{mackgroup}.
For its derivation see appendix \ref{scalarcalc}.

That $u^{-1}= \A^\dagger v^{-1} \A$ 
is truly inverse to $u$ can be shown using the
definition of $\A$.

The generating functional now factorizes
\be
\label{szblock}
  Z_0[j] &=& \sqrt{\det v^{-1}} \int \D \Phi 
  e^{- \frac12 \langle \Phi,u^{-1}\Phi\rangle  + \langle \A \Phi,j\rangle}
  \int \D \phi^h \delta(C\phi^h) 
  e^{- \frac12 \langle \phi^h,v^{-1}\phi^h\rangle  + \langle \phi^h,j\rangle}
\\
  &=& \sqrt{\det v^{-1}} \int \D \phi^s 
  e^{- \frac12 \langle \phi^s,C^\dagger u^{-1} C \phi^s\rangle  
              + \langle \phi^s,j\rangle}
  \int \D \phi^h \delta(C\phi^h) 
  e^{- \frac12 \langle \phi^h,v^{-1}\phi^h\rangle   
              + \langle \phi^h,j\rangle} \quad .
\nonumber
\ee
We use $\Phi = C \phi^s$ and $\phi^s = \A \Phi$ to
introduce the background field instead of the blockspin.

The kinetic operator for the background field is
\be
  v^{s^{-1}} = C^\dagger u^{-1} C = C^\dagger \A^\dagger v^{-1} \A C
  = v^{-1} \A C = C^\dagger \A^\dagger v^{-1} \quad .
\label{sgs-1block}
\ee
It is immediately clear that it cannot have a true inverse since it
contains the projector $\A C$.
But we can calculate the propagator $v^s$ for the background field 
by evaluating the first integral in eq.(\ref{szblock}).
Note that the source $j$ is not coupled to the blockspin, but to
the background field $\phi^s=\A\Phi$.  
\be
  Z_0^s[j] &=& \int \D\Phi  
  e^{-\frac12 \langle \Phi,u^{-1}\Phi\rangle  + \langle \A\Phi,j\rangle} \\
  &=& \frac1{\sqrt{\det u^{-1}}}
  e^{\frac12 \langle j,\A u \A^\dagger j\rangle} 
\ee
We read off the background propagator and use again
eq.(\ref{sinterpol}) to write it in a more suggestive form
\be
  v^s = \A u \A^\dagger =\A C v C^\dagger \A^\dagger 
  = \A C v = v C^\dagger\ \A^\dagger \quad .
\label{sgsblock}
\ee
The background propagator is equal to the fundamental propagator
preceded and/or
followed by the projector to the low frequency fields.
\be 
  v^{s^{-1}} v^s &=& C^\dagger \A^\dagger v^{-1}  v C^\dagger\ A^\dagger
  =  C^\dagger \A^\dagger
\nn
  v^s v^{s^{-1}} &=& \A C v v^{-1} \A C = \A C
\ee
This shows that $v^s$ is the pseudoinverse of $v^{s^{-1}}$.
In the subspace of low frequency fields it is the true inverse.
  
The last step is to calculate the fluctuation propagator.
Because of the delta-functional in the integral it is not possible
to read off the kinetic term as for the other fields.
In place of the delta-function one could use a Gaussian
\be
  \delta_\kappa (C\phi^h) 
  = \prod_{x\in\Lambda} \left( \frac{2\pi}\kappa \right)^{-\frac12} 
  e^{-\frac12 \kappa (C\phi^h(x))^2} \quad .
\ee
Hasenfratz and Niedermayer \cite{hasenfratz} pointed out that optimal
locality properties of the perfect action $L_\Lambda$ are obtained for
a preferred finite value of $\kappa$.
But for finite $\kappa$ the relation $\Phi = C \phi$ is lost and only 
restored in the limit $\kappa \to \infty$.
This blurrs the clearness of the real space renormalization. 

For this reason we perform the integration using the Fourier representation
of the delta-functional.
\be
  {Z_0}^h[j] &=& \int \D \phi^h \delta(C\phi^h) 
                 e^{- \frac12 \langle \phi^h,v^{-1}\phi^h\rangle   
                             + \langle \phi^h,j\rangle} 
\nn
              &=& \frac1{\sqrt{\det v^{-1}}} \frac1{\sqrt{\det u}}
                e^{\frac12 \langle j,(v-\A u\A^\dagger)j\rangle}
\ee
The calculation is done in appendix \ref{scalarcalc}.

We can read off the fluctuation propagator
\be
  v^h = v-\A u\A^\dagger = v (1-C^\dagger \A^\dagger) = (1- \A C) v
  = (1- \A C) v (1-C^\dagger \A^\dagger)
\ee
It is equal to the fundamental propagator preceded and/or followed by the
projector to the high frequency fields
so that $v^h$ cannot have a true inverse.
In analogy to $v^s$ we find its pseudoinverse
\be
  v^{h^{-1}} =  (1-C^\dagger \A^\dagger) v^{-1} (1- \A C)
  = (1-C^\dagger \A^\dagger) v^{-1}
  =  v^{-1} (1- \A C) 
  \quad .
\ee
Because $v^s$ and $v^h$ have only pseudoinverses we obtain the following  
two identities
\be
  v = v^s + v^h 
  \quad \mbox{and} \quad  
  v^{-1} = v^{s^{-1}} + v^{h^{-1}}
  \quad .
\label{spropsum}
\ee
Using the fluctuation propagator we can reexpress the fluctuation
field
\be
  \phi^h = (1 - \A C)\phi = v^h v^{-1} \phi
  \quad .
\label{shfield}
\ee
The same can be done for the background field
\be
  \phi^s = \A C \phi = v^s v^{-1} \phi
  \quad .
\label{ssfield}
\ee

The generating functional factorizes
\be
  Z_0[j]={\sqrt{\det v^{-1}}} {Z_0}^h[j] Z_0^s[j]
  =e^{\frac12 \langle j,v^h j\rangle  + \frac12 \langle j,v^sj\rangle} 
\ee
From eqs.(\ref{shfield}) and (\ref{ssfield}) we see how the hard and soft
fields are defined in terms of the fundamental field once the
propagator split is given.
But even then the averaging procedure is not uniquely
defined as any
averaging operator $C$ with $C \phi^h = 0$ and $C \phi^s \neq 0$ is
allowed.

\subsection{The invariance fixing method}
\label{ssinvariance}

The third method we present is introduced by Mitter and 
Valent \cite{mitter}.
It makes no reference to a lattice.
We split the fundamental field $\phi$ into a hard and a soft part
\be
  \phi = \phi^s + \phi^h
  \quad .
\label{ssplit}
\ee
The nomenclature will be justified later by the properties of the 
respective propagators.
A second condition 
\be
  F[\phi^s,\phi^h] = 0
\label{sfix}
\ee
serves to make the split unique.
The field split is implemented in the generating functional
by insertion of following unity into the partition function
\be
  1 = \int \D\phi^s \D\phi^h 
  \delta(\phi-\phi^s-\phi^h) \delta(F[\phi^s,\phi^h])
  \det (\frac{\delta F}{\delta \phi^h} - \frac{\delta F}{\delta \phi^s} )
  \quad .
\label{sone}
\ee
As eqs.(\ref{ssplit}) and (\ref{sfix}) give the hard and the soft field in
terms of the fundamental field
this approach is similar to the blockspin method.
Up to now the only difference is that we  avoid the intermediate
blockspin and concentrate immediately
on the soft field which is of course the background field.

Now we  change the order of integration.
Integrating the fundamental field results in 
\be
  Z_0[j]= {\sqrt{\det v^{-1}}} \int && \D\phi^s \D\phi^h
  \delta(F[\phi^s,\phi^h])
  \det (\frac{\delta F}{\delta \phi^h} - \frac{\delta F}{\delta \phi^s} ) 
\nn
  && e^{- \frac12 \langle (\phi^s+\phi^h),v^{-1}(\phi^s+\phi^h)\rangle}
  e^{\langle j,(\phi^s+\phi^h)\rangle}
  \quad .
\ee
This form suggests the following alternative interpretation.
We have split the fundamental field into a sum of two new ones.
Since this is not unique the fundamental field and hence the action
are invariant under the following 
hard-soft transformation.
\be
  \phi^h &\to& \phi^h + \Lambda \nn
  \phi^s &\to& \phi^s - \Lambda
  \quad .
\label{shstrafo}
\ee
$\Lambda$ is a field living on the same space as $\phi^s$ and $\phi^h$.
We call this new symmetry of the fundamental theory hard-soft invariance.
One deals with this symmetry as with a gauge symmetry, fixing it with the 
condition $F=0$ eq.(\ref{sfix}).
Using the hard-soft transformation eq.(\ref{shstrafo}) the determinant can
be written as
\be
  \det (\frac{\delta F}{\delta \phi^h} - \frac{\delta F}{\delta \phi^s}) 
  = \det 2 \frac{\delta F} {\delta \Lambda}
  \quad .
\ee
This is the Faddeev-Popov determinant of the hard-soft fixing.
To evaluate $Z_0[j]$ we proceed as in textbooks handling a gauge field
action with the Faddeev-Popov trick.

We remind the reader that $\delta(F)$ need not to be fulfilled 
literarily if we use the invariance fixing interpretation.
We could equally well use $\delta(F-B)$ 
with arbitrary $B$ to obtain a field split.
Of course inversion of this condition yields a functional connection
of $\phi^s$ to $\phi$ dependent on $B$.
In the final step we may integrate with some convergence-producing weight
over all possible $B$.
During this procedure we loose our previous knowledge of $\phi^s$ in
terms of $\phi$.
\be
  1 &=& \int \D B \D\phi^s \D\phi^h \delta(\phi-\phi^s-\phi^h) 
  e^{-\frac12 \frac1{M^2} B^2} \delta(F[\phi^s,\phi^h]-B)
  \det (\frac{\delta F}{\delta \phi^h} - \frac{\delta F}{\delta \phi^s}) 
\nn
  &=& \int \D\phi^s \D\phi^h \delta(\phi-\phi^s-\phi^h) 
  e^{-\frac12 \frac1{M^2} F[\phi^s,\phi^h]^2} 
  \det (\frac{\delta F}{\delta \phi^h} - \frac{\delta F}{\delta \phi^s}) 
\label{sunity}
\ee
The result is that the fixing term $F$ is raised quadratically 
into the exponent. 
No limiting process for $M$ is involved.
The determinant is regarded as a hard-soft ghost kinetic term.
It is this treatment of the second condition (\ref{sfix}) on the two
new fields which makes the difference to the blockspin method.

Eq.(\ref{sunity}) is the unity which is inserted into the generating
functional in the present approach
\be
  Z_0[j] = \int \D\phi^s \D\phi^h \D\bar\eta \D\eta
  e^{-\frac12 \langle (\phi^s+\phi^h), v^{-1} (\phi^s+\phi^h)\rangle  
  - \frac12 \frac1{M^2} F^2[\phi^s,\phi^h] 
  + \langle  (\phi^s+\phi^h),j\rangle  
  + \langle \bar \eta, 2 \frac{\delta F}{\delta \Lambda} \eta\rangle}
  \quad .
\ee
$1 / M^2$ is the hard-soft equivalent to the gauge parameter
with the limit $M^2 \to \infty$ removing the fixing.

Now we have to choose $F$.
The conditions on $F$ can be grouped into necessary ones and those
imposed only to ease further calculations.
We list them below starting with the necessary ones.

\begin{itemize}
\item
  {$\det (\frac{\delta F}{\delta \Lambda})$ must not be zero because
  otherwise $F$ would be hard-soft invariant.
  Then $F=0$ would not be the second condition for the split, but a
  restriction of the fundamental field $\phi=\phi^s+\phi^h$.}
 
\item
  {$\frac12 \langle (\phi^s+\phi^h), v^{-1} (\phi^s+\phi^h) \rangle 
  + \frac12 \frac1{M^2} \langle F[\phi^s,\phi^h],F[\phi^s,\phi^h]\rangle$ 
  may not have a mixed term connecting the hard and soft field.
  Otherwise a hard field could turn into a soft field without interaction
  contradicting the idea of a field split.}

\item
  {The resulting kinetic terms for the hard and soft fields must be
  invertible.
  The one for the hard field must be IR-regulated.
  The soft propagator must be UV-regulated, i.e. fall off faster 
  than $1 / {k^2}$.}
\end{itemize}

Concerning the conditions for convenience
we know that it is possible to split the fundamental field without
invoking ghosts.
This was demonstrated using the other two explained methods.
Therefore it should be possible to find an $F$ such that the ghosts decouple.
We want $\det (\frac{\delta F}{\delta \Lambda})$ to be independent of
the fields $\phi^s$ and $\phi^h$
so that $F$ should be linear in $\phi^s$ and $\phi^h$. 

The linear Ansatz for $F$ is the following
\be
  F[\phi^s,\phi^h] = F^s\phi^s + F^h\phi^h
  \quad \mbox{with} \quad \det (F^s-F^h) \neq 0
  \quad .
\label{sfix2}
\ee
We insert this into the kinetic term 
\be
      \frac12 \langle (\phi^s+\phi^h),v^{-1}(\phi^s+\phi^h) \rangle  
    + \frac12 \frac1{M^2} F^2
  &=& \frac12 \langle \phi^s,v^{-1}\phi^s\rangle  + 
      \frac12 \frac1{M^2} \langle \phi^s,F^{s \dagger} F^s \phi^s\rangle  
\nn
  &+& \frac12 \langle \phi^h,v^{-1}\phi^h\rangle  + 
      \frac12 \frac1{M^2} \langle \phi^h,F^{h \dagger} F^h \phi^h\rangle 
\nn
  &+& \langle \phi^s,v^{-1}\phi^h\rangle  
   + \frac1{M^2} \langle \phi^s,F^{s \dagger} F^h \phi^h\rangle 
  \quad .
\ee
The separation condition is 
\be 
  v^{-1} + \frac1{M^2} F^{s \dagger} F^h 
  = v^{-1} + \frac1{M^2} F^{h \dagger} F^s = 0 
  \quad .
\label{ssep}
\ee  
For the following considerations we
assume $F^s,F^h$ and $v^{-1}$ to be invertible.
The separation condition (\ref{ssep}) gives
\be
  \frac1{M^2} F^{s \dagger} = -v^{-1}F^{h -1} 
  \qquad \mbox{and} \qquad  
  \frac1{M^2} F^{h \dagger} = -v^{-1}F^{s -1}
\ee
leading to the following propagators
\be
  v^h &=& (v^{-1} + \frac1{M^2} F^{h \dagger} F^h)^{-1} 
  = (F^s-F^h)^{-1}F^sv  
\nn
  v^s &=& (v^{-1} + \frac1{M^2} F^{s \dagger} F^s)^{-1} 
  = - (F^s-F^h)^{-1}F^hv
\label{shsprop}
\ee
hence
\be 
  v^s + v^h = v  \quad .
\ee
For every linear choice of $F$ with invertible $F^s$ and $F^h$ 
the sum of the new propagators is the fundamental one.
For any given propagator split $v = v^s + v^h$ inversion of
eqs.(\ref{shsprop}) leads to the coefficients $F^h$ and $F^s$ of the
suitable fixing condition $F$.

The special solution we have in mind and which justifies to call $v^h$
hard and $v^s$ soft is
\be
  F^{s \dagger}=v^{-1} \quad\quad F^h =- M^2  \quad .
\ee
The resulting kinetic terms for the hard and soft fields are
\be
  v^{h^{-1}} &=& v^{-1} + \frac1{M^2} F^{h \dagger} F^h =  v^{-1} + M^2
\nn
  v^{s^{-1}} &=& v^{-1} + \frac1{M^2} F^{s \dagger} F^s 
  =  v^{-1} + \frac1{M^2} (v^{-1})^2 \quad .
\ee
Indeed the `gauge parameter' $M$ plays the role of a mass for the hard field.
The hard propagator is IR-regulated with a mass term.
The soft propagator is UV-regulated with higher derivatives.
This corresponds to a propagator split of the Pauli-Villar type.

\subsection{Comparison}
\label{sscomparison}

To use the convolution theorem for Gaussian measures the fundamental
propagator must exist.
Moreover the hard and soft propagator must be invertible.
This forbids to use a strict frequency separation, since this would
imply the use of projectors.
We have obtained the field split $\phi = \phi^s + \phi^h$, but we do
not know $\phi^s$ in terms of $\phi$ within this approach.

In the blockspin approach we know $\phi = \phi^s + \phi^h$ beforehand.
Both new fields and the blockspin $\Phi$ are defined in terms of the
fundamental field.
The field split is achieved by using projectors so that
the method is complementary to the split of the
Gaussian measure.
During the calculation we obtain the propagators of blockspin,
background and fluctuation field separately.
In addition we find the identity $v^s + v^h = v$.
It is thus shown that for every choice of the background field the
propagator split used in the split of the Gaussian measure is
regained.
The difference is that here $v^s$ and $v^h$ are not truly invertible.
Due to the use of pseudoinverses we have 
$v^{s^{-1}} + v^{h^{-1}} = v^{-1}$ as well.
Alternative definitions of background and fluctuation fields
can be given
\be
  \phi^s = v^s v^{-1} \phi
  \quad \quad \mbox{and} \quad \quad
  \phi^h = v^h v^{-1} \phi \quad .
\ee 
This shows how  background and
fluctuation field are defined for a given  propagator split.
The blockspin can still be chosen at will subject only to the
condition $C\phi^h = 0$.

The hard-soft invariance fixing method starts like the blockspin
method with given functional relationships of $\phi^s$ and $\phi^h$ to
$\phi$.
But these relations are lost during the course of the calculation 
because we use the Faddeev-Popov trick.
We know from the input $\phi = \phi^s + \phi^h$.
Calculation shows the identity $v^s + v^h = v$. 
We end up with an elaborate way of obtaining the convolution theorem
of the Gaussian measure.
The advantage of this method compared to the convolution theorem
of the Gaussian measure is that it can be adapted to the case
when the fundamental propagator does not exist.
If the fundamental propagator does exist both methods are
equivalent.

Both blockspin method and hard-soft invariance fixing method are
generalizations of the split of the Gaussian measure.
But they are complementary since the propagator split
is achieved with projectors in the former method and in an invertible
way in the latter.

\subsection{Explicit choice of the block spin}
\label{ssblock}

We start on the continuum ${\menge R}^4$ of points $z$.  
The continuum is divided into blocks $x$
identified with the sites at their centers.
They may have different extensions in different directions.
In this way a hyper-rectangular lattice $\Lambda$ of lattice spacing
$L_\mu$ in $\mu$-direction is obtained.
For the explicit calculations we use the following notation
\be
  \int_{x\in\Lambda} 
  = L_1 L_2 L_3 L_4  \sum_{n\in{\menge N}^4}
  \quad \quad , \quad \quad 
  x = (n_1 L_1, n_2 L_2, n_3 L_3,n_4 L_4).
\ee 
With a scalar field $\phi(z)$ one associates a block spin $\Phi(x)$. 
Following Gawedzki and Kupiainen \cite{gawedzki}
we choose them as block averages
\be
  \Phi(x) = C \phi(x) = \av_{z\in x} \phi(z) \quad .
\label{sCDef} 
\ee
The averaging operator has a kernel $C(x,z)$ which
equals the properly normalized characteristic function $\chi_x(z)$ of
the block $x$
\be
   C(x,z) = \prod_{\mu=1}^4 \frac1L_\mu \chi_{x_\mu}(z_\mu)
          = \prod_{\mu=1}^4 C(x_\mu,z_\mu)
\label{sCkern}
\ee
with $\chi_{x_\mu}(z_\mu)$ the characteristic function in $\mu$-direction.

Block propagator and kinetic term, interpolation operator, background
and fluctuation propagator are all calculated using the general
formulae of subsection \ref{ssscalarblock}.

\subsection{`Blocking' to the continuum}
\label{sspoincare}

Up to now the fundamental field could live on a fine lattice or on the
continuum, but the blockspin had to live on a lattice.
This has the disadvantage that the Poincar\'e invariance of the
fundamental theory is broken down to its lattice counterpart.
This problem can be cured if we choose blockspins defined on the
continuum and a blocking operator respecting
the full Poincar\'e invariance.

The simplest possible choice is to define the `block'spin $\Phi(z)$ at
any point $z \in {\menge R}^4$ as the average over a hypersphere with 
radius $L$ and center $z$.
This choice is in the spirit of Wetterich et. al. \cite{wetterich}.
It corresponds to a mollified momentum cutoff.

Despite the fact that the `block'spin $\Phi$ lives now on the continuum, it is
not identical to the background field $\phi^s$.
$\phi^s = \A C \phi$ is a projection of the fundamental field $\phi$ to low
frequencies, i.e $\A C \phi^s = \phi^s$.
This is different for the `block'spin.
$C \Phi \neq \Phi$, but every application of $C$ smoothes the
`block'spin further.

The advantage of this peculiar choice of the `block'spin preserving the 
Poincar\'e invariance is that it is possible to apply it to fermions.
To bring fermions to the lattice is a difficult problem because of the
Nielsen-Ninomiya no-go theorem.
One is either plagued by doublers (staggered fermions) or has to introduce 
a mass term which breaks the chiral invariance (Wilson fermions).
Both approaches make it difficult to take the continuum limes after
the calculations are done.
 
One disadvantage of the continuum `block'spin is that its notion has a
tendency to get
mixed up with that of the background field.
Another one is that we cannot use it for dimensional reduction, as we
do with the lattice blockspin when we deal with finite temperature,
see section \ref{sectemper}.

Moreover for numerical calculations one needs a lattice theory, and we 
want to be able to apply the here developed methods to a fundamental
lattice theory.
Therefore we use in the following always blockspins defined on the
lattice.
The switch to the continuum blockspin can be done at the very end of
the derivation of the effective action.
 
\subsection{The effective action}
\label{sseffective}

After all the preparations for the calculation of the effective action
it remains to show how it can be evaluated perturbatively.
Its definition is 
\be
  e^{-S_{\rm eff}[\Phi]}
  &=& \int \D \phi^h \delta(C \phi^h) e^{-S[\A \Phi + \phi^h]}
\nn
  &=& e^{-\frac12 \langle \A \Phi, v^{-1} \A \Phi \rangle}
      \int \D \phi^h \delta(C \phi^h) 
     e^{-\frac12 \langle \phi^h, v^{-1} \phi^h \rangle - V[\A \Phi + \phi^h]}
\nn
  &=& e^{-\frac12 \langle \A \Phi, v^{-1} \A \Phi \rangle}
      e^{- V[\A \Phi + \frac{\delta}{\delta j}]} 
      \int \D \phi^h \delta(C \phi^h) 
     e^{-\frac12 \langle \phi^h, v^{-1} \phi^h \rangle 
        + \langle \phi^h, j \rangle} \big|_{j=0}
\nn
  &=& e^{-\frac12 \langle \A \Phi, v^{-1} \A \Phi \rangle}
      e^{- V[\A \Phi + \frac{\delta}{\delta j}]} 
      e^{+\frac12 \langle j , v^h j \rangle} \big|_{j=0}\quad .
\label{veff}
\ee
The effective potential, in the sense of the interaction part of the
effective action, is given by the set of connected diagrams of the
above expansion.

\section{Maxwell theory}
\label{secmaxwell}

\subsection{The Maxwell action}
\label{ssmodel}

In the previous section we introduced three methods to split the 
fundamental field into a hard and a soft part and factorize the
generating functional.
Now we want to generalize the procedure to a situation with gauge
symmetry.
Maxwell theory is used as an illustrating example.

The Euclidian action is 
\be
  S_M = - \frac14 \langle F_{\mu\nu}, F_{\mu\nu} \rangle 
    = \frac12 \langle a_\mu, 
       (\dnu \dmu - \partial^2 \delta_{\mu\nu}) a_\nu \rangle \quad .
\ee
Again we suppress the spacetime arguments.
The action is invariant under the following gauge transformation
\be
  a_\mu \trafo_{gauge} a_\mu -   \dmu \lambda  \quad .
\label{mgauge}
\ee
The invariance of $S_M$ under gauge transformations gives rise to zero
modes of the kinetic operator as we can see by applying
the kinetic operator to a pure gauge
\be
  (\dnu \dmu - \partial^2 \delta_{\mu\nu}) \dnu \lambda
  = (\dnu \dmu \dnu - \partial^2 \dmu) \lambda = 0 \quad .
\ee
The kinetic operator cannot be inverted, i.e.
the propagator does not exist without gauge fixing.

Hence the convolution theorem for Gaussian measures cannot be applied
and so we have to look whether at least one of the other methods can be
generalized.

\subsection{Hard-soft invariance fixing}
\label{sshsinvariance}

Like in the scalar case we split the fundamental field into 
two parts, one called hard, the other soft
\be
  a_\mu = a_\mu^h + a_\mu^s \quad .
\ee
As before the names are justified later.
The action is invariant under the following transformations
\be
  a_\mu^h \trafo_{hs} a_\mu^h + \Lambda_\mu 
  \quad \mbox{and simultanously} \quad 
  a_\mu^s \trafo_{hs} a_\mu^s - \Lambda_\mu \quad .
\ee
This is the hard-soft transformation.
$\Lambda_\mu$ is a vector field as are $a_\mu^h$ and $a_\mu^s$.
In addition there are now two independent gauge
invariances of the action
\be
  a_\mu^h && \trafo_{gauge} a_\mu^h - \dmu \alpha
\nn
  a_\mu^s && \trafo_{gauge} a_\mu^s - \dmu \beta \quad .
\label{mhsgauge1}
\ee
The next step is to fix the hard-soft invariance by means of a fixing
condition $F_\mu[a^s,a^h]=0$.
As we need one fixing condition for every spacetime coordinate
we have written the fixing condition as a vector identity.
This hard-soft fixing restricts also the allowed gauge transformations
for the hard and soft field.
Whatever gauge transformation is allowed for the hard field is
forbidden for the soft field because otherwise it would serve as a
hard-soft transformation.

The choice of the fixing condition gives restricted and different gauge
transformation properties to the hard and the soft field.
We want to obtain a low energy theory with the same symmetries as the
fundamental one.
In the literature \cite{mitter} is suggested that only the soft field
should transform as a gauge
field, whereas the hard field should be invariant like a matter field
in adjoint representation.
\be
  a_\mu^s && \trafo_{gauge} a_\mu^s - \dmu \lambda
\nn
  a_\mu^h && \trafo_{gauge} a_\mu^h \quad .
\label{mhsgauge2}
\ee
This means that we have to look for such an $F_\mu$ that 
\be
  \delta(F_\mu[a^s,a^h]) e^{-S_M[a^s+a^h]}
\ee
is invariant under any ordinary gauge transformation of the soft field
but under no transformation of the hard field.
One effect of these transformation properties is that the hard field
can never be a pure gauge which are the 
zero modes of the kinetic operator of the Maxwell action.
The hard field no longer contributes zero modes.
Therefore the hard propagator exists and the hard field might be
integrated to obtain a low energy effective action which is still
gauge invariant by construction.

Now we come to the point of choosing $F_\mu[a^s,a^h]$.
as in subsection \ref{ssinvariance} the conditions on $F_\mu$ are
divided into a set of necessary
conditions and those imposed to ease further calculations.

In complete analogy to the scalar problem we list the necessary
conditions.
\begin{enumerate}
\item
   $S_M[a^s+a^h] + \frac12 m^2 \langle F_\mu, F_\mu \rangle$
   must not have a mixed kinetic term since otherwise a hard field
   could transform into a soft field without interaction
   which contradicts the idea of a field split.
\item
   $F_\mu[a^s,a^h]=0$ must be invariant under 
   any gauge transformation of the soft field, 
   whereas any attempt at a gauge transformation of
   the hard field must violate $F_\mu=0$. 
   This breaks down the gauge invariance of eq.(\ref{mhsgauge1}) to the
   desired gauge transformation properties of eq.(\ref{mhsgauge2}).
\item
   $F_\mu$ must be an independent second condition on the field split.
   This is not the case if $F_\mu$ acts only on the sum of the hard
   and soft field because 
   then $F_\mu$ would be a restriction of the fundamental field $a_\mu$.
   To ensure this independence 
   $\det({\delta F_\mu}/{\delta \Lambda_\nu})$ must be non-zero.
\item
   The hard kinetic operator must be invertible without further amendments.
   It should involve some sort of IR-cutoff to justify its name.
\end{enumerate}
For the soft kinetic operator we cannot demand invertibility since 
gauge transformations are still allowed for the soft field and create
zero modes of the soft kinetic operator.

In the scalar case we had other methods for the field split
and so we knew that a fixing condition decoupling
the hard-soft ghosts must exist.
This resticted our Ansatz to be linear in the hard and the soft field. 
Here we argue similarly.
Since we start from a free theory, it should be possible to find
such an $F_\mu$ that the hard-soft ghosts do not couple to both fields
because otherwise integration of the ghosts would create an interaction.
This means that $F_\mu$ must be linear either in $a^s$ or in $a^h$.

After this long prologue we give a choice of $F_\mu$ that fulfills all
the above conditions:
\be 
F_\mu = a_\mu^h + \frac1{m^2} \dnu F_{\nu\mu}^s
\ee
is linear in both fields, hence the hard-soft ghosts decouple
completely.
We show that the four necessary conditions are fulfilled.
\begin{enumerate}
\item
  The mixed terms are
  \be
    \frac12 a_\mu^h (\dmu \dnu - \partial^2 \delta_{\mu\nu}) a_\nu^s
    &+& \frac12 a_\mu^s (\dmu \dnu - \partial^2 \delta_{\mu\nu}) a_\nu^h
  \nn
    + \frac12 m^2 a_\mu^h \frac1{m^2}
               (\partial^2 \delta_{\mu\nu} - \dmu \dnu) a_\nu^s
    &+& \frac12 m^2 a_\nu^s \frac1{m^2}
               (\partial^2 \delta_{\mu\nu} - \dmu \dnu) a_\mu^h 
  \ee
  and they cancel.
\item
  Since $F_{\nu\mu}^s = \dnu a_\mu^s - \dmu a_\nu^s$ is invariant under
  every gauge transformation of the soft field, so is $F_\mu$.
  On the contrary every attempt to gauge transform the hard field violates
  $F_\mu=0$.
\item
  The determinant of
  \be
    \frac{\delta F_\mu}{\delta \Lambda_\nu}
    = \delta_{\mu\nu} +  \frac1{m^2} (\dmu \dnu - \partial^2 \delta_{\mu\nu})
  \ee
  does not vanish.
\item 
  The hard kinetic term is
  \be
    \frac12 a_\mu^h 
    (\dmu \dnu - \partial^2 \delta_{\mu\nu} + m^2 \delta_{\mu\nu})
    a_\nu^h \quad .
  \ee
  It has a mass as an IR-regulator which removes the zero modes.
\end{enumerate}
 
The soft kinetic term is 
\be
  \frac12 a_\mu^s
  [(\dmu \dnu - \partial^2 \delta_{\mu\nu}) 
  + \frac1{m^2} (\dmu \drho - \partial^2 \delta_{\mu\rho})
                (\drho \dnu - \partial^2 \delta_{\rho\nu})]
  a_\nu^s \quad .
\ee
It has a higher derivative part.
Since it allows the same zero modes as the fundamental kinetic term it
cannot be inverted without prior gauge fixing.

The hard kinetic operator is the kinetic operator of a massive vector field.
It can be inverted to
\be 
  D_{\mu\nu}^h = \frac1{k^2+m^2} (\delta_{\mu\nu}
                                  + \frac{k_\mu k_\nu}{m^2}) \quad .
\ee
The propagator does not fall off at all in its longitudinal part.

Up to now we have dealt with a free field theory
where we could tolerate the bad behaviour of the hard propagator because
the only effect of integrating the hard free field 
is an infinite constant.
However as soon as we use the procedure for interacting theories and
generalize the above choice of $F_\mu$ to
the non-Abelian case we run into problems.
Integrating the hard gauge field gives now rise to higher loop
diagrams and
with a propagator not falling off at all, the higher the loop
order the more divergent the diagram will be.
This is a disaster for the calculation of a low energy effective
theory, because no renormalization of a finite set of constants 
can can render the infinitely many coupling constants of the effective
theory finite.

Choosing $F_\mu$ in such a way has wrecked the renormalizability of
the theory.

To trace back the reason of the ill-behaving hard gauge field
propagator is easier in the Abelian case.
With covariant gauge fixing (with gauge parameter $\alpha$) we could
have obtained a fundamental propagator
\be
  D_{\mu\nu} = \frac1{k^2} (\delta_{\mu\nu}
             + \frac{1-\alpha}{\alpha} \frac{k_\mu k_\nu}{k^2})
\ee
falling off as $1/{k^2}$.

How can it be that the hard propagator behaves worse than the fundamental 
one?

The answer is that we did not start with a gauge fixed theory.
So there is no fundamental propagator to compare with.
But from the above we can see that the gauge fixing is responsible for
the good UV-behaviour of the longitudinal part.
We cannot do the same for the hard field because there is no gauge
invariance to be fixed.
This observation brings us one step closer to the source of the
trouble.

First of all we should stop calling the hard gauge field a `gauge
field' since the action together with the fixing condition allows no gauge
transformation of the hard field.
The hard field can righteously be called `hard' because it has a
mass to prevent IR-problems.

What about the soft gauge field $a_\mu^s$?
It is truely a gauge field, but not necessarily a soft one.
The gauge transformation may again add hard components to this field
so that the gauge transformation properties of
eq.(\ref{mhsgauge2}) are not useful for our purpose.

What we really want to achieve is  
\be
  a_\mu^h && \trafo_{gauge} a_\mu^h - (\dmu \lambda)^h
  \nn
  a_\mu^s && \trafo_{gauge} a_\mu^s - (\dmu \lambda)^s \quad .
\label{mhsgauge3}
\ee
The gauge transformation itself is split in such a way that $(\dmu
\lambda)^h$ is
a hard and $(\dmu \lambda)^s$ a soft contribution to the
gauge transformation
not interfering any more with the hard-soft split.
If we could find a modified $F_\mu$ supporting these transformation
properties we had a good starting point for the generalization of this
method to the non-Abelian case.

The fundamental field is free, so integrating out the high frequency
modes should not affect the low frequency field.
This suggests a linear choice of $F_\mu$ in both the hard and the soft
field since any nonlinearity would give rise to interactions.
In addition we know that the mixed part of 
$\frac12 m^2 \langle F_\mu, F_\mu \rangle$
must cancel the mixed kinetic term and must
therefore have the same symmetries as the fundamental action.
It has to be invariant under gauge transformations as in
eq.(\ref{mhsgauge1}) with $\alpha$ and $\beta$ independent of each
other.
On the other hand the hard part of $\frac12 m^2 \langle F_\mu, F_\mu
\rangle$ must only be invariant under a restricted choice of $\alpha$, 
and contain a mass term.
Within the present approach it seems impossible to fulfill all these
requirements simultaneously.

\subsection{Block spin for free Abelian gauge theory}
\label{ssabelblock}

Since the above method failed we try to generalize the block 
spin method to free Abelian gauge fields.
In this section we use 
the Ba{\l}aban-Jaffe block spin transformation for the free Abelian
gauge field at zero temperature \cite{balaban}.

Let the symbol $\partial$ denote the exterior derivative of a
$p$-form and $\partial^\dagger$ its coderivative.  
For the 1-form $a$ and the 0-form $\lambda$
\be
  \partial a &=& \frac12 
  (\dmu a_\nu - \dnu a_\mu){\d}z_\mu \wedge{\d}z_\nu
\nn
  \partial^\dagger a &=& \dmu a_\mu
\nn
  \partial^\dagger \lambda &=& 0  \quad .
\ee
The Laplacian is given by 
$(\partial \partial^\dagger + \partial^\dagger \partial)$.
The Maxwell action of the electromagnetic field can be
written as
\be
  S_M [a] = \frac12 \langle \partial a, \partial a \rangle \quad .
\ee
The perfect lattice action associated with this is defined by a
formula analoguous to eq. (\ref{veff}) for the scalar field. There is
one difference, however. 
In order to give meaning to the fluctuation 
integral some amount of gauge fixing is necessary.
We wish to obtain a gauge covariant perfect lattice action and therefore
to retain the freedom of gauge transformations on the lattice,
i.e.  one gauge degree of freedom per block. Hence global gauge
fixing is not appropriate. Instead, gauge fixing is only used locally
within each block.

\subsubsection{Covariant averaging operator $C$}
\label{ssCcov}

Given a vector potential $a(z) = a_\mu(z){\d}z_\mu$ on the continuum
we define a block spin ${\bf A}$ living on links $b$ of the block lattice. 

We use the alternative notation
\be
  {\bf A}[b] = {\bf A}_\mu(x)
\ee
where $b$ is the link emanating from $x$ in $\mu$-direction.

The explicit blocking operation is the following:
Given a link $b$ from $x$ in $\mu$-direction and a point $z\in x$,
let ${\cal C}_{z,\mu}$ be the straight path of length $L_\mu$, i.e. one block
lattice spacing, in $\mu$-direction starting from $z$.
The blocking procedure is defined by
\be
\label{mCvDef}
  {\bf A}[b] &=& \av_{z\in x} a[{\cal C}_{z,\mu}]
\\
  a[{\cal C}_{z,\mu}] &=& \frac1{L_\mu} \int_{{\cal C}_{z,\mu}} 
                                           {\d} z'_\nu a_\nu(z') \quad .
\ee
This blocking procedure is covariant under gauge transformations in
the following sense. If we transform $a_\mu$ according to
\be
  a_\mu(z) \rightarrow a_\mu(z) - \dmu \lambda(z) 
  \equiv a_\mu^\lambda(z)
\ee
then the block gauge field transforms as 
\be
  {\bf A}_\mu(x) \rightarrow {\bf A}_\mu(x) - \nabla_\mu \Lambda(x)
\ee
with $\Lambda(x)$ the block average of the gauge transformations
$\lambda(z)$.

To distinguish notationally between the averaging operators
for gauge fields on the one hand, and for scalars and gauge
transformations on the other we
denote the latter by $C_S$ from now on. 
Thus ${\bf A} = Ca$ and the gauge transformation blocked to the lattice is
\be
  \Lambda(x) = \av_{z\in x} \lambda(z)
  \quad \quad \mbox{i.e.} \quad \quad
  \Lambda = C_S \lambda \quad .
\ee
We use the notation $a^\lambda$ for the gauge transform of $a$, etc.
In this notation the covariance property reads
\be
  C a^\lambda = (C a)^{C_S \lambda}
\ee
or written with space-time indices
\be
  C_{\mu\nu} \dnu = \nabla_\mu C_S \quad .
\label{mgtrafo}
\ee
To facilitate the comparison with the scalar case we give the explicit
form of the gauge covariant averaging kernel
\be
  C_{\mu \nu}(x,z)= \delta_{\mu,\nu} 
  \int dz'_\mu C_S(x_\mu,z'_\mu) C_S(z'_\mu + \frac12 L_\mu,z_\mu) 
  \prod_{\rho \neq \mu} C_S(x_\rho,z_\rho)
\label{mCkern}
\ee
where $C_S(x_\rho,z_\rho)$ is the one dimensional scalar averaging kernel.
In $\mu$-direction the covariant averaging kernel is not a simple step
function but a convolution of two of them.

\subsubsection{Covariant interpolation operator $\A$}
\label{ssAcov}

A gauge covariant blocking operator ensures that the blockspin has a
gauge invariant action under block gauge transformations.
To evaluate the effective action we need to define the fluctuation field 
for given blockspin.
This must be done in a covariant way.
It is important to separate the background and fluctuation field in
such way that gauge transformations respect this separation.

The background field is subject to non blocked gauge
transformations. 
The same is true for the fluctuation field.
Choosing the block operator has divided the set of gauge
transformations into two groups.
Those which leave the block gauge field ${\bf A_\mu}$ unaltered and
the others.
Since the background field depends only on the blockspin it must
be invariant under gauge transformations of the first type.
Because it lives on the fundamental space it must transform as a
fundamental gauge fields under gauge transformations of the other
type.
The block average of the fluctuation field must vanish, and this must
remain valid if we apply a gauge transformation to it.
For transformations of the first type this is true by definition.
Transformations with non vanishing block average may not be applied to
a fluctuation field.
  
The definition for the interpolation operator is the following
\be
  v^{-1} \A = C^\dagger u^{-1} \quad .
\label{msep}
\ee
Note that eq.(\ref{msep}) defining the interpolation kernel does
not involve propagators.
Therefore it is possible to remove the mixed kinetic term and
thereby to define background and fluctuation field even if the
propagators do not exist.

The fundamental kinetic operator is $v^{-1} = \partial^\dagger \partial$
and $u^{-1}$ is the kinetic operator for the blockspin.
From the form of $v^{-1}$ it is immedialtely obvious that $\A$ is only
determined up to a summand beginning
with an exterior derivative $\partial$.
This is the gauge freedom of $\A$.

To obtain the correct gauge transformation properties for background
and fluctuation fields we look for an interpolation
operator with the following covariance property analogous to that of
the blocking operator displayed in eq.(\ref{mgtrafo})
\be
  \A_{\nu\mu} \nabla_\mu = \dnu \A_S
  \quad \quad \mbox{or} \quad \quad 
  \A A^\Lambda = (\A A)^{\A_S \Lambda} \quad .
\label{mAgauge}
\ee
$\A_S$ is defined by eqs.(\ref{mAgauge}). 

The hard part of the calculation is to find any $\A$ which solves
eq.(\ref{msep}) since $v^{-1}$ has no inverse without gauge fixing.
As we will see our laboriously preserved gauge invariance does not 
interfere with the existence of $u^{-1}$ or $\A$.
If we had a guess for $\A$ we could insert it into the
eq.(\ref{msep}) and verify it.

We will show now how to obtain such a guess.
In section \ref{ssscalarblock} $\A$ is calculated in a 
straightforward way using the propagators $v$ and $u$.
Here these entities do not exist without gauge fixing.

To remedy this we adapt the procedure as follows.
First we fix the gauge which will ensure the existence of all propagators  
and calculate all the quantities we need with the formulae of
section \ref{ssscalarblock}.
Of course the results then depend on the choice of the gauge fixing and 
on the gauge parameters.
Afterwards we remove the gauge fixing, i.e. we perform a limiting
procedure with the gauge parameters. 
Some quantities remain finite and others become divergent.
The divergent quantities do definitely not exist without gauge fixing.
Finally we use the finite quantities as the guess we mentioned
above and look whether they fulfill the gauge invariant equations.
This final step ensures that the results do not depend on
the specific choice of gauge fixing we used in between.

In short we use the gauge fixing as a regulator which we remove
afterwards.
Therefore we do not need ghosts.
Finally we show the independence of the results from the
regulator scheme.

\subsubsection{Gauge fixing within blocks}
\label{ssproj}

We want to disentangle the results of fixing the gauge within the block from
those of the gauge fixing on the block lattice.

For this purpose we introduce the projector $R$ onto those gauge
transformation 
functions $\lambda$ satisfying the constraint $C_S \lambda=0$.
These are the transformations which leave the block spin 
${\bf A}_\mu$ invariant.

We recall from section \ref{secscalar} that the fluctuation field
$\phi^h$ associated with a scalar field $\phi$ satisfies $C_S \phi^h = 0$
and can be obtained by applying a projector, see eq.(\ref{shfield})
\be
  \phi^h = (1- \A^{(\phi)} C_S) \phi  \quad .
\ee
$\A^{(\phi)} C_S$ is an orthogonal projector with respect to the scalar
product furnished by the kinetic operator for $\phi$.

We want to use covariant gauge fixing.
The covariant gauge fixing condition acts on $\lambda$ as $\Delta \lambda$.
Using the Faddeev-Popov trick the gauge fixing condition is raised
quadratically into the exponent.
The kinetic operator of $\lambda$ with respect to which we want the
projector $R$ to be orthogonal is therefore $\Delta^2$ in place 
of $-\Delta$ as in the scalar case of section \ref{secscalar}.

So we may write
\be
  R = 1- \A^{(\lambda)} C_S
\label{mr0}
\ee
where $C_S$ is the same averaging kernel for scalars as before, while 
$\A^{(\lambda)}$ is chosen to satisfy
\be
  v_{\Delta^2}^{-1} \A^{(\lambda)} = C_S^\dagger u_{\Delta^2}^{-1}
  \quad , \quad \quad 
  u_{\Delta^2} = C_S v_{\Delta^2} C_S^\dagger 
  \quad , \quad \quad 
  v_{\Delta^2}^{-1}=\Delta^2
\label{mr2}
\ee
in complete analogy to eq.(\ref{sinterpol}).

From $R$ we construct another projector $\R = \Delta R \Delta^{-1}$.
$\R$ is the projector onto those $\Delta \lambda$ with $C_S \lambda =
0$.
Hence the covariant gauge fixing restricted to act only within blocks is
\be
\R \partial^\dagger a = 0 \quad .
\ee
There remains one gauge degree of
freedom per block $\Lambda(x) = C_S \lambda(x)$ not affected by
the fixing. It extends to a global gauge transformation per block,
$\lambda(z) = \Lambda(x)$ for $z\in x$.

Now we calculate the projectors $R$ and $\R$ explicitly.
We insert the formula for $\A^{(\lambda)}$ in eq.(\ref{mr0})
\be 
  R = 1 - \Delta^{-2} C_S^\dagger u_{\Delta^2}^{-1} C_S
\label{mr1}
\ee
and we see
\be
  \Delta^2 R = R^\dagger \Delta^2 \quad .
\ee
For the other projector we have
\be
  \R = \Delta R \Delta^{-1} 
     = 1 - \Delta^{-1} C_S^\dagger u_{\Delta^2}^{-1} C_S \Delta^{-1}
     = \R^\dagger
\label{mr3}
\ee
Using eqs.(\ref{mr1}) and (\ref{mr2}) we see 
\be
  C_S R = 0 
  \quad \mbox{and} \quad 
  R \Delta^{-2} C_S^\dagger = 0 \quad .
\ee
and also with eq. (\ref{mr3})
\be
  \R \Delta^{-1} C_S^\dagger = \Delta R \Delta^{-2} C_S^\dagger = 0 
  \quad \mbox{and} \quad
  C_S \Delta^{-1} \R = C_S R \Delta^{-1} = 0 \quad .
\label{mr4}
\ee

\subsubsection{Gauge fixed interpolation operator}
\label{ssAfix}

With the help of the projector $\R$ we now introduce two
gauge fixing terms, one forbidding only gauge transformations
with vanishing block average and the other being just the complement of it.

The kinetic operator for the gauge field is then 
\be
  v_{\alpha \beta}^{-1}  
  &=& \partial^\dagger \partial + \alpha \partial \R \partial^\dagger 
  + \beta \partial (1-\R) \partial^\dagger
\nn
  &=& \Delta +(\alpha-1) \partial \R \partial^\dagger 
  + (\beta-1) \partial (1-\R) \partial^\dagger
\ee
The term proportional to $\alpha$ fixes the gauge transformations
with vanishing block average and the term proportional to $\beta$
fixes the remainder and only the remainder.
Due to the use of the projector there exist no gauge transformations
onto which two possibly contradicting conditions are imposed.
The reason to use this complicated gauge fixing is that during the
following calculations we will see that for some entities partial
gauge fixing (i.e. $\beta= 0$ or $\alpha = 0$) is sufficient.
This way we prepare the evaluation of the fluctuation propagator 
when the gauge is fixed only within the blocks.

Now all entities we need to apply the formulae derived in section
\ref{secscalar} exist.
We recall them
\be
  u_{\alpha\beta} &=& C v_{\alpha\beta} C^\dagger
\nn
  \A_{\alpha\beta} &=& v_{\alpha\beta} C^\dagger u^{-1}_{\alpha\beta} 
\nn
  v^s_{\alpha\beta} &=& \A_{\alpha\beta} C v_{\alpha\beta}
                     = v_{\alpha\beta} C^\dagger \A^\dagger_{\alpha\beta}
  = \A_{\alpha\beta} C v_{\alpha\beta} C^\dagger \A^\dagger_{\alpha\beta}
\nn
  \Gamma_{\alpha\beta} &=& v_{\alpha\beta} - v^s_{\alpha\beta}
                        = (1- \A_{\alpha\beta} C ) v_{\alpha\beta}
  \quad .
\label{mgeneral}
\ee
The suffix reminds us of the values of the two gauge parameters used
in the evaluation of the operators.

Before we begin the calculations we summarize our expectations concerning
the question which of the quantities remain finite if we remove one or
the other part of the gauge fixing.
Blockspin and fluctuation fields are introduced without reference to
any propagator.
Kinetic operators exist without gauge fixing, i.e. we expect
well defined $u^{-1}_{00}$,$v^{s^{-1}}_{00}$
and $\Gamma^{-1}_{00}$.
The interpolation kernel $\A$ is defined in 
eq.(\ref{msep}) via $v^{-1} \A = C^\dagger u^{-1}$.
Only kinetic operators are involved hence $\A_{00}$ should exist.
The fundamental propagator needs full gauge fixing.
Neither $v_{\alpha 0}$ nor $v_{0 \beta}$ nor $v_{00}$ exist.
Blockspin and background field propagator should not mind whether the
gauge within a block is fixed or not, but they need gauge fixing on
the coarse scale.
So for $u_{\alpha\beta}$ and $v^s_{\alpha\beta}$
it is possible to choose $\alpha = 0$ but not $\beta = 0$.
For the fluctuation propagator it should be just the other way round.
We expect $\beta = 0$ to be possible but not $\alpha = 0$.

The calculations we have to do are tedious but straightforward.
In their full beauty they are banned to appendix \ref{alphabeta}.
Their result is summarized in eqs.(\ref{msummary}).

The gauge fixed fundamental propagator is
\be
  v_{\alpha\beta} = \Delta^{-1} 
  - (1-\frac1\alpha) \partial \Delta^{-1} \R \Delta^{-1} \partial^\dagger
  - (1-\frac1\beta) \partial \Delta^{-1} (1-\R) \Delta^{-1} \partial^\dagger
  \quad .
\label{mvab}
\ee
To verify this we use the projector properties of $\R$ and $(1-\R)$ and  
\be
  \partial\partial = 0
  \quad \quad &,& \quad \quad
  \partial^\dagger \partial^\dagger = 0
\nn
  \partial \Delta = \Delta \partial
  \quad \quad &,& \quad \quad
  \partial^\dagger \Delta^{-1} = \Delta^{-1} \partial^\dagger \quad .
\ee 
For the calculation of $u_{\alpha\beta}$ we need the covariance
property of the blocking operator $C\partial = \nabla C_S$, and 
$C_S \Delta^{-1} \R = 0$ from the explicit calculation of the
projector $\R$. 
This yields
\be
  u_{\alpha\beta} =  C v_{\alpha\beta} C^\dagger
  = C \Delta^{-1} C^\dagger - (1-\frac1\beta) \nabla 
  \underbrace{C_S \Delta^{-2} C_S^\dagger}_{u_{\Delta^2}} \nabla^\dagger
  \quad .
\ee
Here we already see that $u_{\alpha\beta}$ is independent of $\alpha$.
We see as well that $u_{\alpha\beta}$ does not know about the projector $\R$.
It looks as if it stems from a fundamental propagator
with one term fixing the gauge without distinguishing between high
and low frequencies.

For notational ease we introduce abbreviations
for operators which act on functions on the block lattice
\be
  () &=& (C\Delta^{-1} C^\dagger)
\nn
  { [] } &=& [\nabla^\dagger ()^{-1} \nabla]
\nn
  \{_\beta\} &=& \{[]^{-1} - (1-\frac1\beta)^{-1} []^{-1}
  u_{\Delta^2}^{-1}[]^{-1}\} \quad .
\ee 
$[]$ and $\{_\beta\}$ act on scalars and $()$ is the block gauge field 
propagator for the special choice of the parameter $\beta = 1$.
Their inverses can be computed in momentum space.
Note that only $\{_\beta\}$ is gauge parameter dependent, and that the
limit $\beta \to 0$ can be taken to be 
\be
  \{_0\} = []^{-1} 
\ee
whereas $\{_\beta\}$ is divergent for $\beta \to 1$.
If one is interested in
the special case $\beta=1$ one inserts this in eq.(\ref{mvab}) of the
fundamental gauge fixed propagator and applies the formulae of
eq.(\ref{mgeneral}) directly.
The calculations are much shorter than in the general case and pose no
problems.

The inverse block propagator is 
\be
  u^{-1}_{\alpha\beta} = 
  ()^{-1} - ()^{-1} \nabla []^{-1} \{_\beta\}^{-1} []^{-1} 
                    \nabla^\dagger ()^{-1}  \quad .
\ee
Now we can calculate the interpolation kernel
\be
  \A_{\alpha\beta} = v_{\alpha\beta} C^\dagger u^{-1}_{\alpha\beta} 
  = \Delta^{-1} C^\dagger u^{-1}_{\alpha\beta} 
  + \partial \Delta^{-2} C_S^\dagger u_{\Delta^2}^{-1}[]^{-1}
  \{_\beta\}^{-1} []^{-1} \nabla^\dagger ()^{-1} \quad .
\ee
The interpolation kernel is also independent of $\alpha$.
The same must then be true for the background field.
Contrary to  $u_{0 0}$ and $v_{0 0}$, $\A_{0 0}$ does exist.
Neither $\A_{\alpha\beta}$ nor the propagator of the background field
contain the projector $\R$ explicitly.
We expect that the fluctuation propagator exists
even for $\beta = 0$.
Therefore the background propagator must contain a term proportional to
$1/ \beta$ and $1-\R$
so that the divergence of $v_{\alpha\beta}$ in the limit
$\beta \to 0$ is completely contained in the background propagator and
the fluctuation propagator stays finite.

We recall from the explicit calculation of the projector
\be
  1-\R = \Delta^{-1} C_S^\dagger u_{\Delta^2}^{-1} C_S \Delta^{-1}
  \quad .
\ee
The following expression emerges as the result of the calculation of
the background propagator.
\be
  v^s_{\alpha\beta} 
  &=& \A_{\alpha\beta} C v_{\alpha\beta}
  = v_{\alpha\beta} C^\dagger u^{-1}_{\alpha\beta} C v_{\alpha\beta}
\\
  &=& \Delta^{-1} C^\dagger ()^{-1} C \Delta^{-1}
  - (1-\frac1\beta) \partial \Delta^{-1} (1 - \R) 
                      \Delta^{-1} \partial^\dagger
\nn
  &-& \left( \partial \Delta^{-2} C_S^\dagger u_{\Delta^2}^{-1}
        - \Delta^{-1} C^\dagger ()^{-1} \nabla \right)
           []^{-1} \{_\beta\}^{-1} []^{-1}
  \left( u_{\Delta^2}^{-1} C_S \Delta^{-2} \partial^\dagger 
         - \nabla^\dagger ()^{-1} C \Delta^{-1} \right)
\nonumber
\ee
Here we finally see that the divergent term $\sim (1-1/ \beta)$
is exactly the one already
present in the fundamental propagator.
Therefore it is obvious that the fluctuation propagator cannot contain
it and is indeed finite for $\beta \to 0$ as expected.
The result for the fluctuation propagator reads
\be
\label{mgammaab}
  \Gamma_{\alpha\beta} &=& v_{\alpha\beta} - v^s_{\alpha\beta} 
\\
  &=& \Delta^{-1} - \Delta^{-1} C^\dagger ()^{-1} C \Delta^{-1}
  - (1 - \frac1\alpha) \partial \Delta^{-1} \R \Delta^{-1} \partial^\dagger
\nn
  &+& \left( \partial \Delta^{-2} C_S^\dagger u_{\Delta^2}^{-1}
        - \Delta^{-1} C^\dagger ()^{-1} \nabla \right)
           []^{-1} \{_\beta\}^{-1} []^{-1}
 \left( u_{\Delta^2}^{-1} C_S \Delta^{-2} \partial^\dagger 
         -  \nabla^\dagger ()^{-1} C \Delta^{-1} \right) \quad .
\nonumber
\ee
The reason that the fluctuation propagator is $\beta$-dependent
at all i.e. it feels the gauge fixing on the coarse
scale is that $\A_{\alpha\beta}$ is chosen to separate the kinetic
term including the gauge fixing.

The dependence of the fluctuation propagator on $\beta$ is disturbing.
If we want a gauge invariant effective action we have to take
the limit $\beta \to 0$.
This removes the gauge fixing term and any $\beta$-dependence of
the loop integrations.
In the course of the further calculations we might want to
reintroduce gauge fixing for the degrees of freedom which are next to 
be integrated out.
But the $\beta$-dependence of the loop integrations is lost and cannot
be restored.
There are two possibilities to explain this.
The first one is that the result of the loop integrations is 
$\beta$-independent despite the fact that the fluctuation propagator is
$\beta$-dependent.
The second and threatening possible explanation is that something is wrong with
taking the limits.
This needs to be ruled out.

In addition the fluctuation propagator depends on the choice of the
gauge fixing within blocks and on the associated gauge
parameter $\alpha$.
It has to be shown that the effective action obtained by integration
of the fluctuation field does not depend on $\alpha$.
If the integration could be done exactly the Slavnov-Taylor identities
guaranteed this.
If approximations are used one has to show the independence of the
effective action on $\alpha$ explicitly.
This is the standard situation in all perturbative calculations 
in gauge theories.

Finally as promised the summary of the results with gauge fixing.
\be
\label{msummary}
  u_{\alpha\beta} &=& C \Delta^{-1} C^\dagger 
  - (1-\frac1\beta) \nabla u_{\Delta^2} \nabla^\dagger
\\
\nn
  \A_{\alpha\beta} 
  &=& \Delta^{-1} C^\dagger u^{-1}_{\alpha\beta} 
  + \partial \Delta^{-2} C_S^\dagger u_{\Delta^2}^{-1}[]^{-1}
  \{_\beta\}^{-1} []^{-1} \nabla^\dagger ()^{-1} 
\nn
\nn
  v^s_{\alpha\beta} 
  &=& \Delta^{-1} C^\dagger ()^{-1} C \Delta^{-1}
  - (1-\frac1\beta) \partial \Delta^{-1} (1 - \R) 
                      \Delta^{-1} \partial^\dagger
\nn
  &-& \left( \partial \Delta^{-2} C_S^\dagger u_{\Delta^2}^{-1}
        - \Delta^{-1} C^\dagger ()^{-1} \nabla \right)
           []^{-1} \{_\beta\}^{-1} []^{-1}
  \left( u_{\Delta^2}^{-1} C_S \Delta^{-2} \partial^\dagger 
         - \nabla^\dagger ()^{-1} C \Delta^{-1} \right)
\nn
\nn
  \Gamma_{\alpha\beta} 
  &=& \Delta^{-1} - \Delta^{-1} C^\dagger ()^{-1} C \Delta^{-1}
  - (1 - \frac1\alpha) \partial \Delta^{-1} \R \Delta^{-1} \partial^\dagger
\nn
  &+& \left( \partial \Delta^{-2} C_S^\dagger u_{\Delta^2}^{-1}
        - \Delta^{-1} C^\dagger ()^{-1} \nabla \right)
           []^{-1} \{_\beta\}^{-1} []^{-1}
 \left( u_{\Delta^2}^{-1} C_S \Delta^{-2} \partial^\dagger 
         -  \nabla^\dagger ()^{-1} C \Delta^{-1} \right)
\nonumber
\ee

\subsubsection{Covariant interpolator}
\label{ssAcov2}

To be on the safe side we go back to the quantities $v^{-1}_{\alpha\beta}$, 
$u^{-1}_{\alpha\beta}$ and $\A_{\alpha\beta}$ take the limits $\alpha
\to 0$ and $\beta \to 0$
and show explicitly that $v^{-1}_{00}$, $u^{-1}_{00}$ and $\A_{00}$ fulfill
eq.(\ref{msep})
\be
   v^{-1}_{00} &=& \partial^\dagger \partial
\nn
  \A_{00} &=& \Delta^{-1} C^\dagger u^{-1}_{00} 
  + \partial \Delta^{-2} C_S^\dagger u_{\Delta^2}^{-1}
  []^{-1} \nabla^\dagger ()^{-1} 
\nn
  u^{-1}_{00} &=& ()^{-1} - ()^{-1} \nabla []^{-1} \nabla^\dagger ()^{-1}  
  \quad .
\label{mA00}
\ee
It is obvious from the gauge invariance of $v^{-1}_{00}$ that the
second summand of $\A_{00}$ is not determined by eq.(\ref{msep}).
This is the gauge invariance of the interpolation kernel
\be  
  v^{-1}_{00} \A_{00} &=& \partial^\dagger \partial 
         \Delta^{-1} C^\dagger u^{-1}_{00} 
\nn
  &=& (1 - \partial \partial^\dagger \Delta^{-1}) C^\dagger u^{-1}_{00} 
\nn
  &=& (C^\dagger - \Delta^{-1} \partial \partial^\dagger C^\dagger) 
      u^{-1}_{00} 
\nn
  &=& (C^\dagger - \Delta^{-1} \partial C_S^\dagger \nabla^\dagger) 
      u^{-1}_{00}
\nn
  &=& C^\dagger u^{-1}_{00}
\ee
since 
\be
  \nabla^\dagger u^{-1}_{00} = \nabla^\dagger (()^{-1} - ()^{-1} \nabla
  []^{-1} \nabla^\dagger ()^{-1} ) = 0 \quad .
\label{mugauge}
\ee
One sees immediately that
\be 
  \A_{00} \nabla 
  &=& \Delta^{-1} C^\dagger \underbrace{u^{-1}_{00} \nabla}_{0}  
  + \partial \Delta^{-2} C_S^\dagger u_{\Delta^2}^{-1}
  []^{-1} \underbrace{\nabla^\dagger ()^{-1} \nabla}_{[]}
\nn
  &=& \partial \underbrace{\Delta^{-2} C_S^\dagger
  u_{\Delta^2}^{-1}}_{=:\A_S} \quad .
\ee
The above defined $\A_S$ is identical to $\A^{(\lambda)}$ of
eqs.(\ref{mr0}) and (\ref{mr2}) and fulfills $C_S \A_S = 1$.
We use it to rewrite the projectors $R$ and $\R$
\be
  R &=& 1-\A_S C_S
\nn
  \R &=& \Delta R \Delta^{-1} 
  = 1 - \Delta \A_S C_S \Delta^{-1} = \R^\dagger \quad .
\ee
Explicit calculation confirms as well
\be
  C \A_{00} &=& \underbrace{ C \Delta^{-1} C^\dagger}_{()} u^{-1}_{00}
  + C \partial \Delta^{-2} C_S^\dagger u_{\Delta^2}^{-1}
  []^{-1} \nabla^\dagger ()^{-1} 
\nn
  &=& 1 - \nabla []^{-1} \nabla^\dagger ()^{-1} 
  + \nabla \underbrace{C_S \Delta^{-2} C_S^\dagger}_{u_{\Delta^2}}
  u_{\Delta^2}^{-1} []^{-1} \nabla^\dagger ()^{-1}
\nn
  &=& 1 - \nabla []^{-1} \nabla^\dagger ()^{-1} 
  + \nabla []^{-1} \nabla^\dagger ()^{-1}
\nn
  &=& 1 \quad .
\ee
We use eq.(\ref{mr4}), the explicit form of $\A_{00}$ from eq.(\ref{mA00}) 
and eq.(\ref{mugauge}) to show that $\A_{00}$ fulfills the block
Landau gauge condition
\be
      \R \partial^\dagger \A_{00}
  &=& \R \partial^\dagger \Delta^{-1} C^\dagger u^{-1}_{00} 
    + \R \underbrace{\partial^\dagger \partial}_{\Delta} 
         \underbrace{\Delta^{-2} C_S^\dagger u_{\Delta^2}^{-1} []^{-1} 
         \nabla^\dagger ()^{-1}}_{\mbox{\small 0-form if applied to $a$}}
\nn
  &=& \R \Delta^{-1} \partial^\dagger C^\dagger u^{-1}_{00} 
    + \underbrace{\R \Delta^{-1} C_S^\dagger}_{0} 
    u_{\Delta^2}^{-1} []^{-1} \nabla^\dagger ()^{-1}
\nn
  &=&  \R \Delta^{-1} C_S^\dagger \nabla^\dagger u^{-1}_{00} = 0 \quad .
\label{mproj}
\ee
This property of $\A_{00}$ is not a relict of the gauge fixing in the
course of its calculation.
$\A_{00}$ and hence the background field fulfill the
block Landau gauge condition despite the fact that $\A_{00}$ is gauge 
covariant.
The reason is that $\R \partial^\dagger$ acts on $\A_{00}$
as a projector to high frequency fields whereas $\A_{00} C$ is a
projector to low frequencies.

The gauge transformation for the background field $a^s$ is
\be
  a^s = \A_{00} C a \trafo_{gauge} \A_{00} C (a-\partial\lambda) 
      = a^s - \partial \A_S C_S \lambda
     = a^s - \partial (1-\R) \lambda
\ee
and by subtraction it follows for the fluctuation field
\be
 a^h \trafo_{gauge} a^h - \partial \R \lambda \quad .
\ee
We have now successfully factorized the generating functional
into a high and low frequency contribution without fixing the gauge.
\be
  Z[j] &=& \int \D a e^{-S_M[a] + \langle a,j \rangle}
\nn
     &=& \int \D A e^{ - \frac12 \langle A, \A^\dagger v_{00}^{-1} \A A \rangle 
             + \langle j, \A A \rangle }
          \int \D a^h \delta(Ca^h) 
  e^{ - \frac12 \langle a^h, v_{00}^{-1} a^h \rangle + \langle j, a^h \rangle }
\ee

\subsubsection{The fluctuation propagator}
\label{ssfluc}

To obtain the effective action we need to perform the fluctuation integral.
The integrand is invariant under gauge transformations 
\be
  a \trafo_{gauge} a + \partial \lambda 
  \quad \mbox{with} \quad 
  C_S \lambda = 0 \quad .
\label{mflucgauge}
\ee
Gauge transfomations with nonvanishing block average are excluded
since they violate $Ca=0$.
Before we can perform the integration it is necessary to remove also
the  gauge freedom of eq.(\ref{mflucgauge}).

We have chosen the blocking and interpolation operators 
in such a way that the blockspin
and the background field still possess a restricted gauge invariance.
If we now apply a gauge fixing indiscriminately to the frequency of
the gauge transformations the desired gauge invariance of the
blockspin is destroyed.

To avoid this we have already constructed the projector $\R$.
It ensures that only the gauge degrees of freedom
$\lambda$ with $C\lambda = 0$, i.e. vanishing block averages of
$\lambda$, are fixed but the freedom of performing global
gauge transformations on each block remains.
This freedom is reflected by the
fact that $\exp \{-S_{\rm eff}[{\bf A}] \}$ is invariant under gauge
transformations 
${\bf A} \trafo {\bf A}^\Lambda = {\bf A}-\nabla \Lambda$, 
where $\Lambda$ is a scalar function on the lattice and $\nabla$ is the
lattice exterior derivative.

As before the block gauge fixing condition is $\R \partial^\dagger a =
0$, which can be raised with the Faddeev-Popov trick to the exponent.
This time it is a real gauge fixing which will not be removed after
the calculations.
Therefore we should add a ghost term to the exponent.
But since the gauge fixing condition is linear in the hard field 
the ghosts decouple and we suppress them.
So the fluctuation integral is defined as 
\be  
  Z^h[j] &=& \int \D a^h \delta(Ca^h) 
  e^{-\frac12 
  \langle a^h, (v_{00}^{-1} + \alpha \partial \R \partial^\dagger ) 
               a^h \rangle 
  + \langle j, a^h \rangle }
\nn
  &=:& e^{\frac12 \langle j, \Gamma j \rangle} \quad .
\ee
The last expression is the definition of the fluctuation
propagator $\Gamma$ we are looking for.
The integral is only defined due to the 
delta-function.
The partly gauge fixed kinetic operator
$v_{\alpha 0}^{-1} = v_{00}^{-1} + \alpha \partial \R
\partial^\dagger$ is not invertible since it has the zero modes
$\partial \lambda$ with $C_S \lambda \neq 0$.
The constraint $\delta(Ca^h)$ serves two purposes. 
It forces the field $a^h$ to
be a fluctuation field with block average zero and 
furthermore it excludes 
the remaining zero modes of $v_{\alpha 0}^{-1}$ from the range of
integration.

From the preceding calculations we guess that 
$\Gamma = \Gamma_{\alpha 0}$.
To prove this we must show
\be 
  \Gamma_{\alpha 0} v_{\alpha 0}^{-1} = ( 1 - \A_{00} C)
  \quad \mbox{and} \quad
  C \Gamma_{\alpha 0} = 0 \quad .
\ee
For both purposes we use the general formulae (\ref{mgeneral}) and adapt 
them to the case $\beta = 0$.
\be
  \Gamma_{\alpha \beta} &=& ( 1 - \A_{\alpha \beta} C)  v_{\alpha \beta} 
\nn
  \Gamma_{\alpha \beta} v_{\alpha \beta}^{-1} &=& ( 1 - \A_{\alpha \beta} C)
\nn
  C \Gamma_{\alpha \beta} &=& 0
\ee
for $\beta = 0$ this results in
\be 
  \Gamma_{\alpha 0} v_{\alpha 0}^{-1} &=& ( 1 - \A_{\alpha 0} C)
\nn
  C \Gamma_{\alpha 0} &=& 0
\ee
and since $\A_{\alpha \beta}$ is independent of $\alpha$
\be 
  \Gamma_{\alpha 0} v_{\alpha 0}^{-1} = ( 1 - \A_{0 0} C) \quad .
\ee
The fluctuation propagator in special gauges can be taken from 
eq.(\ref{mgammaab}) for special values of the block gauge parameter $\alpha$.
The block Landau gauge (BLG) is $\alpha \to \infty$ and taking the limit
gives no problems.
In addition to the properties of the fluctuation propagator for finite
$\alpha$ one can show that
\be
  \partial^\dagger \Gamma_{BLG} = 0 \quad .
\ee

\subsubsection{The effective action}
\label{sseff}
Starting with a fundamental action of the form $S_M[a] + S_I[a]$ we
insert our results into an equation of the type of eq.(\ref{veff}).
In applications the interaction $S_I[a]$ may depend on other fields as well.
This is the case for scalar electrodynamics which we treat in the 
following section \ref{secsqed}.
We obtain the gauge invariant effective action 
\be
  e^{-S_{\mathrm{eff}}[{\bf A}]}
  = \int Da \delta(Ca-{\bf A}) 
  e^{-S_M[a] - S_I[a]
  - \frac12 \alpha \langle \partial^\dagger a,\R \partial^\dagger a \rangle}
  \quad .
\ee   
With the field split $a=\A_{00} {\bf A} + a^h$ the Maxwell part of the
action is separated.
The gauge fixing part acts only onto the fluctuation field.
The only interaction is via $S_I$
\be
  e^{-S_{\mathrm{eff}}[{\bf A}]}
  = e^{-S_M[\A_{00} {\bf A}]} \int Da^h \delta(Ca^h) 
  e^{-S_M[a^h] - S_I[\A_{00} {\bf A} + a^h]
  - \frac12\alpha\langle\partial^\dagger a^h,\R \partial^\dagger a^h \rangle}
  \quad .
\ee
The integration of the fluctuation field can be done perturbatively
\be
  e^{-S_{\mathrm{eff}}[{\bf A}]}
  &=& e^{-S_M[\A_{00} {\bf A}] 
  - S_I[\A_{00} {\bf A} + \frac\delta{\delta j}]}
  \int Da^h \delta(Ca^h) 
  e^{-S_M[a^h] + \langle j, a^h\rangle
  - \frac12\alpha\langle\partial^\dagger a^h,\R \partial^\dagger a^h \rangle}
\nn
  &=& e^{-S_M[\A_{00} {\bf A}] 
  - S_I[\A_{00} {\bf A} + \frac\delta{\delta j}]}
  e^{\frac12 \langle j,\Gamma j \rangle} \quad .
\ee
In the free case this reduces to 
\be
  e^{-S_{\mathrm{eff}}[{\bf A}]}
  &=& e^{-S_M[\A_{00} {\bf A}]}
\nn
  &=& e^{-\frac12 \langle \partial \A_{00} {\bf A},
  \partial \A_{00} {\bf A} \rangle}
\nn
  &=& e^{-\frac12 \langle {\bf A}, 
  \A_{00}^\dagger \partial^\dagger \partial \A_{00} {\bf A} \rangle}
\nn
  &=& e^{-\frac12 \langle {\bf A},u_{00}^{-1} {\bf A} \rangle} \quad .
\ee

\section{Scalar electrodynamics}
\label{secsqed}
  
\subsection{The action of scalar electrodynamics}
\label{sssqedaction}

The action is defined by
\be
  S[a,\phi] = S_M[a] 
  + \frac12 \langle \phi, (- D_\mu^\dagger[a] D_\mu[a]) \phi \rangle
  + V[\phi] \quad .
\ee
The scalar self-interaction is 
\be
  V[\phi] = \int_z 
  \left( \frac12 m_0^2 \phi \phi^* + \frac g{4!} (\phi \phi^*)^2 \right)
  \quad .
\ee
$\phi$ denotes a complex Higgs field and 
\be
  D_\mu[a] = \dmu - a_\mu
\ee
the covariant derivative.

The action is invariant under the following simultaneous gauge
transformations
\be
 a_\mu \trafo_{gauge} a_\mu^\lambda = a_\mu - \dmu \lambda
 \quad \mbox{and} \quad  
 \phi  \trafo_{gauge} \phi^\lambda = \phi e^{\lambda} \quad .
\ee
To evaluate the low energy effective action we need to define
blockspin and/or background field and fluctuation fields.
The definitions for the gauge fields can be taken literarily from the
preceding section \ref{secmaxwell}.
Since the Higgs field is not invariant under gauge transformations we
cannot simply use the results of section \ref{secscalar} for it.

We will show two different ways to deal with it.
The first regards the Higgs field as a scalar field with a gauge field
dependent kinetic operator.
The proceeding is analogous to section \ref{secscalar}.
The transformation properties for block, background and fluctuation
Higgs fields follow from the choice of averaging and interpolation
operator.

The second way regards the Higgs field as subject to gauge
transformations.
Its emphasis is to make the gauge transformation properties of block,
background and fluctuation Higgs field compatible with those of block,
background and fluctuation Maxwell fields evaluated in section
\ref{secmaxwell}.
The choice of averaging and interpolation operator satisfies this demand.

\subsection{Block spin transformation for the Higgs field at zero
temperature}
\label{sshiggs}

Here we explain the first way to choose the averaging operator.
For a given Higgs field $\phi(z)$ we wish to define a block Higgs field
$\Phi(x)$ on the lattice $\Lambda$ in a gauge covariant way. 
The blocking procedure for the $\phi^4$-theory cannot be used as it
stands because it is not gauge covariant. 
In order to maintain gauge
covariance, we must use averaging kernels depending on the gauge
field $a$.

We denote the averaging operator for the Higgs field by $C_H[a]$.
and choose the block operator to be linear with respect to the Higgs field,
therefore we can write it as an integral operator 
\be
  \Phi(x) = (C_H[a]\phi)(x)
          = \int_z C_H[a](x,z) \phi(z) \quad .
\ee
In the $\phi^4$-theory we used an averaging kernel which was
constant on blocks and vanished outside. 
A constant is the lowest
eigenvector of the Laplacian with Neumann boundary conditions on block
boundaries.
The natural generalization to the gauge covariant situation is as
follows
\footnote{Here we deviate from the work of Ba{\l}aban, Jaffe and
  Imbrie \cite{balimja}}.

Let $-\Delta_a = - (\partial - a)^\dagger (\partial -a)$ denote the
covariant Laplacian and $-\Delta_a^{N,x}$ the covariant Laplacian with
Neumann boundary conditions on the block boundary of $x$. We demand
that
\be
  -\Delta_a^{N,x} C_H[a]^\dagger(z,x)
  = \varepsilon_0[a](x) C_H[a]^\dagger(z,x)
\label{heigen}
\ee
and
\be
  C_H[a]^\dagger(z,x) = 0
  \quad \mbox{for} \quad \quad
  z \notin x
\ee
where $\varepsilon_0[a](x)$ is the lowest eigenvalue of
$-\Delta_a^{N,x}$.  
In addition we impose the normalization condition
\be
  C_H[a] C_H[a]^\dagger  = 1 \quad .
\ee
This leaves the freedom of multiplying $C_H[a]$ with an 
$a$-dependent phase factor
\be
  C_H[a]^\dagger(z,x)
  \to C_H[a]^\dagger(z,x) \eta[a](x) \quad .
\ee
It follows from the gauge covariance of the eigenvalue problem
(\ref{heigen}) that under a gauge transformation
\be
  C_H[a^\lambda] \phi^\lambda 
  = (C_H[a] \phi)^\Lambda \quad .
\label{hginv}
\ee
$\Lambda$ is a gauge transformation on the lattice which depends
on $\lambda$ and on the choice of conventions to fix $C_H$
uniquely.  
The freedom in the choice of an $a$-dependent phase factor
may be exploited to demand that
\be
  \Lambda(x) = (C_S \lambda)(x)
\ee
where $C_S$ is the scalar averaging kernel introduced in section
\ref{secscalar}.  
Eq.(\ref{hginv}) parallels the gauge covariance
property of the blocking procedure for the gauge field.

We can compute the averaging kernel $C_H[a]$ as a solution
of the eigenvalue equation (\ref{heigen}) by standard quantum
mechanical perturbation theory.  
A prototype of such a computation is found in reference \cite{palma}.

\subsection{Interpolation kernel and fluctuation propagator 
for the Higgs field}
\label{ssHinter}

The interpolation kernel $\A_H[a]$ for the Higgs field
will also depend on the gauge field $a$ because the averaging kernel
does. 
In order to split the kinetic term for the Higgs field, we closely follow
the arguments in section \ref{secscalar}.
First we adapt our notation and call the gauge dependent kinetic
operator of the fundamental Higgs field 
\be
v_H[a]^{-1} = - D_\mu^\dagger[a] D_\mu[a] \quad .
\label{hvH}
\ee
The interpolation kernel is then defined via 
\be
  v_H[a]^{-1} \A_H[a] = C_H[a]^\dagger u_H[a]^{-1}
\ee
for some choice of $u_H[a]^{-1}$, and if
\be
  C_H[a] \A_H[a] = 1
\ee
the last condition implies that
\be
  u_H[a] = C_H[a] v_H[a] C_H[a]^\dagger 
  \quad \mbox{and} \quad \quad
  u_H[a]^{-1} = \A_H[a]^\dagger v_H[a]^{-1} \A_H[a]
  \quad .
\ee
See appendix \ref{scalarcalc} for the derivation of these formulae in
the scalar case.

We can now split the fundamental Higgs field into background and
fluctuation fields
\be
  \phi = \phi^s + \phi^h = \A_H[a] \Phi + \phi^h \quad .
\ee
The fluctuation field propagator for the Higgs field is
\be
  \Gamma_H[a]
  = v_H[a] - \A_H[a] C_H[a] v_H[a] C_H[a]^\dagger \A_H[a^\dagger]
\ee
where $v_H[a] = -\Delta_a^{-1}$ is the full gauge covariant
free massless propagator for the Higgs field.

The interpolation operator and the gauge covariant fluctuation field
propagator can also be computed by standard quantum mechanical
perturbation theory \cite{palma}.

\subsection{The perfect action of scalar electrodynamics: Representation
  as a functional integral with gauge fixed fluctuation fields}
\label{sslandau}

We use again the projector $\R$ of section \ref{secmaxwell} to define
the fluctuation integral.
The essential point making this possible is that both delta-functions
\be
  \delta(Ca-{\bf A}) 
  \quad \quad \mbox{and} \quad \quad
  \delta(C_H[a]\phi-\Phi)
\ee
are invariant under gauge transformations 
$a \to a^\lambda$ and $\phi \to \phi^\lambda$ 
which obey the constraint $C_S\lambda = 0$.  
Proceeding as before we obtain
\be
  e^{-S_{\rm eff}(\Phi,{\bf A})}
  = \int \D a \int \D \phi && \delta(Ca-{\bf A}) \delta(C_H[a]\phi-\Phi)
\nn  
  && e^{-S_M[a] - S_{\rm gf}[a]
  -\frac12 \langle \phi, v_H[a]^{-1} \phi \rangle - V[\phi]}
\ee
where
\be
  S_{\rm gf}[a]
  = \frac12 \alpha \langle \partial^\dagger a, \R \partial^\dagger a \rangle
  \quad .
\ee
Now we split the fields into background and fluctuation part
\be
  \phi = \A_H[a]\Phi + \phi^h
  \quad \quad \mbox{and} \quad \quad
  a = \A {\bf A} + a^h \quad .
\ee
The expression for the perfect action becomes
\be
  e^{-S_{\rm eff}[\Phi,{\bf A}]}
  &=& \int \D a^h \int \D \phi^h \delta(C a^h) \delta(C_H[a]\phi^h)
\nn
  & & \exp \{-S_M[\A {\bf A}] - S_M[a^h] - S_{\rm gf}[a^h]
\nn
  & & - \frac12 \langle  \Phi, 
  \underbrace{\A_H^\dagger[\A {\bf A}+a^h] 
   v_H[\A {\bf A}+a^h]^{-1} \A_H[\A {\bf A}+a^h]}_{u_H[\A {\bf A}+a^h]^{-1}}
   \Phi \rangle 
\nn
  & & - \frac12 \langle \phi^h, v_H[\A {\bf A}+a^h]^{-1} \phi^h \rangle 
  - V[\A_H[\A {\bf A}+a^h]\Phi + \phi^h] \} \quad .
\label{hSeff1}
\ee
We used the fact that $S_{\rm gf}[\A {\bf A}+a^h] = S_{\rm gf}[a^h]$ 
since $\R \partial^\dagger \A = 0$, see eq.(\ref{mproj}).

We separate the terms of zeroeth order in $a^h$ and $\phi^h$
\be
  u_H[\A {\bf A}+a^h]^{-1} 
  &=& u_H[\A {\bf A}]^{-1} + E[{\bf A},a^h]
\nn
  V[\A_H[\A {\bf A}+a^h]\Phi + \phi^h] 
  &=& V[\A_H[\A {\bf A}]\Phi] + W[{\bf A}, a^h, \Phi ,\phi^h] \quad .
\label{hent}
\ee
The effective action becomes
\be
  S_{\rm eff}[\Phi, {\bf A}]
  = S_{M,{\rm eff}}[{\bf A}]
  - \frac12  \langle \Phi, u_H[\A {\bf A}]^{-1} \Phi \rangle
  + V[\A_H[\A {\bf A}]\Phi]  
  + \tilde S_{\rm eff}[\Phi, {\bf A}]
\ee
with $\tilde S_{\rm eff}[\Phi, {\bf A}]$ the result of the fluctuation
integral
\be
  e^{-\tilde S_{\rm eff}[\Phi, {\bf A}]}
  &=& \int \D a \int \D \phi \delta(Ca-{\bf A}) \delta(C_H[a]\phi-\Phi)
\nn
  & & \exp \{- S_M[a^h] - S_{\rm gf}(a^h)
  -\frac12 \langle \phi^h, v_H[a]^{-1} \phi^h \rangle
\nn
  & & - \frac12 \langle \Phi, E[{\bf A},a^h] \Phi \rangle
  - W[{\bf A}, a^h, \Phi, \phi^h]\} 
\ee
with $E$ and $W$ defined in eq.(\ref{hent}).
To zeroeth order in the fluctuation field propagators $\tilde S
_{\rm eff}[\Phi, {\bf A}] = 0$ after subtracting a constant.

\subsection{Alternative definition of the block operator}
\label{ssalternative}

An alternative way to define averaging and interpolation operators
for the Higgs field concentrates on its behaviour under gauge
transformations.
We recall from section \ref{secmaxwell} the transformation properties
of the various gauge fields
\be
  a_\mu && \trafo_{gauge} a_\mu^\lambda = a_\mu - \dmu \lambda
\nn 
  {\bf A}_\mu && \trafo_{gauge} {\bf A}_\mu^\Lambda 
               = {\bf A}_\mu - \nabla_\mu \Lambda
\nn
  a^s_\mu && \trafo_{gauge} a_\mu^{s \,\lambda^s} = a^s_\mu - \dmu \lambda^s
\nn
  a^h_\mu && \trafo_{gauge} a_\mu^{h \,\lambda^h} = a^h_\mu - \dmu \lambda^h
\ee
with the definitions of the gauge fields
\be
  {\bf A} = C a
  \quad , \quad
  a^s = \A {\bf A} = \A C a
  \quad , \quad 
  a^h = a - a^s = (1- \A C) a
\ee
and the gauge transformations
\be
  \Lambda = C_S \lambda
  \quad , \quad 
  \lambda^s = \A_S \Lambda = \A_S C_S \lambda
  \quad , \quad 
  \lambda^h = \lambda - \lambda^s = (1- \A_S C_S) \lambda \quad .
\ee
For the Higgs fields we demand gauge transformation properties
analogous to those for the gauge fields
\be
  \phi && \trafo_{gauge} \phi^\lambda = \phi e^\lambda
\nn 
  \Phi && \trafo_{gauge} \Phi^\Lambda = \Phi e^\Lambda
\nn
  \phi^s && \trafo_{gauge} \phi^{s \,\lambda^s} = \phi^s e^{\lambda^s}
  \quad .
\ee
This suggests the following averaging and interpolation procedures
\be
  \Phi(x) &=& (C_H[\phi])(x) = e^{\int_z C_S(x,z) \ln \phi(z)}
\nn
  \phi^s(z) &=& (\A_H[\Phi])(z) = e^{\int_x \A_S(z,x) \ln \Phi(x)}
  \quad .
\ee
Inserting the gauge transformation property for the fundamental Higgs field 
we obtain the desired properties for blockspin and background field.

The branch of the logarithm must be chosen self-consistently.
Since we want to be able to expand in a small fluctuation field, we 
suppose that there exists some smooth field $\psi$ interpolating the
blockspin $\Phi$, so that $\phi$ is
close to $\psi$.
Since $\psi(z)$ shall not vary much within a block, we approximate it 
with its value in the block center $\psi(\hat x)$.
We now precisize the definition of the blockspin to 
\be
  \Phi(x) = \psi(\hat x) e^{\int_z C_S(x,z) \ln \psi(\hat x)^{-1} \phi(z)}
  \quad .
\ee
This way we use $\psi$ as a reference field to determine the branch
of the logarithm.
The definition  of the background field must contain the same
reference field.
\be
  \phi^s(z) = \psi(z) e^{\int_x \A_S(z,x) \ln \psi(\hat x)^{-1} \Phi(x)}
  \quad .
\ee 
In a non-perturbative approach, one cannot rely on the existence of
some smooth field $\psi$ to determine the branch of the logarithm.
One has to split the field space in a small field region, where the
above considerations remain valid, and a large field region where the
above blockspin definition cannot be applied. 
This is discussed in the work of Ba{\l}aban \cite{balaban}.

The difference to the operators $C_H[a]$ and $\A_H[a]$ of the
subsections \ref{sshiggs} and \ref{ssHinter} is that here the operators 
do not depend on the gauge field $a$ and act nonlinearly on
the Higgs field.
  
From $C_S \A_S = 1$ we infer
\be
  C_H[\phi^s] = C_H[\A_H[\Phi]] = \Phi \quad .
\label{hcomp}
\ee
The blockspin is completely defined by the background field.
Since the averaging operator is not linear, but commutes with 
multiplication, it is reasonable to introduce the fluctuation field as
\be 
 \phi = \phi^s e^{\phi^h} \quad .
\ee
Applying the averaging operator to this identity yields
\be
  C_H[e^{\phi^h}] = 1 
  \quad \quad \mbox{i.e.} \quad \quad
  C_S \phi^h = 0 \quad .
\ee
The last equation is the reason to define the fluctuation field 
in the exponential,
so that it makes sense to expand around vanishing fluctuation field to
evaluate the fluctuation integral.
With this definition the gauge transformation property of the
fluctuation field is
\be
 \phi^h \trafo_{gauge} \phi^h +\lambda^h \quad .
\ee
Note that $\phi^h$ is complex as is $\phi$ whereas $\lambda$ and
$\lambda^h$ are purely imaginary functions.

We are now prepared to define the effective block action.
\be
  e^{-S_{\rm eff}[\Phi,{\bf A}]} 
  = \int \D a \D \phi & & \delta(\Phi - C_H[\phi]) \delta({\bf A} - Ca)
\nn
  & & e^{- S_M[a] - S_{\rm gf}[a] 
     - \frac12 \langle \phi, v_H[a]^{-1} \phi \rangle - V[\phi] }
\ee
$v_H[a]^{-1}$ is the covariant kinetic term of the Higgs field given in
eq.(\ref{hvH}).
Since blockspin, background and fluctuation Higgs fields are defined
nonlinearly it is advisable to be careful with their introduction to
keep track of the functional determinants connected with these
nonlinearities.
Therefore we insert background and fluctuation fields with their
defining delta-functions into the definiton of the effective action
\be
  e^{-S_{\rm eff}[\Phi,{\bf A}]}
  = \int && \D a \D a^s \D a^h 
      \delta({\bf A} - Ca) \delta(a^s - \A {\bf A}) \delta(a^h+a^s-a)
\nn
  && \D \phi \D \phi^s \D \phi^h
     \delta(\Phi - C_H[\phi]) \delta(\phi^s - \A_H[\Phi])
     \delta(\phi^h + \ln \phi^s - \ln \phi)
\nn
  && e^{- S_M[a] - S_{\rm gf}[a]
      - \frac12 \langle \phi, v_H[a]^{-1} \phi \rangle - V[\phi]}
  \quad .
\ee
The next step is to change the order of integration to remove the
fundamental fields $\phi$ and $a$.
Since $\delta(\phi^h + \ln \phi^s - \ln \phi)$ is nonlinear in the
fundamental Higgs field $\phi$ this gives rise to a functional
determinant
\be
    \det \left( \frac{\delta \phi(z)}{\delta \ln \phi(z')} \right)
  = \det \left(\delta(z-z')\phi(z)\right)  
  = \det \left({\bf 1}\phi\right) \quad .
\ee
For the integration of the gauge field $a$ this complication does not exist.
\be
  e^{-S_{\rm eff}[\Phi,{\bf A}]} 
  = \int && \D a^s \D a^h 
      \delta({\bf A} - Ca^s - Ca^h) \delta(a^s - \A {\bf A})
\nn     
  && \D \phi^s \D \phi^h 
     \delta(\Phi - C_H[\phi^s] C_H[e^{\phi^h}]) \delta(\phi^s - \A_H[\Phi])
  \det({\bf 1}\phi^s e^{\phi^h})
\nn
  && e^{- S_M[a^s+a^h] - S_{\rm gf}[a^s + a^h]
  -\frac12 \langle \phi^s e^{\phi^h}, 
  v_H[a^s+a^h]^{-1} \phi^s e^{\phi^h} \rangle 
     - V[\phi^s e^{\phi^h}] }
\ee
We now integrate the background field.
This is possible without introduction of a further functional
determinant
\be
  e^{-S_{\rm eff}[\Phi,{\bf A}]} 
  &=& \int \D a^h \delta(Ca^h) 
     \D \phi^h \delta(\Phi - C_H[\A_H[\Phi]] C_H[e^{\phi^h}]) 
     \det({\bf 1}\A_H[\Phi] e^{\phi^h})
\\
 &&  e^{- S_M[\A {\bf A} + a^h] - S_{\rm gf}[\A {\bf A} + a^h]
  -\frac12 \langle \A_H[\Phi] e^{\phi^h}, 
   v_H[\A {\bf A} + a^h]^{-1} \A_H[\Phi] e^{\phi^h} \rangle 
   - V[\A_H[\Phi] e^{\phi^h}] }
\nonumber
\ee
We simplify this result using eq.(\ref{hcomp})
and $\det = e^{{\rm tr} \ln}$.
Writing the trace as the appropriate integral we obtain
\be 
  \delta(\Phi - C_H[\A_H[\Phi]] C_H[e^{\phi^h}]) \det(\A_H[\Phi] e^{\phi^h})
  &=& \delta(C_H[e^{\phi^h}]) \det({\bf 1}\A_H[\Phi]) \det({\bf 1}e^{\phi^h})
\nn
  &=& \delta(e^{C_S \phi^h}) e^{\int_z \A_S \ln \Phi} e^{\int_z \phi^h}
\nn
  &=& \delta(C_S \phi^h) \frac1{\det({\bf 1}e^{C_S \phi^h})}
      e^{\int_x C_S \A_S \ln \Phi} e^{\int_x C_S \phi^h}
\nn
  &=& \delta(C_S \phi^h) e^{\int_x \ln \Phi} 
\nn
  &=& \delta(C_S \phi^h) \det({\bf 1}\Phi)
\ee
The unit matrix ${\bf 1}$ must be choosen according to the space-time
in which the following diagonal element is defined.

In addition we again use $S_{\rm gf}[\A {\bf A} + a^h] = S_{\rm gf}[a^h]$
to obtain
\be
\label{hSeff2}
  e^{-S_{\rm eff}[\Phi,{\bf A}]} 
  &=&  \det({\bf 1}\Phi) e^{- S_M[\A {\bf A}]} 
       \int \D a^h \delta(Ca^h) \D \phi^h \delta(C_S \phi^h)  
\\
   && e^{- S_M[a^h]- S_{\rm gf}[a^h]
  -\frac12 \langle \A_H[\Phi] e^{\phi^h}, 
   v_H[\A {\bf A} + a^h]^{-1} \A_H[\Phi] e^{\phi^h} \rangle 
     - V[\A_H[\Phi] e^{\phi^h}] } \quad .
\nonumber
\ee
To evaluate the fluctuation integral perturbatively we expand
$e^{\phi^h}$ in the fluctuation field $\phi^h$.

Finally we compare the result of this alternative blockspin
definition in eq.(\ref{hSeff2}) with the effective action of eq.(\ref{hSeff1}).
They differ only in the Higgs field part of the action.
The former definition has the advantage that the operators $C_H[a]$
and $\A_H[a]$ are linear, hence they have an integral representation.
This is not the case for the nonlinear operators $C_H$ and $\A_H$.
The advantage of using the latter is that the interaction between
Higgs and gauge fields remains restricted to the covariant kinetic
operator of the Higgs field and does not invade the self-interaction
of the Higgs as it has happened in eq.(\ref{hSeff1}).

Moreover if we deals with non-Abelian gauge fields we have to use
nonlinear blockspin definitions anyway \cite{yoma}.

\section{Adaption to finite temperature}
\label{sectemper}

\subsection{Notational preliminaries}
\label{ssnotation1}

At finite temperature $\T>0$ we have periodicity in time with period
$\beta = 1 / \T$.  
The extension $L_4$ of blocks in time direction
must be chosen commensurate with $\beta$ i.e. $\beta = NL_4$.
A great simplification results if we chose $L_4=\beta$ so
that only one block fits in time direction. 
The lattice $\Lambda$ then becomes three dimensional.
Our three-dimensional fields still have the same dimension as the original
four-dimensional ones.  
This could be remedied by a rescaling by $\beta^{(1/2)}$.  
We prefer not to do so because without rescaling we can keep track of
the temperature dependence with dimensional considerations.
Since we do not rescale the block fields
in position space the appropriate three-dimensional integration
measure contains a factor $\beta$ and the three-dimensional
delta-functions a corresponding factor $1 / \beta$.

Continuum coordinates are denoted by $z$, lattice coordinates by 
$x$.
In the finite temperatur case it makes sense to distinguish the space
coordinates from the time coordinate 
\be
  z &=& ({\vec z},z_4) 
\nn
  x &=& ({\vec x},x_4) = (n_1L_1, n_2L_2, n_3L_3, n_4L_4)
\ee
If $L_4=\beta$ then $ x=({\vec x},0)$.
Despite the fact that in the latter case $x_4$ is always equal to 0 we do not
drop this superfluous coordinate because our theory is not truly
three-dimensional.

Integration over coordinate space is  a
summation in case of the lattice.
For the sake of readability we define a symbolic notation:
\be
  && \T=0 \mbox{ continuum} \quad \quad 
  \int_z := \int_{{\vec z}} \int_{-\infty}^{\infty} dz_4 
          = \int_{-\infty}^{\infty} d^4z 
\nn
  && \T>0 \mbox{ continuum} \quad \quad 
  \int_z := \int_{{\vec z}} \int_{0}^{\beta} dz_4
          = \int_{-\infty}^{\infty} d^3z \int_{0}^{\beta} dz_4
\nn
  && \T=0 \mbox{ lattice} \quad \quad 
  \int_x := \int_{{\vec x}} L_4 \sum_{n_4 \in {\menge Z}}
          = L_1 L_2 L_3 L_4 \sum_{n \in {\menge Z}^4}
\nn  
  && \T>0 \mbox{ lattice} \quad \quad
  \int_x := \int_{{\vec x}} L_4 \sum_{n_4=0}^{N-1}
          = L_1 L_2 L_3 L_4 \sum_{{\vec n} \in {\menge Z}^3}
                            \sum_{n_4=0}^{N-1}
\nn  
  && \T>0 \mbox{ 3-dim lattice} \quad \quad
  \int_x := \int_{{\vec x}} \beta
          = \beta L_1 L_2 L_3 \sum_{{\vec n} \in {\menge Z}^3}
  \quad .
\ee
with $N=\beta / {L_4}$.

Remark that in the special case $\beta = L_4$ a factor $\beta$ 
remains as a relict of the fourth dimension.
This reflects the fact that the blockfield has the same canonical
dimension as the fundamental field
and this remains true even if we block one direction completely.

Correspondingly we subsume delta-functions on the continuum and
Kronecker symbols on the lattice under the same notation $\delta\{\}$
\be
   && \T=0 \mbox{ continuum} \quad \quad 
  \delta\{z-z'\} := \delta({\vec z}-{\vec z'}) \delta(z_4-z'_4)
                = \prod_{i=1}^4 \delta(z_i-z'_i)
\nn
  && \T>0 \mbox{ continuum} \quad \quad 
  \delta\{z-z'\} := \delta({\vec z}-{\vec z'}) \sum_n \delta(z_4-z'_4-\beta n)
\nn
  && \T=0 \mbox{ lattice} \quad \quad
  \delta\{x-x'\} := \delta({\vec x}-{\vec x'}) \frac1{L_4} \delta_{n_4,n'_4}
                = \prod_{i=1}^4 \frac1{L_i} \delta_{n_i,n'_i}
\nn
  && \T>0 \mbox{ lattice} \quad \quad
  \delta\{x-x'\} := \delta({\vec x}-{\vec x'}) 
                              \frac1{L_4} \sum_m \delta_{n_4-n'_4,mN}
                = \prod_{i=1}^3 \frac1{L_i} \delta_{n_i,n'_i} 
                              \frac1{L_4} \sum_m \delta_{n_4-n'_4,mN}
\nn
  && \T>0 \mbox{ 3-dim lattice} \quad \quad
  \delta\{x-x'\} := \delta({\vec x}-{\vec x'}) \frac1\beta
                = \prod_{i=1}^3 \frac1{L_i} \delta_{n_i,n'_i} \frac1\beta
  \quad .
\ee
For finite extension of the coordinate space the
delta-functions are periodized.
In the special case $L_4 = \beta$ the factor $1 / \beta$ is the
remaining coefficient of the necessarily fulfilled Kronecker delta
in time direction.

If one uses corresponding integrations and delta-functions 
\be
\int_z \delta\{z-z'\} = 1
\quad \quad \mbox{and} \quad \quad
\int_x \delta\{x-x'\} = 1
\ee
is true for all five cases.

\subsection{Scholium}
\label{ssscholium}

Our aim is to adapt the method of block spin transformations 
to the treatment of quantum field theories at finite temperature.
To prepare for this we first
recall the relation between propagators at zero
and at finite temperature.

At finite temperature $\T=1 / \beta$ a field theory lives on an
Euclidean space time
which is periodic in time direction with period $\beta=1 / \T$, 
i.e. on a tube.  
Propagators admit a random walk representation. 
The difference between zero and finite temperature stems from the
possibility that these random walks may wind around the tube several
times. 
Accordingly, the finite temperature propagator
$v_\T$ is obtained from $v_0$ by periodization in time
\be
  v_\T({\vec z},z_4,{\vec z'},z'_4)
  = \sum_{n\in{\menge Z}} v_0({\vec z},z_4+n\beta,{\vec z'},z'_4)
  \quad .
\label{tv}
\ee
The same method is used to obtain the block propagator for finite
temperature from the zero-temperature block propagator.
On the other hand we want to calculate $u_\T$ from $v_\T$
using a thermalized averaging operator $C_\T$
\be
  u_\T({\vec x},x_4,{\vec x'},x'_4) 
  &=& \sum_{n\in{\menge Z}} 
    (C_0 v_0 C^\dagger_0)({\vec x},x_4+n\beta,{\vec x'},x'_4)
\nn
  &=& (C_\T v_\T C^\dagger_\T) ({\vec x},x_4,{\vec x'},x'_4) 
  \quad .
\ee
This equation determines how the averaging operator must be
thermalized, see appendix \ref{tempercalc}.
\be
  C_\T({\vec x},x_4,{\vec z},z_4)
  = \sum_{n\in{\menge Z}} C_0({\vec x},x_4+n\beta,{\vec z},z_4)
\label{tCT}
\ee
The same arguments hold for other operator products.
We conclude that all operators for finite temperature are gained by
periodization from their zero temperature counterparts.

\subsection{Scalar theory at finite temperature}
\label{ssscalarT}

The averaging kernel of eq.(\ref{sCkern}) factorizes into parts 
for each direction.
According to eq.(\ref{tCT}) the kernel for the finite temperature
averaging operator is obtained by periodization in time of the zero
temperature kernel.
Therefore it is convenient to factorize the zero temperature kernel in a
space and a time part
\be
  C_0({\vec x},x_4,{\vec z},z_4) 
  = C_{{\rm time}}(x_4,z_4) C_{{\rm space}}({\vec x},{\vec z})
  = \frac1L_4 \chi_{x_4}(z_4) C_{{\rm space}}({\vec x},{\vec z})
\ee
\be
  C_T({\vec x},0,{\vec z},z_4)
  &=& \sum_{n\in{\menge Z}} C_{{\rm time}}(n\beta,z_4)
      C_{{\rm space}}({\vec x},{\vec z}) 
\nn
  &=& \sum_{n\in{\menge Z}} \frac1\beta \chi_{n\beta}(z_4)
      C_{{\rm space}}({\vec x},{\vec z}) 
\nn
  &=&  \frac1\beta C_{{\rm space}}({\vec x},{\vec z})
\ee
where we have used the fact that $z_4$ must lie in exactly one block.

The factor for the blocking in time direction is altered from
$1 / L_4 \chi_{x_4}(z_4)$ to $1 / \beta$.
It is now independent of the fine scale time coordinate $z_4$.
This facilitates the calculation of the block propagator $u_{\T}$.

Due to the translational invariance of the fundamental propagator $v_0$
one of the time integrations is trivial and gives a factor $\beta$,
see appendix \ref{tempercalc}
\be
  u_{\T}({\vec x},0,{\vec y},0) 
  &=& (C_\T v_\T C^\dagger_\T) ({\vec x},0,{\vec y},0)
\nn
  &=& \frac1\beta 
      \int_{{\vec z}} \int_{{\vec z'}} C_{{\rm space}}({\vec x},{\vec z})
      \int_{-\infty}^{\infty} dz_4 v_0({\vec z},z_4,{\vec z'},0) 
      C_{{\rm space}}^\dagger({\vec z'},{\vec y})
\nn
  &=&: \frac1\beta u_{FT}({\vec x},{\vec y}) \quad .
\label{tufinite}
\ee
Since we have a single block in time direction the factor $1 / \beta$
contains the only temperature dependence.
$u_{FT}({\vec x},{\vec y})$ is
the temperature- and time-independent part of the finite temperature block
propagator.

Taking care of the $\beta$ factors in the integration measure and
the delta-function we find for the block kinetic term
\be
  u_{\T}^{-1}({\vec x},0,{\vec y},0)
  =\frac1\beta u_{FT}^{-1}({\vec x},{\vec y}) \quad .
\ee
We use these simplified (due to $\beta = L_4$) averaging operator and
block kinetic term to obtain a time independent interpolation kernel,
see appendix \ref{tempercalc}
\be
  \A_{\T}({\vec z},z_4,{\vec x},0) 
  &=& \int_{{\vec z'}} \int_0^\beta dz'_4 
  \beta \int_{{\vec x'}} 
  v_{\T}({\vec z},z_4,{\vec z'},z'_4) 
  \frac1\beta C_{{\rm space}}^\dagger({\vec z'},{\vec x'})
  \frac1\beta u_{FT}^{-1}({\vec x'},{\vec x})
\nn
  &=&\A_{\T}({\vec z},0,{\vec x},0) 
  =: \frac1\beta \A_{FT}({\vec z},{\vec x}) \quad .
\label{tAfinite}
\ee
Again the factor $1 / \beta$ is the only temperature dependence, and
$\A_{FT}$ is time independent.

It is immediately clear that
\be
  \A_{\T} C_{\T}({\vec z},z_4,{\vec z'},z_4')
  &=& \beta \int_{{\vec x}} \A_{\T}({\vec z},z_4,{\vec x},0)
                          C_{\T} ({\vec x},0,{\vec z'},z_4')
\nn
  &=& \frac1\beta \int_{{\vec x}} \A_{FT}({\vec z},{\vec x})
                          C_{{\rm space}} ({\vec x},{\vec z'})
\ee
is time independent as well as are all background quantities.

The above simplifications, i.e. time independence of $C_{\T}$ and
$\A_{\T}$ and the simple
structure of the temperature dependence, result from the fact that
we use a single block in time direction.
As mentioned above the resulting lattice is only three-dimensional, so
time does not exist anymore.
The temperature dependence can be inferred from dimensional
considerations.

\subsection{Gauge fields at finite temperature}
\label{ssgaugeT}

As in the scalar case the operators are adapted to finite temperature
by periodization.

In principle we could perform the same calculations as in the scalar
case.
However the space-time indices $\mu,\nu$ and indications of possible
gauge fixing
(indices $\alpha,\beta$ of section \ref{secmaxwell}) in addition to
the subscripts indicating finite temperature, scalar averaging
operator and spatial part of whatever averaging operator would make
these calculations extremely ugly.

Therefore we go again through the arguments for the scalar case
written explicitly in appendix \ref{tempercalc}
to extract the necessary conditions for their validity.

These are two conditions.
The first is that the fundamental zero temperature
propagator must be translationally invariant in time.
This is of course the case for the gauge field propagator.
The second is that the finite temperature averaging kernel is time
independent and has a factor $1 / \beta$ as sole temperature
dependence.
We now show this explicitly.

We recall the covariant averaging kernel from section \ref{secmaxwell}
\be 
  C_{\mu\nu}(x,z) 
  = \delta_{\mu,\nu} \prod_{i\neq\mu} C(x_i,z_i) 
  \int_{z'_\mu} C(x_\mu,z'_\mu) C(z'_\mu+\half L_\mu,z_\mu) \quad .
\ee
Before we periodize in time we factorize it in a time and a space part.
To do that we must consider two different cases.
The first is $\mu = 4$
\be
  C_{44}({\vec x},x_4,{\vec z},z_4) 
  = \int_{-\infty}^{\infty} dz'_4 C(x_4,z'_4) C(z'_4+\half L_4,z_4) 
  C_{{\rm space}}({\vec x},{\vec z}) 
\ee
where $C_{{\rm space}}({\vec x},{\vec z})$ is the same as in the
scalar case.

The second case is $\mu \neq 4$
\be 
  C_{\mu\mu}({\vec x},x_4,{\vec z},z_4) 
  =  C(x_4,z_4) \prod_{{i\neq4},{i\neq\mu}} C(x_i,z_i) 
  \int_{z'_\mu} C(x_\mu,z'_\mu) C(z'_\mu+\half L_\mu,z_\mu)
\ee
Periodization involves only the first factor in both cases.
Again we use only one block in time direction i.e. we set $x_4=0$
and $L_4=\beta$.

For the second case we need
\be
  \sum_{n\in{\menge Z}} C(n\beta,z_4) 
  = \sum_{n\in{\menge Z}} \frac1\beta \chi_{n\beta}(z_4)
  = \frac1\beta
\ee
and for the first case
\be
  &&\sum_{n\in{\menge Z}} \int_{-\infty}^{\infty} dz'_4 
  C(n\beta,z'_4) C(z'_4+\half \beta,z_4) 
\nn  
  &=& \int_{-\infty}^{\infty} dz'_4 
    \underbrace{\sum_{n\in{\menge Z}} C(n\beta,z'_4)}_{\frac1\beta}
    C(z'_4+\half \beta,z_4) 
\nn  
  &=& \frac1\beta \int_0^\beta dz'_4 
    \underbrace{\sum_{n\in{\menge Z}} C(z'_4+n\beta+\half\beta,z_4)}
     _{\frac1\beta}
\nn
  &=& \frac1\beta \quad .
\ee
The finite temperature covariant averaging kernels are
for $\mu\neq4$
\be 
  C_{\T \,\, \mu\mu}({\vec x},0,{\vec z},z_4) 
  = \frac1\beta \prod_{{i\neq4}\top{i\neq\mu}} C(x_i,z_i) 
  \int_{z'_\mu} C(x_\mu,z'_\mu) C(z'_\mu+\half L_\mu,z_\mu)
\ee
and for $\mu=4$
\be
  C_{\T \,\, 44}({\vec x},0,{\vec z},z_4) 
  = \frac1\beta C_{{\rm space}}({\vec x},{\vec z}) \quad .
\ee
We conclude
that formulae similar to eqs.(\ref{tufinite}) and (\ref{tAfinite}) 
hold for the gauge field case as well.

But we can see something else from the explicit calculation.
The finite temperature kernel in time direction 
is the finite temperature scalar kernel.
It has lost its covariance.
This is no problem since on the lattice there is no time direction
and hence there can be no lattice derivative in time direction of a
lattice gauge function. 
Because of the
anisotropy, the time component of the lattice gauge fields ${\bf A}_4$
has lost its gauge field property.
The dependence of the resulting action on ${\bf A}_4$ and
on ${\bf A}_i, \, i=1,2,3$ is different. 
The ${\bf A}_4$ field behaves
like an extra scalar field in the effective three-dimensional theory.

\subsection{Scalar electrodynamics at finite temperature}
\label{ssscqedt}

We wish to adopt the considerations of the preceding subsections to 
scalar electrodynamics.
The blocking of the gauge field at finite temperature was discussed
before.
So we are only concerned with the Higgs sector.

First we consider the gauge field dependent linear averaging and
interpolation operators.
Because of this gauge field dependence
a covariant momentum space description does not exist, and we cannot
apply exactly the same periodization procedure as for a purely scalar
theory.  
Nevertheless, the transition to finite temperature by
periodization in time is straightforward.  
The formulae of the
preceding section remain valid as they stand when they are properly
interpreted.

Since averaging and integration operators are integral operators the
arguments of subsection \ref{ssscholium} can be applied.
All operators and propagators are adapted to finite temperature by
periodization in time.

For the gauge field dependent covariant propagator $v_H[a]$ this is
done as in eq.(\ref{tv}).

Finding $C_{H,\T}$ is more involved since here the time dependence is
not explicitly known.
Instead it is contained in the definition of the gauge field dependent
Laplacian with Neumann boundary conditions whose eigenfunction
to the lowest eigenvalue we chose as $C_H^\dagger$.

Here the periodization must take place in the definition of the
Laplacian.
We regard the blocks as coming equipped with
periodic boundary conditions in time direction. 
That means we deal immediately with the three-dimensional lattice with
$L_4 = \beta$.
The Neumann boundary
conditions apply only to those boundaries of the block which remain,
after periodic boundary conditions in time direction are imposed.

This finite temperature Laplacian defines the finite temperature block
operator as the zero temperature Laplacian the zero temperature
block operator.

All other operators can be obtained from these two explicitly
thermalized ones using the zero temperature relations of section
\ref{ssHinter} for the finite temperature operators.

Since the temperature dependence of $C_{H,\T}$ is more complicated
than that of the scalar block operator, we can not extract it as
simple factors of $\beta$.
This implies a hidden temperature dependence also for the block Higgs
field.

The effective action is defined as before, with
$C_{H,\T}$ substituted for $C_H$ , etc..
It lives on a three-dimensional lattice $\Lambda$.  

The gauge field independent nonlinear averaging and interpolatoin
operators $C_H$ and $\A_H$ are adapted to finite temperature by the
substitution $C_S \to C_{S, \T}$ and $\A_S \to \A_{S, \T}$ in their
respective definitions.
The fundamental gauge field dependent propagator $v_H[a]$ is
periodized in time as before.
Now we insert these finite temperature quantities into
eq.(\ref{hSeff2}) and continue as in the zero temperature case keeping
in mind that all time integrations are restricted to the range from
$0$ to $\beta$.

\section{Propagators, averaging operators, and
interpolators in momentum space}
\label{secfourier}

\subsection{More notational preliminaries}
\label{ssnotation2}

Before we calculate the various Fourier transforms,
we give a short review of the
definition of Fourier transformations for operators on different
supports.
In the previous sections we dealt with functions and operators living
either on the continuum or on a lattice $\Lambda$ with lattice spacing
$L_i$ in $i$-direction.

As explained in section \ref{sectemper} 
at finite temperature we have periodicity in time with period
$\beta = 1 / \T$.  
For our simplifying choice $L_4=\beta$ the lattice $\Lambda$ becomes
three-dimensional.

The type of coordinate space determines the type of momentum space.
To the infinite continuum as coordinate space 
belongs the infinite continuum as momentum space.
An infinite lattice with lattice constant $L$ as coordinate space
has as corresponding momentum space the continuous first Brillouin
zone (BZ) $]-\frac \pi L ,\frac \pi L]$.
This is shown in any textbook of solid state physics.
As shown in subsection \ref{ssscholium} the result of periodization
in time with period $\beta$ 
is discretization in energy with lattice constant $(2\pi) / \beta$.
If a lattice coordinate is periodized the corresponding momentum
component is restricted to a discrete first BZ.
If our block lattice is three-dimensional, so is the momentum space, i.e.
the fourth component is always zero.
But as in coordinate space factors of $\beta$ remain to remind us that
we did not rescale the block field.

We will call momenta restricted to the first Brillouin zone $p$ and
nonrestricted momenta $k$.
Sometimes the decomposition of not confined momenta $k$
into a part $k_{BZ}$
lying in the first BZ and a discrete summand which shifts
the momentum to the other Brillouin zones is convenient
\be
k = k_{BZ} + \frac{2\pi}L n \quad .
\ee
where both $ k_{BZ}$ and $n$ are uniquely determined by $k$.
Again we define a uniform notation for integration
\be
  && \T=0 \mbox{  continuum} \quad \quad 
  \int_k := \int_{\vec k} \int_{-\infty}^{\infty} \frac{dk_4}{2\pi}
          = \int_{-\infty}^{\infty} \frac{d^4k}{(2\pi)^4} 
\nn
  && \T>0 \mbox{  continuum} \quad \quad
  \int_k := \int_{\vec k} \frac1\beta \sum_{n=-\infty}^{\infty}
          = \int_{-\infty}^{\infty} \frac{d^3k}{(2\pi)^3} 
            \frac1\beta \sum_{n=-\infty}^{\infty}
\nn
  && \T=0 \mbox{  lattice} \quad \quad
  \int_p := \int_{\vec p} 
            \int_{-\frac{\pi}L_4}^{\frac{\pi}L_4} \frac{dp_4}{2\pi}
          = \prod_{i=1}^4
            \int_{-\frac{\pi}L_i}^{\frac{\pi}L_i} \frac{dp_i}{2\pi}            \nn
  && \T>0 \mbox{  lattice} \quad \quad
  \int_p := \int_{\vec p} \frac1\beta \sum_{n=0}^{N-1}
          = \prod_{i=1}^3
            \int_{-\frac{\pi}L_i}^{\frac{\pi}L_i} \frac{dp_i}{2\pi}  
            \frac1\beta \sum_{n=0}^{N-1}
\nn
  && \T>0 \mbox{  3-dim lattice} \quad \quad
  \int_p := \int_{\vec p} \frac1\beta 
          = \prod_{i=1}^3
            \int_{-\frac{\pi}L_i}^{\frac{\pi}L_i} \frac{dp_i}{2\pi}  
            \frac1\beta 
\ee
and for the delta-function in momentum space
\be
  && \T=0 \mbox{  continuum} \quad \quad 
  \delta\{k-k'\} := (2\pi)^3 \delta({\vec k}-{\vec k'}) 2\pi \delta(k_4-k'_4)
                = (2\pi)^4 \prod_{i=1}^4 \delta(k_i-k'_i)
\nn
  && \T>0 \mbox{  continuum} \quad \quad
  \delta\{k-k'\} := (2\pi)^3 \delta({\vec k}-{\vec k'}) \, \beta \delta_{n,n'}
                = (2\pi)^3 \prod_{i=1}^3 \delta(k_i-k'_i)
                   \, \beta \delta_{n,n'}
\nn
  && \T=0 \mbox{  lattice} \quad \quad
  \delta\{p-p'\} := (2\pi)^4 \prod_{i=1}^4 
                  \sum_{n_i} \delta(p_i-p'_i -\frac{2\pi}L n_i)
\nn
  && \T>0 \mbox{  lattice} \quad \quad
  \delta\{p-p'\} :=  (2\pi)^3 \prod_{i=1}^3 
                     \sum_{n_i} \delta(p_i-p'_i -\frac{2\pi}L n_i)
                 \, \beta \sum_m \delta_{n-n',mN}
\nn
  && \T>0 \mbox{  3-dim lattice} \quad \quad
  \delta\{p-p'\} :=  (2\pi)^3 \prod_{i=1}^3 
                     \sum_{n_i} \delta(p_i-p'_i -\frac{2\pi}L n_i)
                 \, \beta \quad .
\ee
The factors and periodizations are chosen such that 
\be
f(k) = \int_{k'} \delta\{k-k'\} f(k')
\quad \quad\ \mbox{and} \quad \quad 
f(p) = \int_{p'} \delta\{p-p'\} f(p') \quad .
\ee

Moreover with the above defined symbolic notation we regain the
familiar Fourier representation of the delta-function
\be
\delta\{k\} = \int_z e^{-ikz}
\quad \quad\ \mbox{and} \quad \quad 
\delta\{p\} = \int_x e^{-ipx}
\nn
\delta\{z\} = \int_k e^{ikz}
\quad \quad\ \mbox{and} \quad \quad 
\delta\{x\} = \int_p e^{ipx} \quad .
\ee
One can show the validity of these formulae with
the above definitions and the formulae for finite and infinite
geometric series.

In the following it is often necessary to
give up the symbolic notation during the course of the calculation
and to use the explicit expressions.

\subsection{Translational invariance}

Translationally invariant operators are trivially to invert in momentum
space, moreover the successive application of translationally
invariant operators
in coordinate space reduces to the multiplication of their Fourier
transforms in momentum space.
Here most of the operators are only lattice translational invariant,
which means that their Fourier transforms  have a simpler structure than
that of a non-invariant operator, but are more complicated than that of
a fully translational invariant operator.

To keep the formulae simple we write them for the one dimensional case.
The generalization to four dimensions must allow different lattice
constants in different directions.

An operator kernel $O(z,z')$ has the Fourier representation
\be
  O(z,z') = \int_k \int_{k'} e^{ikz + ik'z'} \tilde O(k,k') \quad .
\ee
If $O$ is translational invariant  $O(z,z')=O(z+a,z'+a)$.
Then the  momenta are restricted by
\be
  \frac{2\pi}a m = k + k' 
  = k_{BZ} + k'_{BZ} + \frac{2\pi}Ln + \frac{2\pi}Ln'
\ee
with the decomposition of momenta.

If $z$ and $z'$ are both in the continuum there are two types of
translational invariance.
Full translational invariance allows any $a\in R$.
The only solution for the momenta is $n+n'=0$ and $k_{BZ} + k'_{BZ}=0$
hence $k+k'=0$.
For $a\in \Lambda$ there are solutions for every $m\in N$.
They are $k_{BZ} + k'_{BZ}=0$ and $n+n'=m$.  
Hence the sum of the momenta is only fixed up to jumps from one 
Brillouin zone to another.

If $z\in R$ and $z'\in \Lambda$ (or the other way round) one of the
momenta is not restricted to the first brillouin zone and
$k+k' = (2\pi / L)m$ can be fulfilled for $m\neq 0$ as 
$k_{BZ} + k'_{BZ}=0$ and $n=m$ (or $n'=m$).

If $z$ and $z'$ are both block lattice coordinates $k$ and $k'$ are
both restricted to the first Brillouin zone hence $n$ and $n'$ are zero.
The solution $k_{BZ} + k'_{BZ}=0$ exists only for $m=0$.
That means for operators living entirely on the block lattice, lattice
translational invariance is full translational invariance.

There are two possibilities to use this relation of the momenta.
One can either display
the delta-function explicitly in the Fourier transform, or one can
use it to restrict the range of integration in the Fourier
representation of the translational invariant operator.
As an example we suppose the operator $O$ to be lattice translational
invariant
\be
  O(z,z') &=& \int_k \int_{k'} e^{ikz + ik'z'} \tilde O(k,k')
\nn
  &=& \int_k \int_{k'} e^{ikz + ik'z'}
    \underbrace{\sum_n (2\pi)^4 \delta(k+k'-\frac{2\pi}Ln) O(k,k')}
    _{\tilde O(k,k')}
\nn
  &=& \sum_n \int_k e^{ik(z-z') +i\frac{2\pi}Ln z'} O(k,-k+\frac{2\pi}Ln)
  \quad
\ee
We prefer the notation of the second line with the openly displayed
delta-function.
The missing tilde over $O(k,k')$ denotes that the
momentum delta-function is not any more implicitly contained in the
transform.

\subsection{Concatenation of two operators}
\label{ssconcat}

In concatenations the resulting operator has the translation
invariance property common to both parent operators.
In general the Fourier transform of the resulting operator is not the
product of the Fourier transforms of the single operators but
\be
  \widetilde{(AB)} (k,k'') = \int_{k'} \tilde A(k,k') \tilde B(-k',k'')
  \quad .
\ee
If the operators are translational invariant the equation
simplifies.
To show this we use the explicit expressions for the momentum
delta-functions.
For example if $A$ is fully and $B$ lattice translational invariant we
can observe how the delta-function corresponding to the lattice
translational invariance of the product appears
\be
  \widetilde{(AB)} (k,k'') 
  &=&  \int_{k'} (2\pi)^4 \delta(k+k') 
       \sum_n (2\pi)^4 \delta(-k'+k''+\frac{2\pi}Ln) A(k,k') B(-k',k'')
\nn
  &=& \sum_n (2\pi)^4 \delta(k+k''+\frac{2\pi}Ln) 
      \underbrace{A(k,-k) B(k,k'')}_{(AB)(k,k'')} \quad .
\ee

\subsection{Fourier transformed operators of the scalar model at zero
temperature}

Now we are prepared to calculate the Fourier transforms of all the
operators we used in section \ref{secscalar}.
We have to evaluate only two of them, the transforms of the
fundamental kinetic term and of the averaging operator, from their
respective coordinate representation.
All the others are constructed from these two and their Fourier
transforms can be obtained from the momentum representation of the
former using the rule for concatenation. 

The fundamental kinetic term 
$v^{-1}(z,z')= \delta\{z-z'\} (-\Delta + m^2)$ 
is continuum translational invariant hence
\be
  \tilde v^{-1} (k,k')
  &=& \delta\{k+k'\} v^{-1}(k,k')= \delta\{k+k'\} (k^2 + m^2) \quad .
\ee
Inversion in momentum space is trivial
\be 
  \tilde v(k,k') 
  &=& \delta\{k+k'\} v(k,k') = \delta\{k+k'\} \frac1{(k^2 + m^2)}
  \quad .
\ee

\subsubsection{The averaging kernel}

We consider the kernel $C(x,z)$ of the averaging operator $C$ which is
given in eq.(\ref{sCkern}).
\be
  C(x,z) = \frac1{\prod_{i=1}^4 L_i} \chi_x(z) 
  = \int_k \int_p e^{ikz + ipx} \tilde C(p,k)  
\ee
Here the integrals are four-dimensional, and $kz=\prod_{i=1}^4 k_i z_i$.
Because $C$ is lattice translational invariant under simultaneous translations
of $z$ and $x$ by lattice vectors in $\Lambda$, its Fourier
transform includes a delta-function relating the momenta $p$ and $k$.
The product form of $C$ allows to perform the Fourier transformation
independently for each dimension so that the actual calculation 
can be done in one dimension
\be
  \tilde C(p,k) &=& \int_x \int_z e^{-ipx - ikz} \frac1L \chi_x(z)
\nn
  &=& \frac1L \int_x e^{-ipx} \int_{x-\frac L2}^{x+\frac L2} dz e^{-ikz}
\nn
  &=& \frac1L \int_x e^{-ipx} \frac1{-ik} e^{-ikx}
      \underbrace{[e^{-ik\frac L2} - e^{ik\frac L2}]}_{-2i\sin{\frac{kL}2}}
\nn
  &=& \underbrace{\int_x e^{-i(p+k)x}}_{\delta\{p+k\}}
      \underbrace{\frac2{Lk} \sin{\frac{kL}2}}_{C(p,k)} \quad .
\ee
Note that the emerging delta-function is to be understood in the
symbolic sense since it is the result of an integration in
symbolic notation.
The translation to the explicit calculation is 
\be
  \int_x e^{-i(p+k)x}
  = L \sum_{n=-\infty}^{\infty} e^{i(p+k)Ln} 
  = 2\pi \sum_{n=-\infty}^{\infty} \delta(p+k-\frac{2\pi}Ln)
  = \delta\{p+k\} \quad .
\ee
$p$ lies in the first Brillouin zone and because the delta-function is
periodized $k$ may take values outside of it.

The four-dimensional result is
\be 
  \tilde C(p,k) = \prod_{i=1}^4 \tilde C(p_i,k_i)
\label{fCtilde}
\ee

\subsubsection{The block propagator and block kinetic operator}

The block spin propagator was defined in eq.(\ref{suDef}) as $u=C v C^\dagger$.
In momentum space this results in
\be
  \tilde u(p,p') = \int_{k} \int_{k'} \tilde C(p,k) 
  \tilde v(k,k')
  \tilde C^\dagger(k',p') \quad .
\ee
Here all the Fourier transforms on the right hand side contain
delta-functions.
We use them to obtain
\be 
  \tilde u(p,p') 
  &=& \delta\{p+p'\} \int_{k}  \frac1{k^2} \mid C(p,k) \mid^2 \delta\{p+k\}
\nn
  &=& \delta\{p+p'\} \sum_n \frac1{(p-\frac{2\pi}Ln)^2} 
                 \mid C(p,-p+\frac{2\pi}L n) \mid^2
\nn
  &=&: \delta\{p+p'\} u(p,-p) \quad .
\label{futilde}
\ee
We note that
\be
  u(p,-p) \sim \frac1{p^2}
  \quad \quad \mbox{for} \quad \quad 
  p^2 \to 0
\ee
because only the $n=0$ term in eq.(\ref{futilde}) is singular at $p=0$.

The kinetic operator of the effective action is obtained as the inverse 
of the lattice translational invariant block propagator $u$.
In momentum space the inversion is trivial:
\be
  \tilde u^{-1} (p,p') = \delta\{p+p'\} \frac1{u(p,-p)} \quad .
\ee

\subsubsection{The interpolation kernel, the background and
  fluctuation propagators}

We recall the definitions of these operators in coordinate space in
eqs.(\ref{sinterpol}), (\ref{sgsblock}) and (\ref{spropsum})
\be 
  \A &=& v C^\dagger u^{-1}
\nn
  v^s &=& \A C v 
\nn
  v^h &=& v - v^s  \quad .
\ee
All of them are only lattice translational invariant, despite the fact
that $v^s$ and $v^h$ live on the continuum.

The kernel $\A(z,x)$ has as Fourier transform
\be
  \tilde \A(k,p) &=& \int_{k'} \int_{p'} \tilde v(k,k') 
           \tilde C^*(k',p') \tilde u^{-1}(p',p)
\nn
  &=& \delta\{k+p\} 
      \underbrace{v(k,-k) C^*(p,k) \frac1{u(-p,p)}}_{\A(k,p)} \quad .
\label{fatilde} 
\ee
The momentum representation of the background propagator is
\be 
  \tilde v^s(k,k') 
  &=& \int_p \int_{k''} \tilde \A(k,p) \tilde C(p,k'') \tilde v(k'',k')
\nn
  &=& \int_p \int_{k''} \delta\{k+p\} \delta\{k''+p\} \delta\{k''+k'\}
      \A(k,p) C(p,k'') v(k'',k')
\nn
  &=& \delta\{k+k'_{BZ}\} 
      \underbrace{\A(k,-k_{BZ}) C(k'_{BZ},-k') v(-k',k')}_{v^s(k,k')}
  \quad .
\label{fgstilde}
\ee
The symbolic delta-function $\delta\{k+k'_{BZ}\}$ is to be read as
periodized so that $k$ can take values outside of the first BZ.
To obtain this we have used the fact that 
\be
\delta(k+p-\frac{2\pi}Lm) = \delta(k_{BZ}+\frac{2\pi}Ln + p-\frac{2\pi}Lm)
\ee
for a given $k$, i.e. given $k_{BZ}$ and $n$ both momenta $p$ and
$(2\pi / L)m$
are determined since $p$ is confined to the first Brillouin zone.
$\tilde v^s(k,k')$ does not contain the ordinary continuum
delta-function $\delta\{k+k'\}$ because $v^s(z,z')$ is not continuum-
but only lattice translational invariant.

The fluctuation propagator in momentum space is given by
\be
  \tilde v^h(k,k') &=& \tilde v(k,k') -\tilde v^s(k,k') 
\nn
  &=& \delta\{k+k'\} v(k,k') - \delta\{k+k'_{BZ}\} v^s(k,k')
\nn
  &=& \delta\{k+k'_{BZ}\} 
  \underbrace{[\delta_{n,0} v(k,k') - v^s(k,k')]}_{v^h(k,k')} \quad .
\ee

%
%
%
%

\subsection{Scalar fields at finite temperature}

To adjust to the finite temperature situation we need to periodize
the zero temperature quantities in time in the manner described in section
\ref{sectemper}. 

We exhibit the result of this periodization for the Fourier representation 
of a the fundamental propagator $v$
\be
  && v_T({\vec z},z_4,{\vec z'},z'_4) 
  = \sum_{n\in{\menge Z}} v_0({\vec z},z_4+n\beta,{\vec z'},z'_4) 
\nn 
  &=& \int_{-\infty}^{\infty} \frac{d^4k}{(2\pi)^4} 
      \int_{-\infty}^{\infty} \frac{d^4k'}{(2\pi)^4} 
    e^{i {\vec k}{\vec z} + i k_4 z_4 + i {\vec k'}{\vec z'} + i k'_4 z'_4 } 
    \tilde v_0({\vec k},k_4,{\vec k'},k'_4)
    \sum_{n\in{\menge Z}} e^{i n\beta k_4} \quad .
\ee
One may use Poisson's resummation formula
\be
  \sum_{n\in{\menge Z}} e^{i n\beta k_4}
  = \sum_{n\in{\menge Z}} \frac{2\pi}\beta \delta(k_4 - \frac{2\pi}\beta n)
  \quad .
\ee
As a result for the momentum component $k_4$
the finite temperature Fourier expansion is obtained
as a sum over Matsubara frequencies $(2\pi / \beta) n$.
Since all operators are invariant under translation by $\beta$ in time
direction all Fourier transfomed contain 
\be
  \sum_{n \in {\menge Z}} \delta(k_4+k'_4 - \frac{2\pi}\beta) \quad .
\ee
Through this the restriction of $k_4$ to the Matsubara frequencies is
imposed on $k'_4$ as well
\be
  && v_T({\vec z},z_4,{\vec z'},z'_4)  
\\ 
  &=& \int_{-\infty}^{\infty} \frac{d^3k}{(2\pi)^3} 
      \frac1\beta \sum_{n\in{\menge Z}}
      \int_{-\infty}^{\infty} \frac{d^3k'}{(2\pi)^3} 
      \frac1\beta \sum_{n'\in{\menge Z}}    
  e^{i {\vec k}{\vec z} + i \frac{2\pi}\beta nz_4 
   + i {\vec k'}{\vec z'} + i \frac{2\pi}\beta n' z'_4 }
  \tilde v_0({\vec k},\frac{2\pi}\beta n,{\vec k'},\frac{2\pi}\beta n')
  \quad .
\nonumber
\ee
The momentum components corresponding to the periodized coordinate are
discretized with lattice constant $2\pi / \beta$.
We anticipated this in our symbolic notation so that we can write
\be
  v_T({\vec z},z_4,{\vec z'},z'_4)  
  = \int_k \int_{k'} e^{ikz+ik'z'} \tilde v_0(k,k')
\ee
for the finite temperature case as well.
We only have to keep in mind what kind of momentum space is appropriate in
this formula.

In case of a lattice operator the momentum is in addition confined to
the first BZ.
If we have only one block in time direction, i..e. $L_4 = \beta$,
the fourth lattice momentum component is set to zero.
This is the effect of the dimensional reduction.
The factor $1 / \beta$ in the measure of integration and
the factor $\beta$ in the definition of the symbolic
delta-function are the remainders of the fourth dimension.

To summarize, all the effect of the finite temperature on the
operators is incorporated in the integration measure in momentum
space.
The kernels of the operators remain unchanged, provided one never
forgets which values of the fourth momentum component are allowed.


\subsection{Free Abelian gauge field at zero temperature}

In analogy to the scalar theory there are two operators whose Fourier
transforms
we have to calculate from the coordinate representation:
the fundamental kinetic operator and the averaging operator.

In addition there exist the projectors $R$ and $\R$ defined in
subsection \ref{ssproj}.
They need 
the momentum representation of $v_{\Delta^2}^{-1}$ and the scalar
averaging kernel for their evaluation in Fourier space.

\subsubsection{The kinetic terms}

Since both kinetic operators are fully translational invariant their
respective Fourier representations are momentum conserving.

In coordinate space 
\be
  v_{00}^{-1}(z,z') 
  &=& \delta(z-z') (\dmu \dnu - \partial^2   \delta_{\mu,\nu})
\nn
  v_{\Delta^2}^{-1} (z,z') &=& \delta(z-z') \Delta^2 \quad .
\ee
Their Fourier transforms are
\be
  \tilde v_{00}^{-1}(k,k') 
  = \delta(k+k') (k^2 \delta_{\mu,\nu} - k_\mu k_\nu) \quad ,
\ee
which is not invertible due to gauge invariance and
\be
  \tilde v_{\Delta^2}^{-1} (k,k') = \delta(k+k') k^4 \quad .
\ee

\subsubsection{The gauge covariant averaging kernel}

As the last building block we need the Fourier transform of the 
averaging operator $C$ for Abelian gauge fields. 
Its coordinate space kernel is given in eq.(\ref{mCkern}) as 
\be
  C_{\mu \nu}(x,z)= \delta_{\mu,\nu} 
  \int_{z'} C_S(x_\mu,z'_\mu) C_S(z'_\mu + \frac12 L_\mu,z_\mu) 
  \prod_{\rho \neq \mu} C_S(x_\rho,z_\rho)  \quad .
\ee
Like the scalar block operator it factorizes in the different
directions.
For three directions we can directly use the scalar result.
To evaluate the Fourier transform in $\mu$-direction we 
use
\be
  C_S(z'_\mu + \frac12 L_\mu,z_\mu) 
  = C_S(z'_\mu,z_\mu - \frac12 L_\mu) 
\ee
and then apply the formula for concatenation of operators.
The resulting expression is
\be
  \tilde C_{\mu\nu}(k,p) = \delta\{k+p\} \delta_{\mu\nu}
  \exp^{-i\frac{k_\nu L_\nu}2}
  \frac2{L_\nu k_\nu} \sin \frac{k_\nu L_\nu}2 C_S(k,p)
\ee
where $C_S(k,p)$ is the scalar kernel of eq.(\ref{fCtilde}), and no sum
over $\nu$ is implied.

\section{Computation of effective actions by perturbation theory}
\label{secperturbation}

We know from eq.(\ref{veff}) that the effective actions are obtained as 
logarithms of partition functions of auxiliary field theories whose 
free propagator ist the fluctuation field propagator $\Gamma$ and which have
$\Phi$-dependent coupling constants 
\be
  e^{-V_{\rm eff}[\Phi]}
  = \int \D \phi^h \delta (C \phi^h) 
  e^{-\frac12 \langle \phi^h, v^{-1} \phi^h \rangle -V[\A \Phi + \phi^h]}
  \quad .
\label{pveff1}
\ee
The fluctuation integral can be evaluated perturbatively
\be
  e^{-V_{\rm eff}[\Phi]}
  &=& \int \D \phi^h \delta (C \phi^h) 
  e^{-\frac12 \langle \phi^h, v^{-1} \phi^h \rangle 
  -V[\A \Phi + \frac{\delta}{\delta j}]
   + \langle \phi^h, v^{-1} \phi^h \rangle} |_{j=0}
\nn
  &=& e^{-V[\A \Phi + \frac{\delta}{\delta j}]}
   e^{\frac12 \langle j, \Gamma j \rangle} |_{j=0} \quad .
\label{pveff2}
\ee
This form is appropriate if one wants to obtain $V_{\rm eff}[\Phi]$ as an
expansion in the interaction.
If one is interested in its expansion in powers of the fluctuation
propagator $\Gamma$ it is easier to use an alternative formula.

The perturbative expansion to all orders is given by Wick's theorem
in the form
\be
  e^{-V_{\rm eff}[\Phi]}
  &=& e^{\frac12 \langle \frac{\delta}{\delta\phi^h},
  \Gamma \frac{\delta}{\delta\phi^h} \rangle}
  e^{-V[\A \Phi + \phi^h]} |_{\phi^h=0}
\nn
  &=& e^{\frac12 \langle \frac{\delta}{\delta\A \Phi},
  \Gamma \frac{\delta}{\delta\A \Phi} \rangle}
  e^{-V[\A \Phi]} \quad .
\label{pveffsc}
\ee
That both equations give the same result can be seen by the
following short consideration
\be
  f(\frac d{dx}) g(x)|_{x=0} 
  &=& \sum_n f_n (\frac d{dx})^n \sum_m g_m x^m |_{x=0}
\nn  
  &=& \sum_n f_n g_n (\frac d{dx})^n x^n
\nn  
  &=& \sum_n f_n g_n n!
\nn
  &=& g(\frac d{dy}) f(y)|_{y=0} 
\ee
since the result is symmetric in $f$ and $g$.

In both expansions the effective potential itself contains only the 
contributions corresponding to the connected diagrams.

The effective action at finite temperature is given by the same
equation provided one substitutes 
\be
  C & & \to C_\T
\nn
  v^{-1} & & \to v_\T^{-1}
\nn
  \A & & \to \A_\T
\nn
  \Gamma & & \to \Gamma_\T
\label{psubst}
\ee
The effective interaction, including mass terms, is temperature
dependent because the fluctuation field propagator
$\Gamma_\T$ and the interpolation kernel 
$\A_\T$ are both temperature dependent.  
This temperature dependence is weak and disappears in a zeroth order local
approximation.

\subsection{Leading terms in the perturbative expansion of the perfect
  action for $\phi^4$-theory}

Let us consider $\phi^4$-theory with bare mass $m_0$
\be
  S_0[\phi] 
  &=& \frac12 \langle \dmu \phi, \dmu \phi \rangle
\nn
  V[\phi] &=& \int_z \left( \frac12 m_0^2 \phi^2 
  + \frac g{4!} \phi^4 \right)  
  + \mbox{wave function renormalization term.}
\ee
It is the same action we used in section \ref{secscalar} but here the
mass term is considered as a part of the interaction.

Of course without a mass term the fundamental kinetic operator will have
zero modes and the fundamental propagator is in momentum space divergent 
for $k \to 0$.
In coordinate space it does not decay exponentially.
How to circumvent this problem we have learned in section
\ref{secmaxwell}, where we had to impose a regulator to define the
fundamental gauge field propagator.
Here we could introduce an auxiliary mass as a
regulator into
the fundamental kinetic term and remove this mass at the very end of the
calculations.
The in this way obtained fluctuation propagator remains finite for 
$k \to 0$, because as a fluctuation propagator it is
IR-regulated by definition.
  
Inserting the field split $\phi = \phi^s + \phi^h$, we obtain
\be
  V[\phi^s + \phi^h] 
  = U_{\rm cl}[\Phi]
  + \sum_{n=1}^4 \int_z \frac1{n!} g_n[\phi^s] \phi^h(z)^n
\ee
with
\be
  g_1 &=& \frac g{3!} \phi^s(z)^3 + m_0^2 \phi^s(z)
\nn
  g_2 &=& \frac g{2!} \phi^s(z)^2 + m_0^2
\nn
  g_3 &=& g \phi^s(z)
\nn
  g_4 &=& g
 \nn
  \phi^s &=& \int_x \A(z,x) \Phi(x)
\nn
  U_{\rm cl}[\Phi] &=& V[\A \Phi]
\ee
$V_{\rm eff}$ can be calculated in a loop expansion.  
When the starting point is a lattice theory, 
one can use Mayer expansions instead.  
They are convergent for weak coupling, and the asymptotic
expansion of individual terms in powers of the bare coupling constant
contains infinite sets of diagrams \cite{pordt}.

We write the perturbative expansion of $V_{\rm eff}$ as
\be
  V_{\rm eff}[\Phi] 
  = U_{\rm cl}[\Phi] 
  + \sum_{N\geq 1} V_{\rm eff}^{(N)}[\Phi]
\ee
where $V_{\rm eff}^{(N)}$ scales as $\gamma^N$ when the
fluctuation field propagator 
$\Gamma \longrightarrow \gamma \Gamma$.

Let us compute $V_{\rm eff}^{(1)}$. 
\be
  V_{\rm eff}^{(1)} 
  &=& \frac1{2!} \int_z (\frac g{2!} \phi^s(z)^2 + m_0^2) \Gamma(z,z)
\nn
  - \frac1{2!} \!\!\!\!\!\!\! && \int_z \int_z' 
  (\frac g{3!} \phi^s(z)^3 + m_0^2 \phi^s(z))
  \Gamma(z,z')
  (\frac g{3!} \phi^s(z')^3 + m_0^2 \phi^s(z'))
\label{pvefftsc}
\ee
Higher order calculations can be found in the literature \cite{yoma},
\cite{max}.

The effective potential up to second order in the fluctuation
propagator is displayed in appendix \ref{max} and given in diagrammatic 
form in eq.(\ref{maxbild}).
Factors $\phi^s$ are indicated by solid external lines and a
fluctuation field propagator by a dotted inner line.
\newlength{\widtwo} \setlength{\widtwo}{1cm}
\newlength{\widthree} \setlength{\widthree}{1.5cm}
\newlength{\widfour} \setlength{\widfour}{2cm}
\be
   V_{\rm eff} &=& 
      \ 
      \raisebox{0.04\widtwo}{\epsfig{file=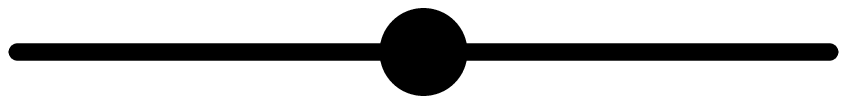,width=\widtwo}} \ 
      \ + \ 
      \raisebox{-0.3\widtwo}{\epsfig{file=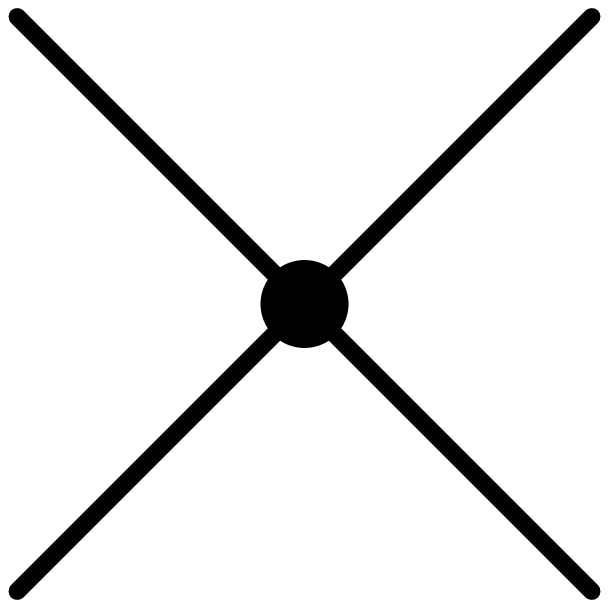,width=0.8\widtwo}}\ 
      \ +\  
      \raisebox{0.02\widthree}{\epsfig{file=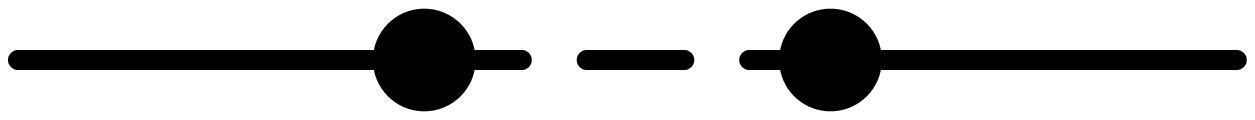,width=\widthree}}\ 
      \ +\  
      \raisebox{0.045\widtwo}{\epsfig{file=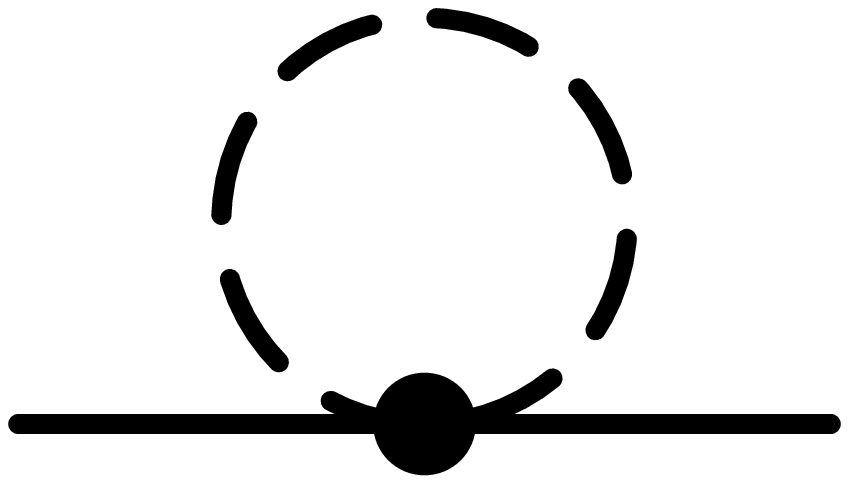,width=\widtwo}} \nn
    & &  \ +\  
      \raisebox{0.02\widthree}{\epsfig{file=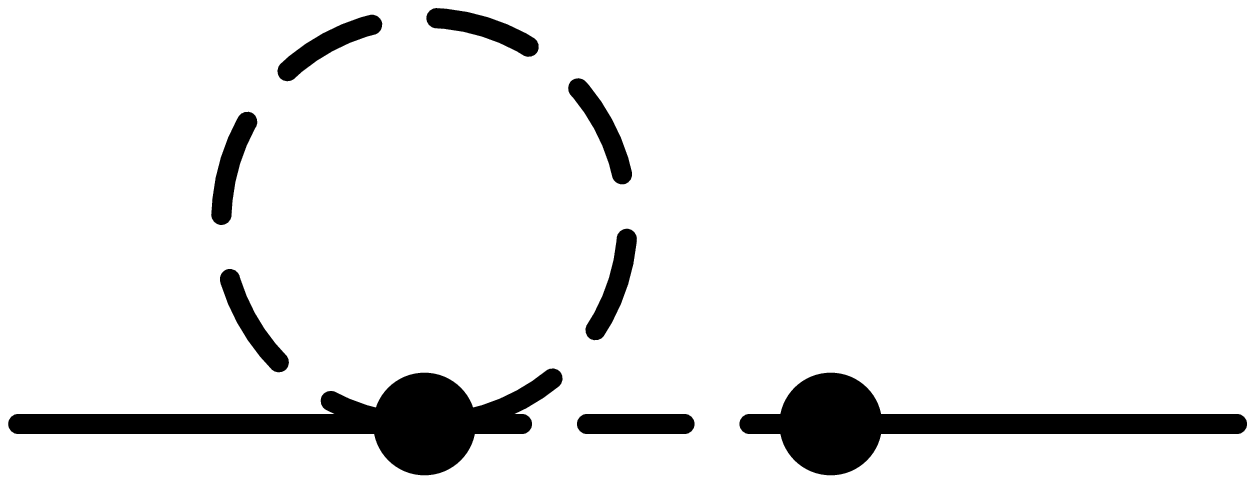,width=\widthree}}\ 
      \ +\  
      \raisebox{0.045\widtwo}{\epsfig{file=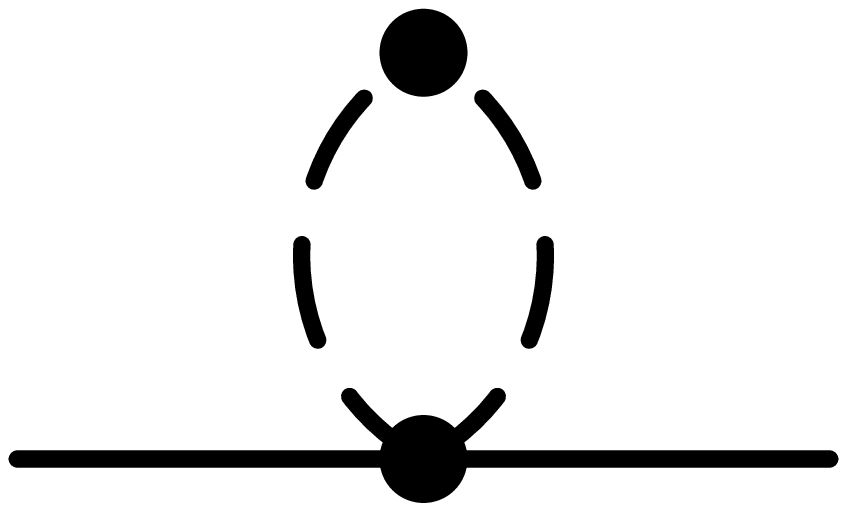,width=\widtwo}} 
      \ + \ 
      \raisebox{0.01\widfour}{\epsfig{file=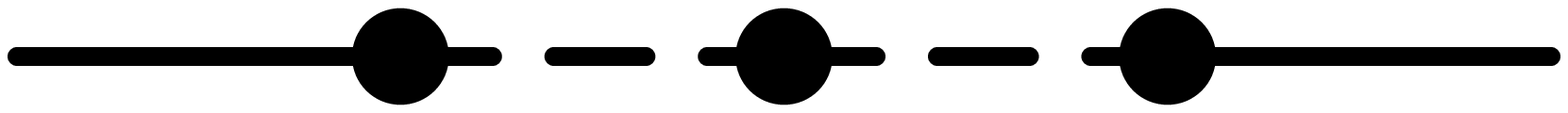,width=\widfour}} \nn
    & &  \ +\  
      \raisebox{-0.16\widthree}{\epsfig{file=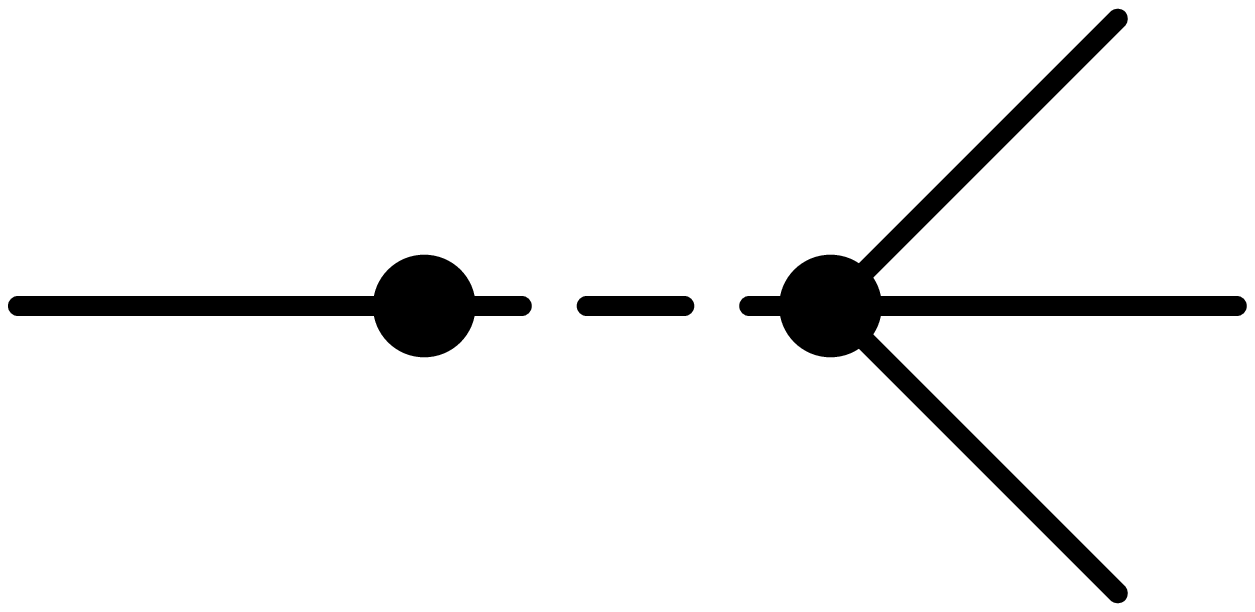,width=\widthree}}\  
      \ +\  
      \raisebox{-0.155\widthree}{\epsfig{file=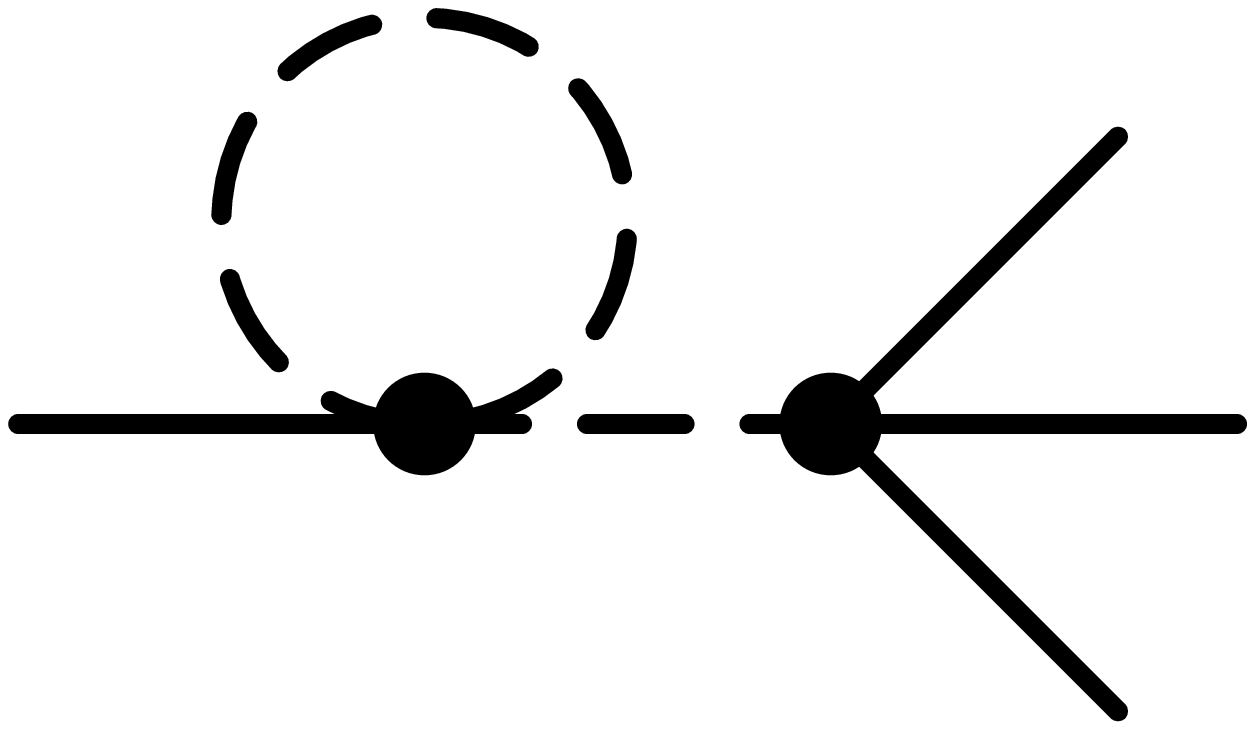,width=\widthree}} 
      \ + \ 
      \raisebox{-0.12\widfour}{\epsfig{file=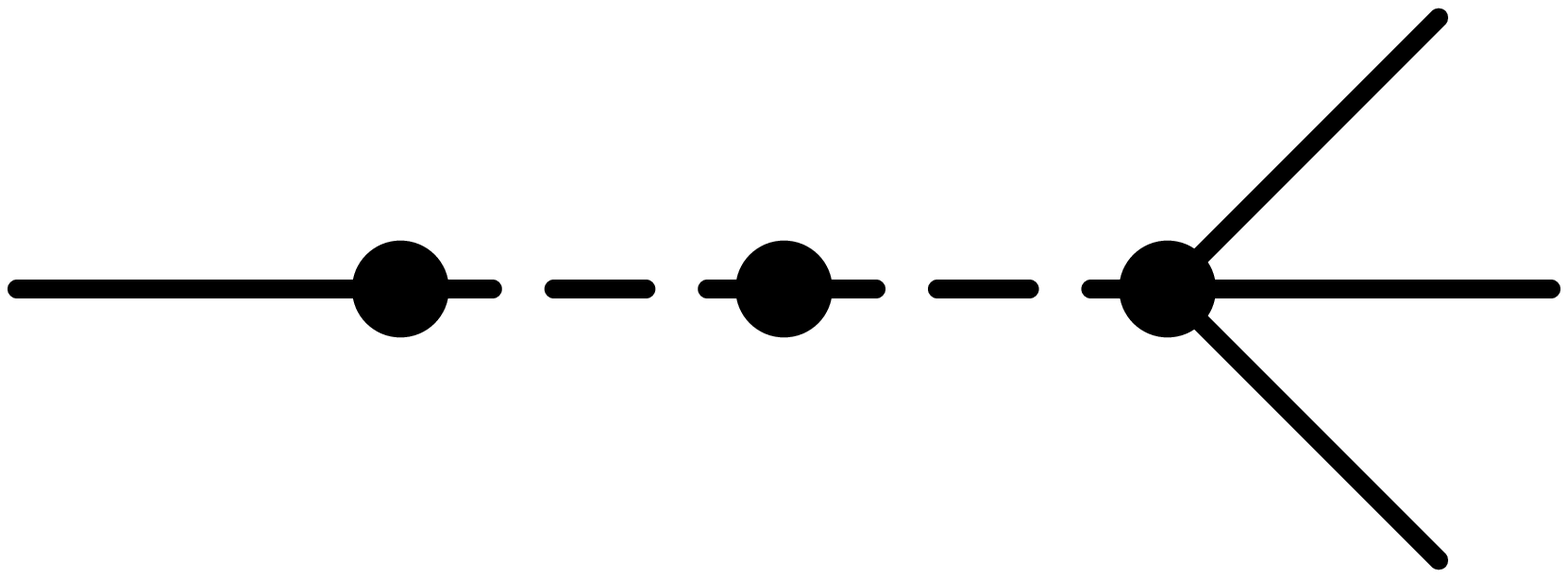,width=\widfour}} \nn
    & & \ +\  
      \raisebox{-0.16\widthree}{\epsfig{file=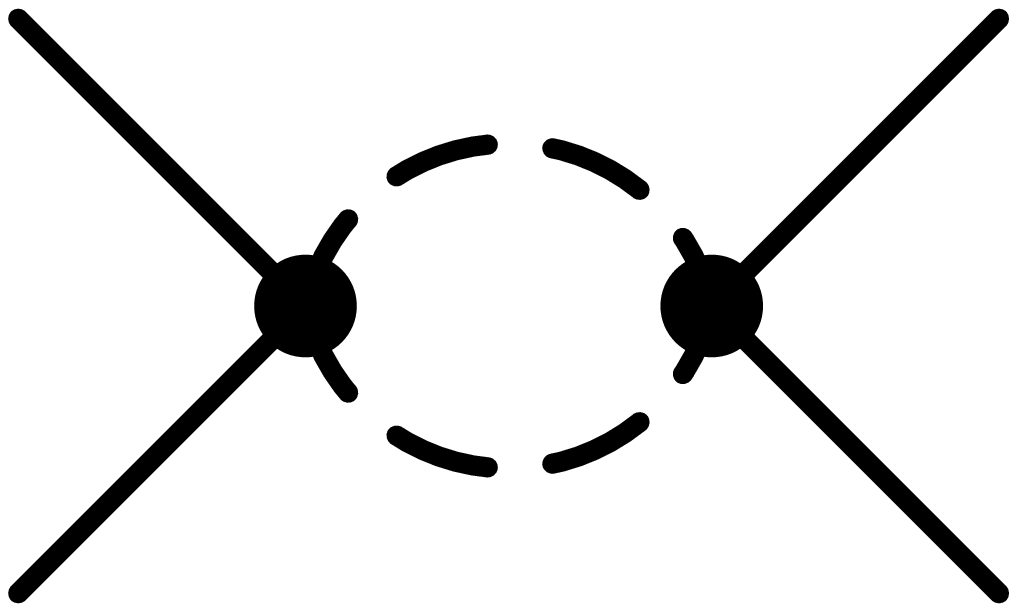,width=0.8\widthree}}\
      \ + \ 
      \raisebox{-0.2\widfour}{\epsfig{file=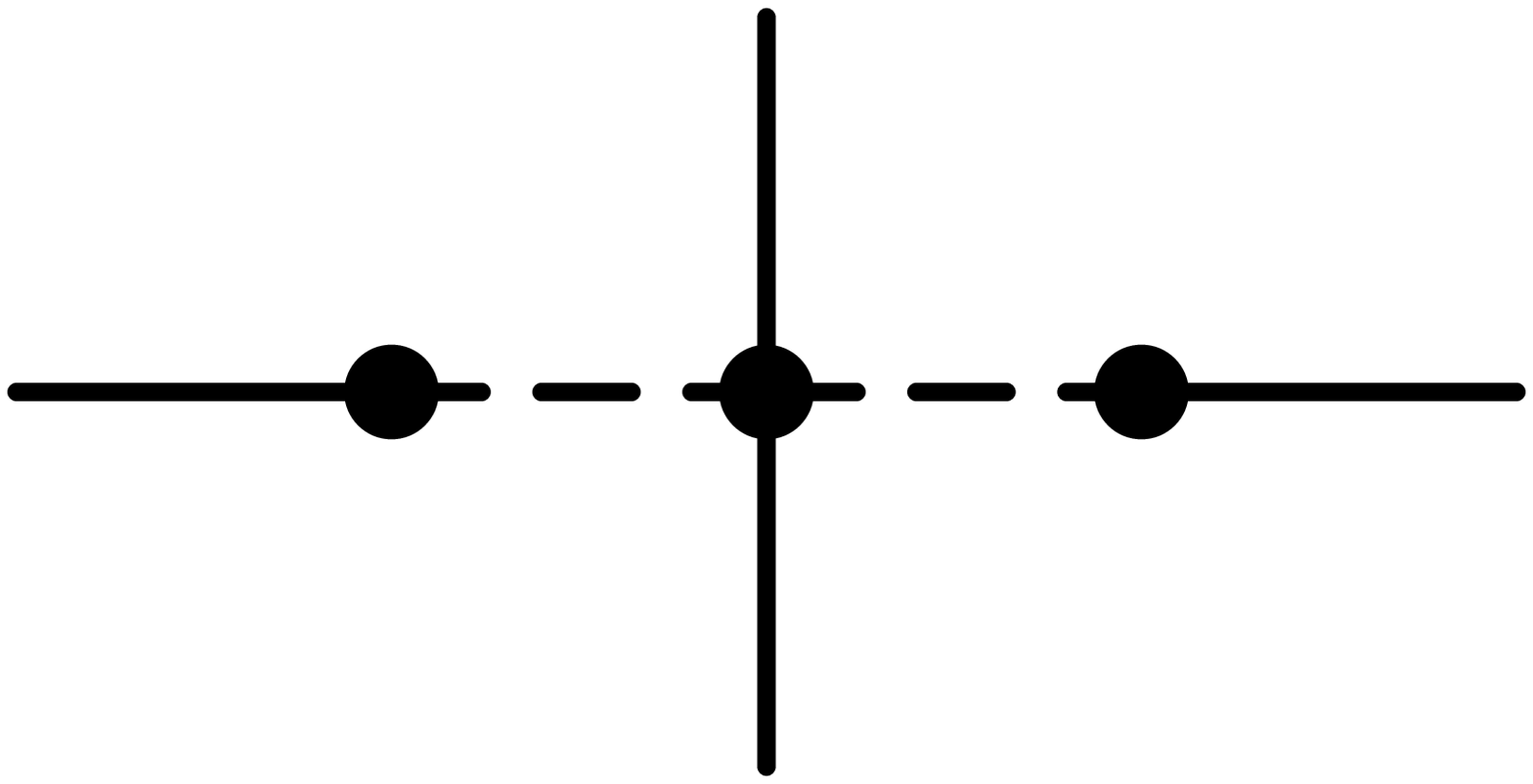,width=\widfour}} 
      \ +\  
      \raisebox{-0.16\widthree}{\epsfig{file=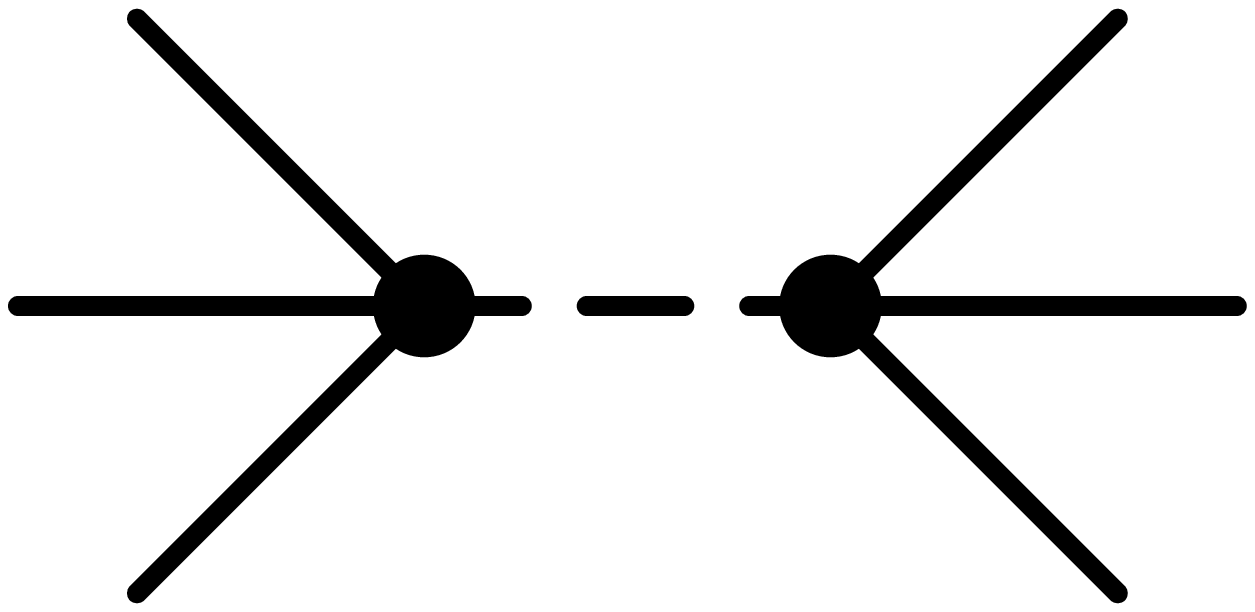,width=\widthree}} \nn
    & &  \ + \ 
      \raisebox{-0.12\widfour}{\epsfig{file=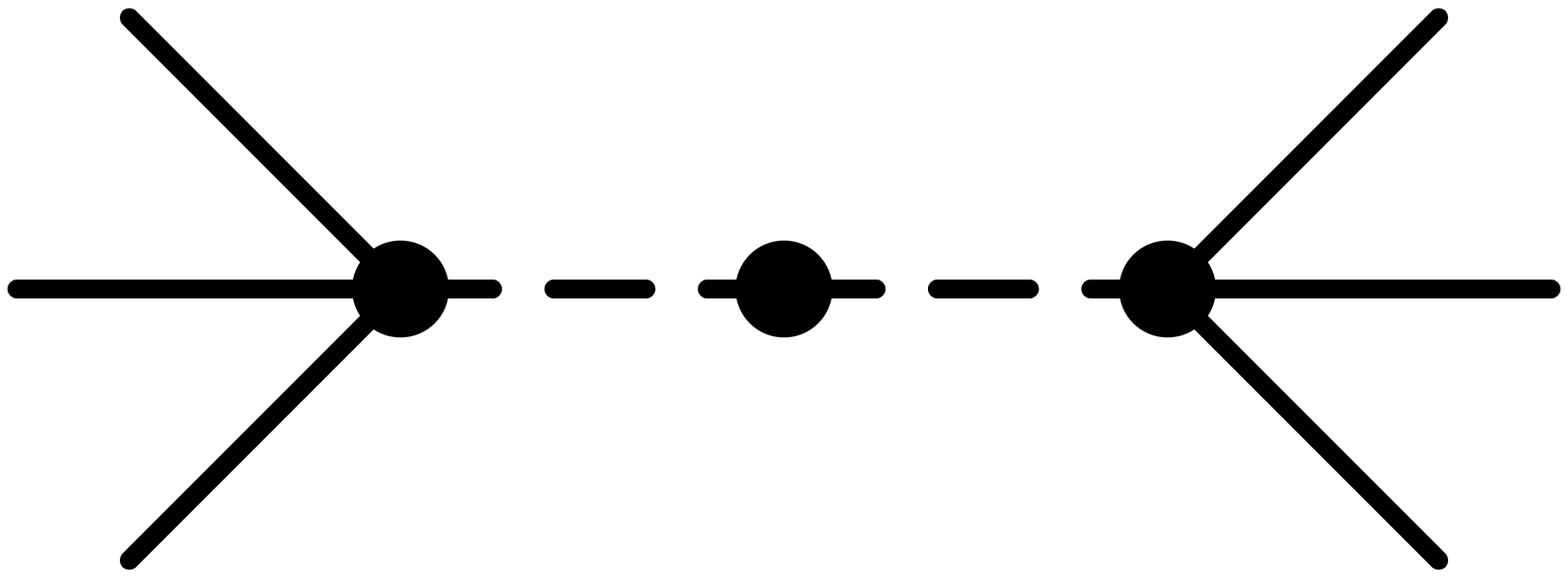,width=\widfour}} \
      \ + \ 
      \raisebox{-0.195\widfour}{\epsfig{file=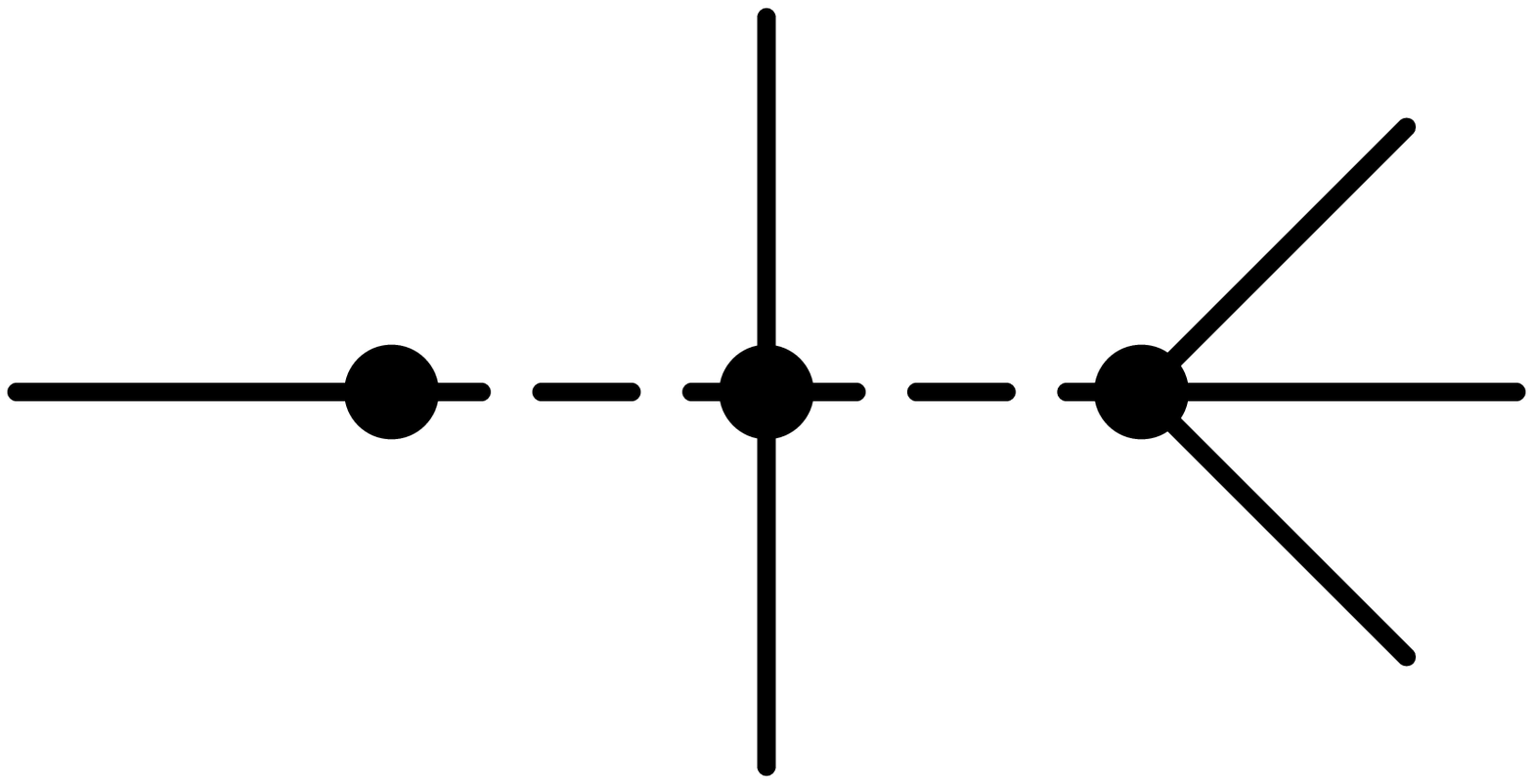,width=\widfour}} \
      \ + \ 
      \raisebox{-0.195\widfour}{\epsfig{file=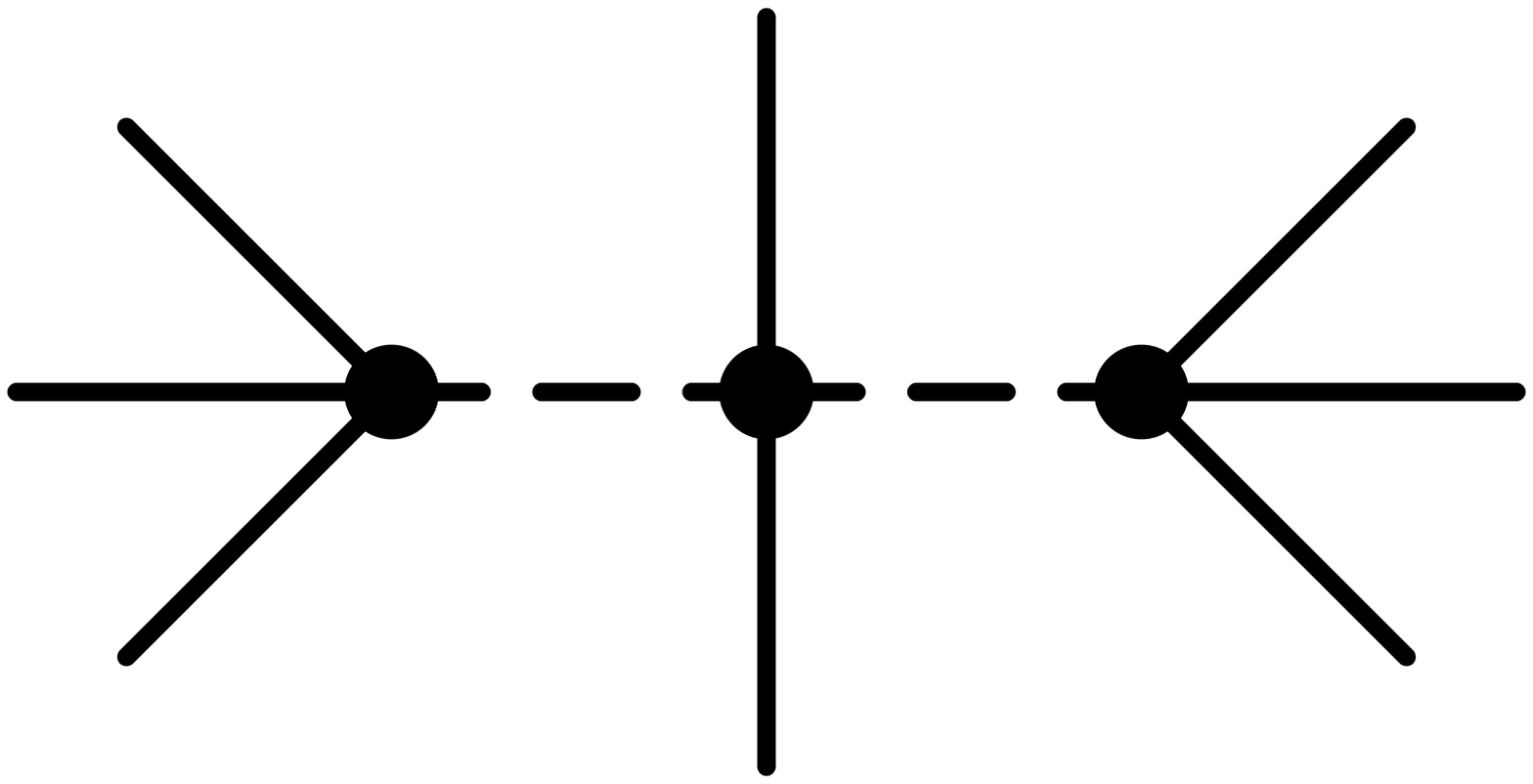,width=\widfour}} \
      \ + \ \mbox{constant}
\label{maxbild}
\ee
For computer simulations it is appropriate to consider
$V_{\rm eff}$ as a function of $\Phi(x)$. 
For analytical computations it is more convenient to regard 
it as a function of $\phi^s(z)$.

The fluctuation field propagator $\Gamma(z,z')$,
decays exponentially with distance $\left| z-z' \right|$ with
decay length one block lattice spacing. 
That is, it decays with
$\left| z_i - z'_i \right|$ with decay length $L_i$. 
Similarly, $\A (z,x)$ decays
exponentially in $\left| z_i -x_i \right|$ with decay length
$L_i$.  
As a result, each term in $V_{\rm eff}$ is
local in $\Phi$ modulo exponential tails with decay length of one
lattice spacing.

By a process of partial integration, or rather summation, one can
exhibit each term in $V_{\rm eff}$ as a sum of local terms and
small remainders which represent the exponential tails (see appendix
\ref{local}).  
The terms with coefficients of dimensions up to
(mass$)^{-2}$ are

\be
  V_{\rm eff}[\Phi] 
  \! = \! \int_x \! \left( \half m^2 \Phi^2 
  + \delta_z (\nabla_\mu \Phi)^2 
  + \frac{g_r}{4!} \Phi^4
  + \frac{g_6}{6!} \Phi^6 
  + \tilde \gamma (\nabla_\mu \Phi)^2 \Phi^2 + \cdots \right)
\ee
Again the adaption to finite temperature is made with the
substitutions of eq.(\ref{psubst}).
Then all the coefficients are T-dependent and finite.
Since $u_{\rm FT}^{-1}(p) \sim p^2$ as $p \to 0$,
we may write the finite temperature expansion equally well in the form
\be
  V_{{\rm eff},\T}[\Phi] 
  &=& \int_x \left( \half m_\T^2 \Phi^2
  + \frac{g_{r, \T}}{4!} \Phi^4 
  + \frac{g_{6, \T}}{6!} \Phi^6 \right)    
\nn
  &+& \frac \beta 2 \int_{\vec x} \int_{\vec y} \Phi(x)
  u_{\rm FT}^{-1}(x,y) \Phi(y) 
  \left( 2 \delta_{z, \T} + \gamma_\T \Phi(x)^2 \right) + \cdots
\label{psefft}
\ee

\subsection{Note on the cancellation of temperature dependent
UV-divergent diagrams}

The UV-convergence of a quantum field theory concerns its local
behaviour and is therefore not temperature dependent. 
If the proper
choice of counter terms makes the theory finite at zero temperature,
then also at finite temperature. 
The cancellation occurs order by order in perturbation theory. 
This is well known.

Problems can occur when one sums selected classes of diagrams. 
We do not propose to do so when deriving the perfect action, but stay
strictly within the realm of standard perturbation theory. 
So there can be no problem.

It is nevertheless appropriate to point out that there do exist
individual diagrams with temperature dependent UV-divergent pieces.
They cancel. 
We give an example.

\begin{figure}
  \centerline{\epsfxsize=250pt \epsfbox{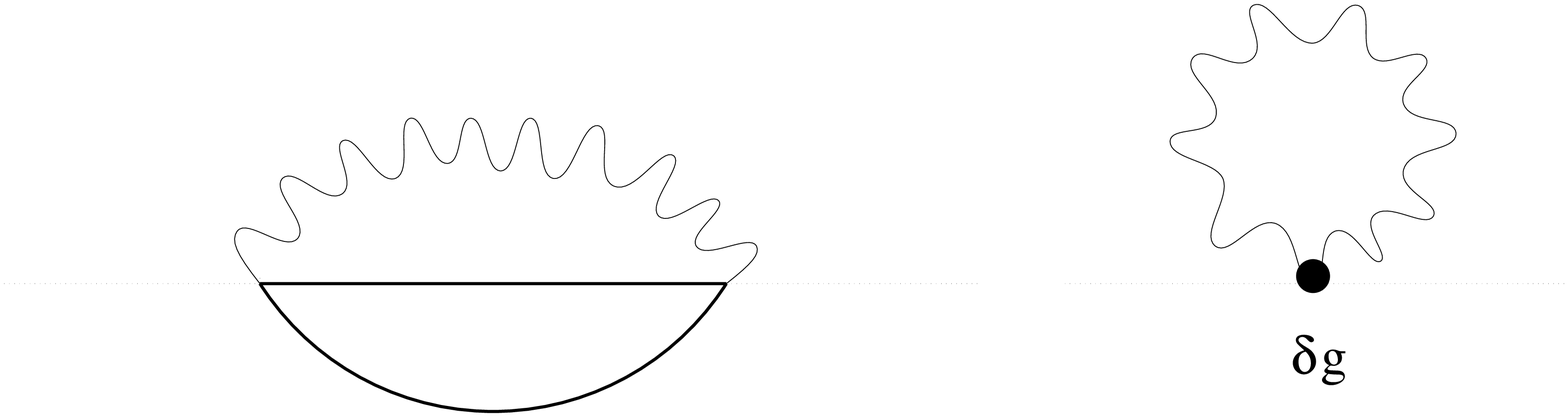}}
  \refstepcounter{bild}
  \caption{\ temperature dependent divergent diagrams}
\label{divergenz}
\end{figure}

Let us split the fluctuation propagator $\Gamma$ into a static
part (heavy line) $\Gamma_{\rm stat}$ and a non static part
$\Gamma_{\rm ns}$ (wavy line).  
The non static part represents
random walks which wind several times around the tube; it is not
singular at coinciding arguments: $\Gamma_{\rm ns}(z,z') \leq \infty$.  
But it is not zero and is T-dependent.  
As a result, the diagrams in fig.(\ref{divergenz})
%
are both logarithmically UV-divergent, with a temperature dependent
coefficient $\Gamma_{\rm ns}(z,z')$. 
They must cancel.
$\delta g$ is the logarithmically divergent coupling constant
counter term to second order in $g$.

\section{Selfconsistent improvements of perturbation theory: Gap equations}

\subsection{The Feynman-Bogoliubov Method}
\label{secfeyman}

To obtain information on the nature of a phase transition at finite
temperature, one wants to compute the effective potential, i.e. the
free energy as a function of the magnetization
\be
  M = \int_z \phi(z) = \int_x \Phi(x) \quad .
\ee
Alternatively, one may compute the constraint effective potential
which gives the probability distribution of $M$.

To do so we should apply nonperturbative methods to solve the lattice
theory.  One of these methods consists in the solution of gap
equations. It is very old and known as the Feynman-Bogoliubov method
\cite{feybo}.

The gap equations have `perturbative' solutions which come from
summations of superdaisy diagrams. In principle they might also have
other solutions. We wish to examine this method in order to see what
will be the effect of terms like $(\nabla_\mu \Phi)^2 \Phi^2$ etc. in
the perfect lattice action.

Let us make it clear that it is not sufficient to find the solutions
of these gap equations. In order to justify the perturbative
calculation of the perfect action it will be necessary to investigate
also the stability properties of the solutions of the gap equations
against small perturbations of the lattice action. We will comment on
this, but a thorough treatment of this question is beyond the scope of
this thesis.

Given an action $S[\Phi]$ of some Euclidean field theory, one seeks an
optimal quadratic approximation $S_{\rm free}[\Phi]$ around which
to expand
\be
  S_{\rm free}[\Phi] = \frac12 \int_x \int_y \Phi(x) J(x,y) \Phi(y)
\ee
By Peierls inequality \cite{feybo}, the partition functions obey the
inequality
\be
  \ln(Z) \geq \ln(Z_{\rm free}) 
  - \langle S-S_{\rm free} \rangle_{\rm free}
\label{ppeierls}
\ee
for any choice of the $S_{\rm free}$. 
Herein, $\langle \rangle_{\rm free}$ is the expectation value in the
theory with action $S_{\rm free}$.  
The right hand side is the first order
approximation to the left hand side in the perturbative expansion
around $S_{\rm free}$. 
The optimal choice of $S_{\rm free}$
is that which makes the right hand side of (\ref{ppeierls}) maximal. 
In other words, it makes the first order approximation as good as
possible.  
It is asserted that there exists always a unique optimal
choice of $S_{\rm free}$. 
It is not asserted that the optimal choice is necessarily a good one. 
The optimal $J$ is determined by the extreme value condition
\be
  \langle \frac{\delta^2 S}{\delta \Phi(x) \delta \Phi(y)}
  \rangle_{\rm free} = J(x,y).
\label{pgap}
\ee
This is equivalent to the condition that the right hand side of
(\ref{ppeierls}) is maximal, i.e.
\be
  \frac\delta{\delta J} \left( \ln(Z_{\rm free}) 
  - \langle S-S_{\rm free} \rangle_{\rm free} \right) = 0
\ee
When applied to the standard $\Phi^4$ action, this produces the gap
equation whose perturbative solution is the sum of superdaisy diagrams
(see below).

Let us consider the gap equation which results from the more
complicated finite temperature lattice action.
\be
  S[\Phi] = \frac{\beta}2 \langle \Phi, u_{\rm FT}^{-1} \Phi \rangle
  + V_{{\rm eff},\T}[\Phi]
\label{psphi}
\ee
with $V_{{\rm eff},\T}$ from equation (\ref{psefft}).

We obtain
\be
  \frac{\delta^2 S[\Phi]}{\delta \Phi(x) \delta \Phi(y)}
  &=& \beta u_{\rm FT}^{-1}(x,y) 
  \left( 1 + 2\delta_{z, \T} + 3\gamma_\T \Phi(x)^2 \right)
\nn
  &+& \beta \delta(x-y) 3\gamma_\T \Phi(x) (u_{\rm FT}^{-1} \Phi)(x)
\nn
  &+& \delta(x-y) \left( m_\T^2 + \frac{g_{r, \T}}{2!} \Phi(x)^2
  + \frac{g_{6, \T}}{4!} \Phi(x)^4 \right)
\ee
The expectation value in the theory with action 
$S_{\rm free} = \int \frac12 \Phi J \Phi$ is
\be
  \langle \frac{\delta^2 S[\Phi]}{\delta \Phi(x) \delta \Phi(y)}
  \rangle_{\rm free} 
  &=& \beta u_{\rm FT}^{-1}(x,y) 
  \left( 1 + 2\delta_{z, \T} + 3\gamma_\T J^{-1}(0) \right)
\nn
  &+& \beta \delta(x-y) 3\gamma_\T (u_{\rm FT}^{-1} J^{-1})(0)
\nn
  &+& \delta(x-y) \left( m_\T^2+ \frac{g_{r, \T}}2 J^{-1}(0) 
  + \frac{g_{6, \T}}{2^2 2!} J^{-1}(0)^2 \right)
\ee
where $J^{-1}(0) = J^{-1}(x,x)$ is independent of $x$ by translation
invariance, assuming that we seek a translation invariant solution.
Other solutions could be of interest.

The gap equation (\ref{pgap}) can be solved by the Ansatz
\be
 J(x,y) = A u_{\rm FT}^{-1}(x,y) + B \delta(x-y)
\ee
Inserting the Ansatz results in two transcendental equations for $A,B$
whose solutions depend on the coefficients 
$\beta, \delta_z, \gamma, g_i$.

Basically, the inclusion of the $(\nabla_\mu \Phi)^2 \Phi^2$-term
results in a system of two equations for mass and wave functions
renormalization. In standard $\Phi^4$-theory there is only one
equation for the mass.

\subsection{Self-consistent calculation of the effective action}

For a given blockspin $\Phi$ one chooses a background field
$\phi^s$ which minimizes the full action.
This corresponds to a nonlinear choice of the interpolator $\A$.

The fluctuation integral is performed by a saddle point method with
blockspin-dependent saddle point, but only after  the action is normal
orderd with respect to the fundamental propagator.
If one performs consecutive renormalization group transformations 
the effective action of one step will be the fundamental one of the
next.
So normal ordering with respect to the fundamental
propagator is required anew after each evaluation of the effective action.

This method developed by Griessl, Mack, Palma and Xylander \cite{yoma} 
preserves the stability properties of the Boltzmann factor.

\section{Conclusions}
\label{secconclusion}

We compared three methods to perform a field split of  a scalar field 
into a high and a low frequency part.
We showed that the blockspin method and the hard-soft invariance
fixing method are two complementary generalizations of the split of
the Gaussian measure. 
The blockspin method is a real space method.
It uses projectors to separate the fields and therefore the operators
containing these projectors do not have inverses anymore but only
pseudoinverses.
The hard-soft invariance fixing method is a momentum space method.
It introduces the split as a Pauli-Villar regularization for the hard
field and as a regularization with higher derivatives for the soft
field.
It renders invertible operators.

Both methods could be adapted to preserve the gauge invariance if
applied to an abelian gauge field.
However, the hard-soft invariance fixing method destroyed the
perturbative renormalizability of the fundamental theory
which ruled out its
use for perturbative calculations of the effective action.

The adaption of the blockspin method required an intermediate gauge
fixing as a means of regularization which could be removed at the
end.
A new set of projectors were introduced to split the gauge
transformations themselves into a soft and a hard part.

We applied the blockspin method to scalar electodynamics and presented
two ways to adapt it to the needs of the Higgs field.
One involved gauge field dependent projectors the other nonlinear
operators.

The generalization to finite temperature was done by restricting the
time to a finite range with periodic boundary conditions.
The operators were adapted to finite temperature by
periodization in time.
Within the developed formalism we could mimick dimensional reduction
by choosing a block lattice with a single block in time direction.
The temperature dependence of the block lattice quantities could then
simply be inferred by purely dimensional considerations.

To prepare for perturbative calculations of the effective action we
calculated the Fourier transforms of the various operators.
Here adaption to finite temperature resulted in a discretization of
energy to the Matsubara frequencies.

Applying the blockspin method to a scalar $\phi^4$ action we presented
the expansion in the fluctuation propagator up to second order.
We explained why the UV-divergencies must be temperature independent
and how they are removed by counterterms of the fundamental theory.
We discussed possible resummations to approximate the effective action
non-perturbatively.

\section{Outlook}
\label{secoutlook}

There are several points which still need to be studied.
One of them is the inclusion of non-abelian gauge fields into the
outlined formalism.
To deal with them requires the use of nonlinerar averaging and
interpolation operators.
Closely related to this problem is the question whether an
perturbative expansion in the fluctuation field makes sense.
It might be better to use non-perturbative approximations if one
evaluates the fluctuation integral numerically.
These questions are addressed in \cite{yoma} and are under further study.
 
A next source of trouble is the inclusion of fermions.
In section \ref{secscalar} we have already mentioned that 
by a special choice of the averaging operator 
the blockspin method can work entirely in the continuum, 
but then one cannot use
MC methods to solve the effective action.
One has to think about a hybrid approach. 
First all fields of the theory could be separated into background and
fluctuation fields using the continuum `block'spins.
This would guarantee a consistent implementation of the cutoff.
Fluctuation fields must then be integrated out.
The fermionic background fields could then be handled in the
continuum, 
whereas the other background fields might be blocked to a lattice
without lowering the cutoff further.
It needs careful considerations to show that this way gauge fields and
fermions interacting via minimal coupling are treated consistently.
To facilitate the problem one could start again with an abelian gauge
field, attacking QED this way.

\newpage

\centerline{\bf Acknowledgements}

Its a pleasure to thank my supervisor Gerhard Mack for his patient and
thorough guidance.
I am very much indebted for his constant encouragement and innumerable
discussions.

I would like to thank Markus Grabowski and York Xylander for the many
helpful discussions during the preparation of this work.

Also many thanks to Martin B\"aker and Max Griessl for pointing out
misprints, misthoughts and other mischief.

Financal support from the Deutsche Forschungsgemeinschaft is
gratefully acknowledged.

\newpage

\begin{appendix}

\section{Interpolator for scalar fields}
\label{scalarcalc}

The Ansatz for the interpolation operator
\be
  v^{-1} \A = C^\dagger \tilde{\A}
\ee
removes the mixed kinetic term since $C\phi^h=0$.

The kinetic term for the blockspin is
\be
  \frac12 \langle \Phi,
  \underbrace{\A^\dagger v^{-1}\A }_{=:u^{-1}}\Phi\rangle \quad .
\ee

The definition of $u^{-1}$ helps us to get rid of the auxiliary operator
$\tilde \A$
\be
  u^{-1}:=\A^\dagger v^{-1}\A = \A^\dagger C^\dagger \tilde \A 
  = \tilde \A
\ee
hence
\be
  v^{-1}\A = C^\dagger u^{-1}  
  \quad \mbox{with} \quad
  u^{-1}:=\A^\dagger v^{-1} \A 
  \quad \mbox{and} \quad
  u = CvC ^\dagger \quad .
\label{ainterpol}
\ee

The derivation of the fluctuation propagator is done using the Fourier
representation of the $\delta$-functional
\be
  {Z_0}^h[j] &=&\int  \D \phi^h \delta(C\phi^h) 
  \e^{-\frac12 \langle \phi^h,v^{-1}\phi^h \rangle 
             + \langle \phi^h,j \rangle} 
\nn
  &=& \int \D \phi^h \D \sigma 
  \e^{-\frac12 \langle \phi^h,v^{-1}\phi^h \rangle  
             + \langle \phi^h,j \rangle +i \langle \sigma,C\phi^h \rangle} 
\nn
  &=& \frac1{\sqrt{\det v^{-1}}} \int \D\sigma 
  \e^{-\frac12 \langle (C ^\dagger \sigma-ij),v(C ^\dagger \sigma-ij)\rangle} 
\nn
  &=& \frac1{\sqrt{\det v^{-1}}} \e^{\frac12 \langle j,vj \rangle} 
      \int \D \sigma \e^{-\frac12 \langle \sigma,CvC ^\dagger \sigma \rangle 
                               + i\langle \sigma,Cvj \rangle} 
\nn 
  &=& \frac1{\sqrt{\det v^{-1}}} \e^{\frac12 \langle j,vj \rangle}
      \frac1{\sqrt{\det (CvC^\dagger)}} 
  \e^{-\frac12 \langle j,vC^\dagger(CvC^\dagger)^{-1}Cvj \rangle}
  \quad .
\ee
From eq.(\ref{ainterpol}) we know
\be
  vC^\dagger = \A u
  \quad \quad \mbox{and} \quad \quad
  Cv = u \A^\dagger \quad .
\ee
Using the definition of the blockspin propagator
\be
  u = CvC^\dagger 
\ee
the above expression simplifies to
\be
  {Z_0}^h[j] = \frac1{\sqrt{\det v^{-1}}} \frac1{\sqrt{\det u}}
  \e^{\frac12 \langle j,(v-\A u\A^\dagger)j \rangle} \quad .
\ee

\newpage

\section{Propagators and interpolation kernel in 
the gauge fixed case}
\label{alphabeta}

Given
\be
  v^{-1}_{\alpha\beta} 
  = \Delta +(\alpha-1) \partial \R \partial^\dagger 
           + (\beta-1) \partial (1-\R) \partial^\dagger
\ee
one verifies explicitly that 
\be
  v_{\alpha\beta} = \Delta^{-1} 
  - (1-\frac1\alpha) \partial \Delta^{-1} \R \Delta^{-1} \partial^\dagger
  - (1-\frac1\beta) \partial \Delta^{-1} (1-\R) \Delta^{-1} \partial^\dagger
\ee
is its inverse.

For this one uses the projector properties of $\R$ and $(1-\R)$ and  
\be
  \partial\partial = 0
  \quad \quad &,& \quad \quad
  \partial^\dagger \partial^\dagger = 0
\nn
  \partial \Delta = \Delta \partial
  \quad \quad &,& \quad \quad
  \partial^\dagger \Delta^{-1} = \Delta^{-1} \partial^\dagger \quad .
\ee 
One computes
\be
  v^{-1}_{\alpha\beta}v_{\alpha\beta} &=&
  \Delta \Delta^{-1} 
  - (1-\frac1\alpha) \Delta \partial \Delta^{-1} 
       \R \Delta^{-1} \partial^\dagger
  - (1-\frac1\beta) \Delta \partial \Delta^{-1} 
      (1-\R) \Delta^{-1} \partial^\dagger
\nn
  &+& (\alpha-1) \partial \R \partial^\dagger \Delta^{-1}  
   - (\alpha-1) (1-\frac1\alpha) \partial \R \partial^\dagger
                 \partial \Delta^{-1} \R \Delta^{-1} \partial^\dagger
\nn
  &-& (\alpha-1) (1-\frac1\beta) \partial \R \partial^\dagger
                 \partial \Delta^{-1} (1-\R) \Delta^{-1} \partial^\dagger
\nn
  &+&  (\beta-1) \partial (1-\R) \partial^\dagger \Delta^{-1} 
   -  (\beta-1) (1-\frac1\alpha) \partial (1-\R) \partial^\dagger
                 \partial \Delta^{-1} \R \Delta^{-1} \partial^\dagger
\nn
  &-& (\beta-1) (1-\frac1\beta) \partial (1-\R) \partial^\dagger
                 \partial \Delta^{-1} (1-\R) \Delta^{-1} \partial^\dagger
\nn
  &=& 1 - (1-\frac1\alpha) \partial \R \Delta^{-1} \partial^\dagger
  - (1-\frac1\beta) \partial (1-\R) \Delta^{-1} \partial^\dagger
\nn
  &+& (\alpha-1) \partial \R \partial^\dagger \Delta^{-1}
   + (\beta-1) \partial (1-\R) \partial^\dagger \Delta^{-1}
\nn
  &-&  (\alpha-1) (1-\frac1\alpha) \partial \R \Delta^{-1} \partial^\dagger
  -  (\beta-1) (1-\frac1\beta) \partial (1-\R) \Delta^{-1} \partial^\dagger
\nn
  &=& 1 + \underbrace{[(\alpha-1) - (1-\frac1\alpha) 
  - (\alpha-1) (1-\frac1\alpha)]}_{0} 
  \partial \R \Delta^{-1} \partial^\dagger
\nn
  &+&\underbrace{[(\beta-1) - (1-\frac1\beta) - (\beta-1) (1-\frac1\beta)]}_{0}
  \partial (1-\R) \Delta^{-1} \partial^\dagger 
\nn
  &=& 1 \quad .
\ee
The block propagator is
\be
  u_{\alpha\beta} =  C v_{\alpha\beta} C^\dagger
  = C \Delta^{-1} C^\dagger
  - (1-\frac1\beta)
  \nabla \underbrace{C_S \Delta^{-2} C_S^\dagger}_{u_{\Delta^2}} \nabla^\dagger
  \quad .
\ee
We use the abbreviations
\be
  () &=& (C\Delta^{-1} C^\dagger)
\nn
  { [] } &=& [\nabla^\dagger ()^{-1} \nabla]
\nn
  \{_\beta\} &=& \{ []^{-1} 
            - (1-\frac1\beta)^{-1} []^{-1} u_{\Delta^2}^{-1} []^{-1} \}
\ee 
with the identities
\be
  \left( [] \{_\beta\} -1 \right) 
  &=& 1 - (1-\frac1\beta)^{-1} u_{\Delta^2}^{-1} []^{-1} - 1
\nn
  &=& - (1-\frac1\beta)^{-1} u_{\Delta^2}^{-1} []^{-1} 
\label{aid1}
\ee
and
\be
  []^{-1} \{_\beta\}^{-1} 
  &=& 1 + \left( []^{-1} - \{_\beta\} \right) \{_\beta\}^{-1}
\nn
  &=& 1 + (1-\frac1\beta)^{-1} []^{-1} u_{\Delta^2}^{-1} []^{-1} 
                                                 \{_\beta\}^{-1} \quad .
\label{aid2}
\ee
The inverse block propagator is
\be
  u^{-1}_{\alpha\beta}  
  &=& \underbrace{(C \Delta^{-1} C^\dagger)^{-1}}_{()^{-1}}
  + (1-\frac1\beta) ()^{-1} \nabla \{ u_{\Delta^2}^{-1} - (1-\frac1\beta)^{-1}
  \underbrace{[\nabla^\dagger ()^{-1} \nabla]}_{[]} \}^{-1} 
  \nabla^\dagger ()^{-1}
\nn
  &=& ()^{-1} - ()^{-1} \nabla []^{-1}
  \underbrace{\{[]^{-1} - (1-\frac1\beta)^{-1} []^{-1}
  u_{\Delta^2}^{-1}[]^{-1}\}^{-1}}_{\{_\beta\}^{-1}} []^{-1} 
                                        \nabla^\dagger ()^{-1} \quad .
\ee
This is again verified explicitly
\be
  u_{\alpha\beta} u^{-1}_{\alpha\beta}  
  &=& \{ () - (1-\frac1\beta) \nabla u_{\Delta^2} \nabla^\dagger \}
  \{ ()^{-1} - ()^{-1}\nabla[]^{-1} \{_\beta\}^{-1}  []^{-1}
                                                  \nabla^\dagger ()^{-1} \} 
\nn
  &=& 1 + \nabla \{ - (1-\frac1\beta) u_{\Delta^2} 
             - []^{-1} \{_\beta\}^{-1}  []^{-1} 
  + (1-\frac1\beta) u_{\Delta^2} 
  \underbrace{\nabla^\dagger ()^{-1} \nabla}_{[]} []^{-1} \{_\beta\}^{-1}
  []^{-1} \} \nabla^\dagger ()^{-1}
\nn
  &=& 1 + \nabla \{ - (1-\frac1\beta) u_{\Delta^2} 
      ( 1 - \{_\beta\}^{-1} []^{-1} )
  - []^{-1} \{_\beta\}^{-1}  []^{-1} \} \nabla^\dagger ()^{-1}  
\nn
  &=& 1 + \nabla \{ - (1-\frac1\beta) u_{\Delta^2} 
  \underbrace{([] \{_\beta\} -1)}_{-(1-\frac1\beta)^{-1} 
   u_{\Delta^2}^{-1} []^{-1}}
  - []^{-1} \} \{_\beta\}^{-1}  []^{-1} \nabla^\dagger ()^{-1} 
\nn
  &=& 1 \quad .
\ee
Now we come to the interpolation kernel
\be
  \A_{\alpha\beta} &=& v_{\alpha\beta} C^\dagger u^{-1}_{\alpha\beta} 
\nn
  &=& (\Delta^{-1} - (1-\frac1\beta) \partial \Delta^{-2} \partial^\dagger)
  C^\dagger u^{-1}_{\alpha\beta} 
\nn
  &=& (\Delta^{-1} C^\dagger - (1-\frac1\beta) \partial \Delta^{-2} 
  C_S^\dagger \nabla^\dagger) u^{-1}_{\alpha\beta} 
\nn
  &=& \Delta^{-1} C^\dagger u^{-1}_{\alpha\beta}  
  - (1-\frac1\beta) \partial \Delta^{-2} C_S^\dagger \nabla^\dagger
  [ ()^{-1} - ()^{-1} \nabla []^{-1} \{_\beta\}^{-1} []^{-1} 
  \nabla^\dagger ()^{-1}]
\nn
  &=& \Delta^{-1} C^\dagger u^{-1}_{\alpha\beta} 
  - (1-\frac1\beta) \partial \Delta^{-2} C_S^\dagger
  [1 - \{_\beta\}^{-1} []^{-1}] \nabla^\dagger ()^{-1}
\nn
  &=& \Delta^{-1} C^\dagger u^{-1}_{\alpha\beta} 
  - (1-\frac1\beta) \partial \Delta^{-2} C_S^\dagger
  \underbrace{[ [] \{_\beta\} - 1]}_{-(1-\frac1\beta) 
  u_{\Delta^2}^{-1} []^{-1}}
  \{_\beta\}^{-1} []^{-1} \nabla^\dagger ()^{-1}
\nn
  &=& \Delta^{-1} C^\dagger u^{-1}_{\alpha\beta} 
  + \partial \Delta^{-2} C_S^\dagger u_{\Delta^2}^{-1}[]^{-1}
  \{_\beta\}^{-1} []^{-1} \nabla^\dagger ()^{-1}  \quad .
\ee
The background field propagator is
\be
  v^s_{\alpha\beta} 
  &=& \A_{\alpha\beta} C v_{\alpha\beta}
  = v_{\alpha\beta} C^\dagger u^{-1}_{\alpha\beta} C v_{\alpha\beta}
\nn
  &=&\{ \Delta^{-1} - (1-\frac1\beta) \partial \Delta^{-2} \partial^\dagger \}
  C^\dagger u^{-1}_{\alpha\beta} C
  \{ \Delta^{-1} - (1-\frac1\beta) \partial \Delta^{-2} \partial^\dagger \}
\nn
  &=& (1-\frac1\beta)^2 \partial \Delta^{-2} C_S^\dagger
   \left( [] \{_\beta\} -1 \right) 
   \{_\beta\}^{-1} C_S \Delta^{-2} \partial^\dagger
\nn
  &-& (1-\frac1\beta) \partial \Delta^{-2} C_S^\dagger
   \left( [] \{_\beta\} -1 \right) \{_\beta\}^{-1} []^{-1} 
    ()^{-1} \nabla^\dagger C \Delta^{-1}
\nn
  &-& (1-\frac1\beta) \Delta^{-1} C^\dagger \nabla ()^{-1}  
   \left( [] \{_\beta\} -1 \right) 
   \{_\beta\}^{-1} []^{-1} C_S \Delta^{-2} \partial^\dagger
\nn
  &+& \Delta^{-1} C^\dagger 
  \left( ()^{-1} 
 - \nabla ()^{-1} []^{-1} \{_\beta\}^{-1} []^{-1} 
  ()^{-1} \nabla^\dagger \right) C \Delta^{-1}  \quad .
\ee
Using the identity (\ref{aid1})
we see that only the first summand is divergent for $\beta = 0$.
\be
  v^s_{\alpha\beta} &=& - (1-\frac1\beta) \partial \Delta^{-2} C_S^\dagger
    u_{\Delta^2}^{-1} []^{-1} \{_\beta\}^{-1} C_S \Delta^{-2} \partial^\dagger
\nn
  &+& \partial \Delta^{-2} C_S^\dagger
    u_{\Delta^2}^{-1} []^{-1} \{_\beta\}^{-1} []^{-1} 
    \nabla^\dagger ()^{-1} C \Delta^{-1}
\nn
  &+& \Delta^{-1} C^\dagger ()^{-1} \nabla  
  []^{-1} \{_\beta\}^{-1} []^{-1} u_{\Delta^2}^{-1} 
  C_S \Delta^{-2} \partial^\dagger
\nn
  &+& \Delta^{-1} C^\dagger 
  \left( ()^{-1} 
  - ()^{-1} \nabla []^{-1} \{_\beta\}^{-1} []^{-1} 
  \nabla^\dagger ()^{-1} \right)
  C \Delta^{-1}
\ee
We use the identity (\ref{aid2}) to separate the divergent term in
the first line and obtain
\be
  v^s_{\alpha\beta} = &-& (1-\frac1\beta) \partial \Delta^{-2} C_S^\dagger
    u_{\Delta^2}^{-1} C_S \Delta^{-2} \partial^\dagger
\nn
  &-&  \partial \Delta^{-2} C_S^\dagger 
    u_{\Delta^2}^{-1} []^{-1} \{_\beta\}^{-1} []^{-1} u_{\Delta^2}^{-1}
    C_S \Delta^{-2} \partial^\dagger
\nn
  &+& \partial \Delta^{-2} C_S^\dagger
    u_{\Delta^2}^{-1} []^{-1} \{_\beta\}^{-1} []^{-1} 
    \nabla^\dagger ()^{-1} C \Delta^{-1}
\nn
  &+& \Delta^{-1} C^\dagger ()^{-1} \nabla 
  []^{-1} \{_\beta\}^{-1} []^{-1} u_{\Delta^2}^{-1} 
  C_S \Delta^{-2} \partial^\dagger
\nn
&-& \Delta^{-1} C^\dagger 
  ()^{-1} \nabla []^{-1} \{_\beta\}^{-1} []^{-1} \nabla^\dagger ()^{-1} 
  C \Delta^{-1}
\nn
  &+& \Delta^{-1} C^\dagger ()^{-1} C \Delta^{-1}
\nn
  = &-& (1-\frac1\beta) \partial \Delta^{-1} (1 - \R) 
                      \Delta^{-1} \partial^\dagger
\nn
  &-& \left( \partial \Delta^{-2} C_S^\dagger u_{\Delta^2}^{-1}
        - \Delta^{-1} C^\dagger ()^{-1} \nabla \right)
           []^{-1} \{_\beta\}^{-1} []^{-1}
 \left( u_{\Delta^2}^{-1} C_S \Delta^{-2} \partial^\dagger 
         - \nabla^\dagger ()^{-1} C \Delta^{-1} \right)
\nn
  &+& \Delta^{-1} C^\dagger ()^{-1} C \Delta^{-1} \quad .
\ee
Here we finally see that the divergent term is exactly the one already
present in the fundamental propagator.
So it will cancel when the two propagators are subtracted from each
other to construct the fluctuation propagator.

\newpage

\section{Thermalization of concatenated operators}
\label{tempercalc}

Since the difference between zero and finite temperature is only the
extension of the time direction we handle space and time coordinates
in a different way in this appendix.
Space coordinates are supressed whereever possible.
Integration over them is written in symbolic notation.
Time coordinates are always explicitly written, integration over them
is written with limits so that the reader can follow how periodization in
time and restricted and unrestricted time integration conspire to give
finite temperature results.
\be
  & &u_\T(x_4,x'_4) 
\nn
  &=& \sum_{n\in{\menge Z}} u_0(x_4+n\beta,x'_4) 
\nn
  &=& \sum_{n\in{\menge Z}} 
  (C_0 v_0 C^\dagger_0)(x_4+n\beta,x'_4)
\nn
  &=& \int_{{\vec z}} \int_{{\vec z'}} \sum_{n\in{\menge Z}}
  \int_{-\infty}^{\infty} dz_4 \int_{-\infty}^{\infty} dz_4'
    C_0(x_4+n\beta,z_4)
    v_0(z_4,z'_4)
    C^\dagger_0(z'_4,x'_4)
\nn
  &=& \int_{{\vec z}} \int_{{\vec z'}}
  \sum_{n\in{\menge Z}} \sum_{m\in{\menge Z}} \sum_{l\in{\menge Z}}
  \int_0^{\beta} dz_4 \int_0^{\beta} dz_4'
    C_0(x_4+n\beta,z_4+m\beta)
    v_0(z_4+m\beta,z'_4+l\beta)
    C^\dagger_0(z'_4+l\beta,x'_4)
\nn
  &=& \int_{{\vec z}} \int_{{\vec z'}}
  \sum_{n\in{\menge Z}} \sum_{m\in{\menge Z}} \sum_{l\in{\menge Z}}
  \int_0^{\beta} dz_4 \int_0^{\beta} dz_4'
    C_0(x_4+(n-m)\beta,z_4)
    v_0(z_4+(m-l)\beta,z'_4)
    C^\dagger_0(z'_4+l\beta,x'_4)
\nn
  &=& \int_{{\vec z}} \int_{{\vec z'}}
  \sum_{n'\in{\menge Z}} \sum_{m'\in{\menge Z}} \sum_{l\in{\menge Z}}
  \int_0^{\beta} dz_4 \int_0^{\beta} dz_4'
    C_0(x_4+n'\beta,z_4)
    v_0(z_4+m'\beta,z'_4)
    C^\dagger_0(z'_4+l\beta,x'_4)
\nn
 &=& \int_{{\vec z}} \int_{{\vec z'}}
  \int_0^{\beta} dz_4 \int_0^{\beta} dz_4'
  \underbrace    
  {\sum_{n'\in{\menge Z}} C_0(x_4+n'\beta,z_4)}
  _{C_\T(x_4,z_4)}
  \underbrace
  {\sum_{m'\in{\menge Z}} v_0(z_4+m'\beta,z'_4)}
  _{v_\T(z_4,z'_4)}
  \underbrace
  {\sum_{l\in{\menge Z}}C^\dagger_0(z'_4+l\beta,x'_4)}
  _{C^\dagger_\T(z'_4,x'_4)}
\nn
  &=& (C_\T v_\T C^\dagger_\T) (x_4,x'_4) 
\ee
To get from line 5 to line 6 we can use the lattice translation invariance of
$C_0$ and $v_0$ since
$\beta$ is always a multiple of the lattice spacing in time direction.

\newpage

For the calculation of the temperature dependence of the block
propagator we 
use the simple thermalized averaging operator with only one block
in time direction.
Due to the translation invariance of the fundamental propagator $v_0$
one of the time integrations is trivial and gives a factor $\beta$
\be
  u_{\T}(0,0) &=& (C_\T v_\T C^\dagger_\T) (0,0)   
\nn
  &=& \int_{{\vec z}} \int_{{\vec z'}}
      \int_0^\beta dz_4 \int_0^\beta dz'_4 
      \frac1\beta C_{{\rm space}}
      \sum_{n\in{\menge Z}} v_0(z_4+n\beta,z'_4) 
      \frac1\beta C^\dagger_{{\rm space}}  
\nn
  &=& \int_{{\vec z}} \int_{{\vec z'}}
      \frac1\beta C_{{\rm space}} 
      \int_{-\infty}^{\infty} dz_4 \int_0^\beta dz'_4 v_0(z_4,z'_4) 
      \frac1\beta C^\dagger_{{\rm space}}  
\nn
  &=& \int_{{\vec z}} \int_{{\vec z'}}
      \frac1\beta C_{{\rm space}} 
      \int_{-\infty}^{\infty} dz_4 \int_0^\beta dz'_4 v_0(z_4-z'_4,0) 
      \frac1\beta C^\dagger_{{\rm space}}  
\nn
  &=& \int_{{\vec z}} \int_{{\vec z'}}
      \frac1\beta C_{{\rm space}} 
      \int_{-\infty}^{\infty} dz''_4 \int_0^\beta dz'_4 v_0(z''_4,0) 
      \frac1\beta C^\dagger_{{\rm space}} 
\nn
  &=& \frac1\beta \int_{{\vec z}} \int_{{\vec z'}}
      \int_{-\infty}^{\infty} dz_4 
      C_{{\rm space}}({\vec x},{\vec z})  
      \int_{-\infty}^{\infty} dz_4 v_0(z_4,0)
      C^\dagger_{{\rm space}}({\vec z'},{\vec y})
\nn
  &=&: \frac1\beta u_{FT}({\vec x},{\vec y}) \quad .
\ee
The factor $\frac1\beta$ is the only temperature dependence.

We now present the calculation of the finite temperature interpolator
\be
  \A_{\T}(z_4,0) 
  &=& \int_{{\vec z'}} \int_{{\vec x'}}
      \int_0^\beta dz'_4 \beta v_{\T}(z_4,z'_4) 
      \frac1\beta C^\dagger_{{\rm space}}
      \frac1\beta u_{FT}^{-1}
\nn
  &=& \int_{{\vec z'}} \int_{{\vec x'}}
      \int_0^\beta dz'_4 \beta 
      \sum_{n\in{\menge Z}} v_0(z_4,z'_4-n\beta) 
      \frac1\beta C^\dagger_{{\rm space}}
      \frac1\beta u_{FT}^{-1}
\nn
  &=& \frac1\beta \int_{{\vec z'}} \int_{{\vec x'}}
      \int_{-\infty}^{\infty} dz'_4   
      v_0(z_4,z'_4) C^\dagger_{{\rm space}} u_{FT}^{-1}
\nn
  &=& \frac1\beta \int_{{\vec z'}} \int_{{\vec x'}}
      \int_{-\infty}^{\infty} dz'_4 
      v_0(0,z'_4) C^\dagger_{{\rm space}} u_{FT}^{-1}
\nn
  &=& \A_{\T}(0,0) =: \frac1\beta \A_{FT} \quad .
\ee

Again the factor $\frac1\beta$ is the only temperature dependence.

\newpage

\section{The effective potential for $\phi^4$-theory up
to second order in the fluctuation propagator}
\label{max}

\begin{eqnarray*}
  V_{\rm eff}[\phi^s]  
  &=& \frac12 m_0^2 \int_z  \phi^s(z)^2
    + \frac g {4!} \int_z \phi^s(z)^4
\nn
  &-& \frac12 m_0^4 \int_{z_1,z_2} \phi^s(z_1)\Gamma(z_1,z_2)\phi^s(z_2) 
    + \frac g4 \int_{z} \phi^s(z)^2\Gamma(z,z) 
\nn 
  &-& \frac12 m_0^2 g \int_{z_1,z_2}   
      \phi^s(z_1)\Gamma(z_1,z_1)\Gamma(z_1,z_2)\phi^s(z_2) 
\nn
  &-& \frac14 m_0^2 g \int_{z_1,z_2} \phi^s(z_1)^2\Gamma(z_1,z_2)^2 
\nn
  &+& \frac12 m_0^6 \int_{z_1,z_2,z_3}   
      \phi^s(z_1)\Gamma(z_1,z_2) \Gamma(z_2,z_3)\phi^s(z_3) 
\nn
  &-& \frac1{3!} m_0^2 g \int_{z_1,z_2} 
      \phi^s(z_1)\Gamma(z_1,z_2)\phi^s(z_2)^3 
\nn
  &-& \frac12 \frac{g^2}{3!}\int_{z_1,z_2}   
    \phi^s(z_1)\Gamma(z_1,z_1) \Gamma(z_1,z_2)\phi^s(z_2)^3 
\nn
  &+& \frac1{3!} m_0^4 g \int_{z_1,z_2,z_3}   
    \phi^s(z_1)\Gamma(z_1,z_2) \Gamma(z_2,z_3)\phi^s(z_3)^3 
\nn
  &-& \frac{g^2}{2^4}\int_{z_1,z_2}   
    \phi^s(z_1)^2\Gamma(z_1,z_2)^2\phi^s(z_2)^2 
\nn
  &+& \frac14 m_0^4 g \int_{z_1,z_2,z_3}   
      \phi^s(z_1)\Gamma(z_1,z_2) \phi^s(z_2)^2 \Gamma(z_2,z_3)\phi^s(z_3) 
\nn
  &-& \frac{g^2}{2\cdot 3!^2} \int_{z_1,z_2}   
      \phi^s(z_1)^3 \Gamma(z_1,z_2) \phi^s(z_2)^3 
\nn
  &+& \frac{m_0^2 g^2}{2\cdot 3!^2} \int_{z_1,z_2,z_3}   
      \phi^s(z_1)^3\Gamma(z_1,z_2) \Gamma(z_2,z_3)\phi^s(z_3)^3 
\nn
  &+& \frac1{12} m_0^2 g^2 \int_{z_1,z_2,z_3}   
      \phi^s(z_1)\Gamma(z_1,z_2) \phi^s(z_2)^2 \Gamma(z_2,z_3)\phi^s(z_3)^3 
\nn
  &+& \frac{g^3}{4\cdot 3!^2}\int_{z_1,z_2,z_3}   
    \phi^s(z_1)^3\Gamma(z_1,z_2) \phi^s(z_2)^2 \Gamma(z_2,z_3)\phi^s(z_3)^3 
\nn
  &+& \frac12 m_0^2 \int_{z} \Gamma(z,z)
    - \frac14 m_0^4 \int_{z_1,z_2} \Gamma(z_1,z_2)\Gamma(z_2,z_1) 
    + \frac18 g \int_{z} \Gamma(z,z)^2 
\end{eqnarray*}

\newpage

\section{Partial integration of nonlocal terms in the effective
  action to obtain a sum of local and irrelevant terms}
\label{local}

Consider a term which is quadratic in the field such as
\be
  I_2[\Phi] = \int_{x_1} \int_{x_2} \rho_2(x_2 - x_1) \Phi(x_2) \Phi(x_1) 
\ee
We are interested in situations where $\rho_2(x)$ falls off
exponentially with decay length one lattice spacing.  In this case the
sum can be rewritten in the form
\be
  I_2[\Phi] = \mu^2 \int_{x_1} \Phi(x_1)^2 
  + z_{\mu\nu} \int_{x_1} \nabla_\mu \Phi(x_1) \nabla_\nu \Phi(x_1)
  + \mbox{irrelevant term}
\label{integral}
\ee
where the irrelevant term is of the form 
\be
  \gamma_{\mu \nu \rho \sigma} \int_{x_1} \int_{x_2} 
  \nabla_\mu \nabla_\nu \Phi(x_1)
  \rho_2^\prime (x_2 - x_1) \nabla_\rho \nabla_\sigma \Phi(x_2) 
\ee
$\rho_2^\prime$ also decays exponentially with distance $x_2-x_1$.
The coefficients are
\be
  \mu^2 &=& \int_x \rho_2(x)  \\
  z_{\mu \nu} &=& - \half \int_x x_\mu x_\nu \rho_2(x)
\ee
Because of the exponential falloff of $\rho_2$, its Fourier transform
$\tilde \rho_2(p)$ is holomorphic in a strip.  Because of the presence
of a lattice, it is a periodic and even function of $p$.  Therefore
\be
  \tilde \rho_2(p)
  = \mu^2 + z_{\mu \nu} \sin p_\mu \sin p_\nu 
  + \gamma_{\mu \nu \rho \sigma}
  \sin p_\mu \sin p_\nu \sin p_\rho \sin p_\sigma 
  \, \tilde \rho^\prime_2(p)
\ee
where $\tilde \rho^\prime_2(p)$ is also holomorphic, periodic and
even.
\be
  \mu^2 &=& \tilde \rho_2(0)  \\
  z_{\mu \nu} &=& \half \frac\partial{\partial p_\mu}
  \frac\partial{\partial p_\nu} \tilde \rho_2(p)|_{p=0}
\ee
Inserting back one gets eq.(\ref{integral}).

\end{appendix}

\end{document}